\documentclass{article}
\usepackage[utf8]{inputenc}
\usepackage{authblk}
\usepackage{setspace}
\usepackage[margin=1.25in]{geometry}
\usepackage{graphicx}
\usepackage{subcaption}
\usepackage{amsmath}
\usepackage{lineno}
\usepackage{color}
\usepackage{braket}
\usepackage{makecell}

\usepackage{dcolumn}
\usepackage{amsfonts}
\usepackage{amssymb}
\usepackage{amsthm}
\usepackage{latexsym}
\usepackage{epsfig}
\usepackage{lscape}
\usepackage{multirow}
\usepackage{bbm}
\usepackage{mathrsfs}
\usepackage{fancyhdr}
\usepackage{lastpage}
\usepackage{soul}
\usepackage{longtable}
\usepackage{amsmath}
\usepackage{booktabs}

\makeatletter
\newcommand{\smalloplus}{\mathbin{\mathpalette\make@small\oplus}}

\DeclareMathOperator{\poly}{poly}

\title{Review on Quantum Walk Computing: \\Theory, Implementation, and Application}

\author[1*]{Xiaogang Qiang}
\author[1]{Shixin Ma}
\author[1]{Haijing Song}

\affil[1]{National Innovation Institute of Defense Technology, AMS, 100071 Beijing, China.}
\affil[*]{Address correspondence to: qiangxiaogang@gmail.com}

\date{}

\begin{document}

\maketitle

\begin{abstract}

Classical random walk formalism shows a significant role across a wide range of applications. As its quantum counterpart, the quantum walk is proposed as an important theoretical model for  quantum computing. By exploiting the quantum effects such as superposition, interference and entanglement, quantum walks and their variety have been extensively studied for achieving beyond classical computing power, and they have been broadly used in designing quantum algorithms in fields ranging from algebraic and optimization problems, graph and network analysis, to quantum Hamiltonian and biochemical process simulations, and even further quantum walk models have proven their capabilities for universal quantum computation. Compared to the conventional quantum circuit models, quantum walks show a feasible path for implementing application-specific quantum computing in particularly the noisy intermediate-scale quantum era. Recently remarkable progress has been achieved in implementing a wide variety of quantum walks and quantum walk applications, demonstrating the great potential of quantum walks. In this review, we provide a thorough summary of quantum walks and quantum walk computing, including aspects of quantum walk theories and characteristics, advances in their physical implementations and the flourishingly developed quantum walk computing applications. We also discuss the challenges facing quantum walk computing, toward realizing a practical quantum computer in the near future. 

\end{abstract}

\section{Introduction}

Quantum computing is a computing model that is based on quantum mechanics, and it utilizes quantum physics effects such as superposition, interference and entanglement, to achieve computational power beyond classical computing. With the great potential of quantum computing, a large number of tasks that may be intractable to classical computers could be solved with exponential or polynomial speedup by quantum algorithms, ranging from factoring a large prime number~\cite{shor1997polynomial} and searching an unsorted database~\cite{grover1996fast} to simulating quantum systems and solving large systems of linear equations. Toward the ultimate goal of building a quantum computer, enormous efforts have been made and significant progress has been achieved. The developments of quantum computing have passed through the stage of a prototype of demonstrating basic quantum operations and small quantum algorithms, and quantum computational advantages have also been experimentally demonstrated on certain tasks like Boson sampling~\cite{aaronson2011computational}. While the universal fault-tolerant quantum computer remains still underway, quantum computing is now in the noisy intermediate-scale quantum (NISQ) era and possesses possibilities for tackling applications of practical interest. 

The category of NISQ computers can include both analogue and digital quantum devices, depending on the employed quantum computing model. There are various quantum computing models, including the quantum circuit model, quantum Hamiltonian evolution model, quantum walk model, quantum annealing model, and measurement-based quantum computation model, on which quantum algorithms and applications can be designed and implemented in quite different ways. These distinct models can hold the capability for universal quantum computing but possess different requirements for their physical implementations. Considering the properties of specific quantum computing models, and the rapid progress in quantum computing hardware, quantum computing for certain applications can be achieved more quickly by exploiting proper models, particularly in the NISQ era.

Quantum walks (QWs) are the quantum mechanical equivalent of classical random walks. A classical random walk is used to describe a particle named a walker that moves stochastically around a discrete space, while in a quantum walk, the walker is governed by quantum physics effects and shows markedly distinct properties compared to a classical random walk. Similar to the classical random walk that forms the basis for designing many classical algorithms from modelling share's prices to describing the random movement of molecules, quantum walks become an important model for designing quantum algorithms and also provide a promising path for implementing quantum computing~\cite{farhi1998quantum,kempe2003quantum,childs2013universal}. Quantum walks have shown beyond classical potentials in a variety of problems in the fields of quantum computing, quantum simulation, quantum information processing and graph-theoretic applications, e.g., database search~\cite{childs2004spatial}, distinguishing graph isomorphism~\cite{douglas2008classical,gamble2010two,berry2011two}, network analysis and navigation~\cite{berry2010quantum,sanchez2012quantum}, and simulating quantum Hamiltonian dynamics~\cite{lloyd1996universal,mohseni2008environment,berry2012black}. Meanwhile, quantum walks on their own have been proven that they can implement universal quantum computing. Quantum walk models also play a key role in demonstrating quantum supremacy over classical computers, according to that a boson sampling task can be viewed as particular instances of multi-particle quantum walk on specific graphs~\cite{aaronson2011computational}, and even for a single-particle quantum walk case, it has been shown that sampling quantum walks evolution probability on an exponentially large circulant graph is classically intractable~\cite{qiang2016efficient}. Moreover, quantum walk dynamics lead to rich and complicated phenomena to investigate~\cite{menssen2017distinguishability} and present the way for exploring various features of different particles, such as investigating multi-fermion quantum walk dynamics~\cite{melnikov2016quantum} and studying braiding statistics via anyonic walks~\cite{brennen_anyonic_2010,lehman2011quantum}.

With the broad interests of quantum walks and their significant potential, nowadays a variety of quantum physical systems are used for physically implementing quantum walks and their applications, such as superconducting, trapped ions and atoms, bulk optics, and integrated photonics systems. These physical implementation approaches are rapidly developing, showing great progress in both the scalability and programmability of quantum walks. For example, the experimental demonstrations of quantum walks have been realized from simple one-dimensional to high-dimensional cases and even complicated structures. The capability of control over the parameters in quantum walk evolutions such as the evolution time, the underlying Hamiltonian, and the evolving particle features is continuously increasing in physical quantum walk experiments. These significant advances in physical implementations of quantum walks allow various complicated quantum walk based applications and have laid a very solid foundation for implementing specialized quantum walk computing systems for applications of practical interest. 

Here, we present a comprehensive review of quantum walks and quantum walk based computing, including the aspects of quantum walk models and theories, their physical implementation, and potential applications. The outline of this review is as follows: in Section 2, we introduce various quantum walk models including discrete-time quantum walks, and continuous-time quantum walks, and describe the important characteristics of quantum walks that are essential for exploring quantum walks and designing quantum algorithms beyond the classical models, and also compare these quantum walk models and discuss their relationships; in Section 3, we summarize the various physical implementations of quantum walk and quantum walk based applications, together with the latest advancements. We provide a comprehensive overview of both the digital physical implementation of quantum walks, i.e., using quantum circuit models, and analogue physical implementation of quantum walks including in solid-state, bulk optics and integrated photonics systems; in Section 4, we present the advancements in the algorithms and potential applications based on quantum walk models, which are mainly described according to four main categories: quantum computing applications, quantum simulation, quantum information processing, and graph-theoretic applications; and finally, in Section 5, we give an outlook on the future of quantum walk computing and discuss the challenges.

\section{Quantum Walk Theories}

In this section, we will introduce various proposed quantum walk models, compare these different quantum walk models, and discuss some essential quantum walk characteristics.

Quantum walks are first proposed as the counterpart of classical random walks, mainly including discrete-time quantum walks that evolve step by step, and the continuous-time quantum walk evolving continuously with time.
Besides these original models, additions to this field have been disclosed to address more complex scenarios, such as a larger class of discrete models that can perform quantum walk evolution on directed graphs, non-unitary quantum walk for performing evolution in an open quantum system, and discontinuous quantum walks that 
combine both discrete-time and continuous-time quantum walks, and so on. 

There are obvious differences between these models, but relationships between them have also gained much care in recent works, which are reflected in transformation and similar limit distribution between discrete and continuous models, direct equivalence among different discrete models and generalization ability of non-unitary models for describing both quantum and classical random walks.

In various characteristics, quantum walks represent different from classical random walks, mainly involving evolution characteristics and sampling complexity. On the one hand, quantum walks can accelerate the evolution process, especially making a difference in the searching algorithm, and have distinctive probability distribution leading to particular localization. On the other hand, sampling of quantum walks has been certified hard in classical simulation, such as single particle quantum walk composed of IQP circuits and multi-particle quantum walk, one relevant example of which is boson sampling.

\begin{figure}[h]
    \centering
    \includegraphics[width=1\textwidth]{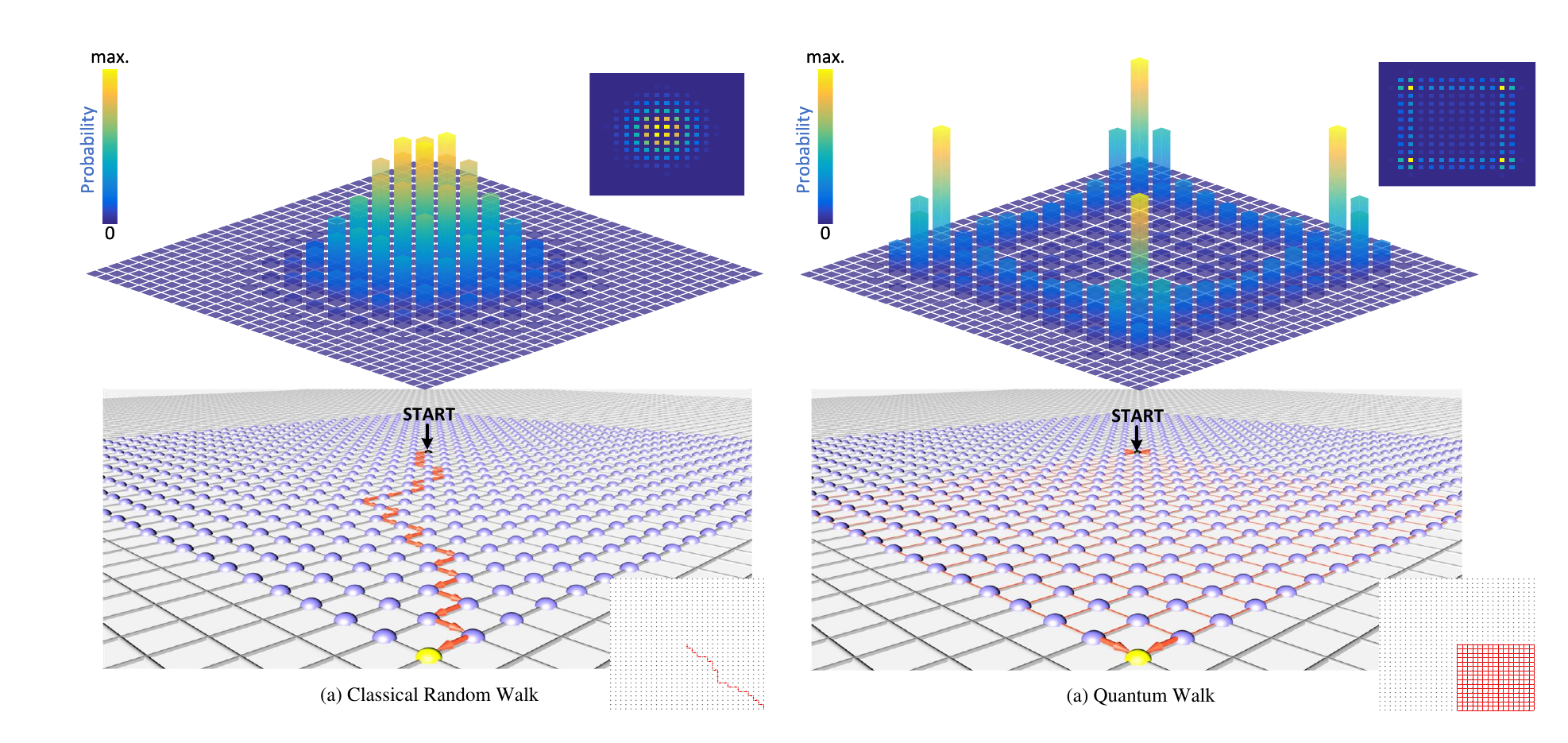}
    \caption{\textbf{Comparison of the probability distributions between the classical random walk and the quantum walk on a large lattice.} The red lines with arrows depict a route that a walker starts from the central point painted black and attains the yellow point after 30 steps. Above the grid is the probability distribution, which represents the possibility that a walker appears in the corresponding lattice site after 15 steps. }
    \label{fig-1}
\end{figure}

\subsection{Quantum walk models}

Most quantum walk models can be divided into two kinds: discrete-time and continuous-time quantum walk. Discrete-time has no time factor and can be realized by several steps, while continuous-time quantum walk can be seen as a special quantum simulation within time $t$ using a graph-related time-independent Hamiltonian. Discontinuous quantum walk composed of several continuous quantum walks, fuses the features of models described above, which provides a hybrid scheme for universal quantum computation. 
Besides these unitary models, two non-unitary models are also covered here. 

\begin{table}[htbp]
    \centering
    \footnotesize
    \caption{A list of random walk models.}\label{tab-1}
    \renewcommand\arraystretch{1.5}
    \begin{tabular}{llll}
    \toprule
     Classical random walk models& \multicolumn{2}{c}{Quantum random walk models} \\
     \midrule
     \multirow{7}{*}{Discrete-time Markov chain}
     & \multirow{3}{*}{Coin-based DTQW}& Hadamard walk~\cite{liang2022hadamard}    \\
     & & Grover walk~\cite{di2011mimicking, ambainis2004coins}  \\
     & & Lackadaisical quantum walk~\cite{wong2018faster, giri2020lackadaisical}\\
     \Xcline{2-3}{0.5pt}
     & \multirow{2}{*}{Other discrete QWs} & Szegedy quantum walk~\cite{szegedy2004quantum}    \\
     & & Staggered quantum walk~\cite{portugal2016staggered-0, portugal2016staggered-1} \\
     \Xcline{1-3}{0.5 pt}
     \multirow{2}{*}{Continuous-time Markov chain} & \multicolumn{2}{c}{Continuous-time quantum walk~\cite{childs2003exponential}}  \\
     &\multicolumn{2}{c}{Discontinuous quantum walk~\cite{christandl2004perfect}}\\
     \Xcline{1-3}{0.5 pt}
     \multicolumn{3}{c}{Quantum stochastic walk~\cite{whitfield2010quantum}} \\
     \Xcline{1-3}{0.5 pt}
     \multicolumn{3}{c}{Open quantum walk~\cite{attal2012open}}\\
     \bottomrule
    \end{tabular}
\end{table}

\subsubsection{Discrete quantum walk models}\label{sec-2.2}

Discrete quantum walk models incorporate a large class of models, where coin-based models, and coinless models consist
ing of Szegedy quantum walk and staggered quantum walk, can be easily distinguished depending on whether there is a coin or not.

\textbf{(a) {coin-based quantum walk}}

The DTQW generally refer to the coin-based quantum walk, which was first proposed as the counterpart of the classical random walk~\cite{aharonov1993quantum}, as both of them are realized by two procedures, namely flipping a coin and moving in the direction decided by the flipping result. Differently, coin-based quantum walk needs two separate spaces to accomplish the above processes, namely coin space $\mathcal{H}_C$ and shift space $\mathcal{H}_S$. In Hilbert space $\mathcal{H}_C\otimes\mathcal{H}_S$, a quantum state $|\phi\rangle$ is defined to describe the probability of a walker arriving at a position, which can be a computational state or a superposition state where every computational state represents one position at one direction or coin side. Due to the superposition and entanglement of different positions, it's shown that this model has a different distribution~\cite{omar2006quantum,tregenna2003controlling} and a quadratically faster spread~\cite{ambainis2001one,kempe2003quantum} compared with the classical model.

Define two unitary transformations, namely the coin operator $C\otimes I_S$ and the shift operator $S$, then it follows that the coin-based quantum walk can be described by a unitary
\begin{equation}
    U = S(C\otimes I_S),
\end{equation}
so the state of the walker will change to $U^n|\phi\rangle$ from the intial state $|\phi\rangle$ after $n$ steps~\cite{aharonov1993quantum}.

The most simple example is about the DTQW on the line, on which occasion the walker uses two coin states $|\uparrow\rangle$ and $\downarrow\rangle$ to represent the left or right direction, so the shift operator can be defined by
\begin{equation}\label{eq-5}
    S = |\uparrow\rangle\langle\uparrow\rangle\otimes\sum_i|i-1\rangle\langle i|+|\downarrow\rangle\langle\downarrow|\otimes\sum_i|i+1\rangle\langle i|.
\end{equation}

Coin operators make a great difference for DTQW models, and its common selection includes Hadamard coin $C_H$ or Grover coin $C_G$, which is separately defined by
\begin{equation}\label{eq-6}
    C_H = \frac{\sqrt{2}}{2}\left[ {\begin{array}{*{20}{c}}
        1&1\\
        1&{ - 1}
        \end{array}} \right], 
    C_G = \left[ {\begin{array}{*{20}{c}}
        0&1\\
        1&{ 0}
        \end{array}} \right].
\end{equation}

DTQW on the line has interesting quantum effects~\cite{nayak2000quantum} and more about this can be found in Refs~\cite{nayak2000quantum,konno2008quantum,chandrashekar2008optimizing}. Moreover, Eq. (\ref{eq-6}) can also be extended to $N$-dimensional version, namely
\begin{equation}\label{eq-7}
    C_H'=\otimes_{i=1}^{\log N} H_C,
\end{equation}

\begin{equation}
    C_G'= \left[ {\begin{array}{*{20}{c}}
        {1 - \frac{2}{N}}&{\frac{2}{N}}&{\dots}&{\frac{2}{N}}\\
        {\frac{2}{N}}&{1 - \frac{2}{N}}&{\dots}&{\frac{2}{N}}\\
        {\vdots}&{\vdots}&{\ddots}&{\vdots}\\
        {\frac{2}{N}}&{\frac{2}{N}}&{\dots}&{1 - \frac{2}{N}}
        \end{array}} \right].
\end{equation}

Generally, the coin-based models have covered most discrete models, which can be easily defined by the coin operators. Taking the Hadamard walk as an example, the Hadamard gate is used as this model's coin operator~\cite{liang2022hadamard}. As is shown in the Table~\ref{tab-1}, the Grover search is also a coin-based model, which is equivalent to the discrete-time quantum walk with the Grover coin~\cite{ambainis2004coins}. Table~\ref{tab-1} also lists the Lackadaisical quantum walk as the coin-based kind because this model still contains the coin operator and shift operator but adds self-loops on the positions or vertices~\cite{wong2018faster}.

In Figure~\ref{fig-1}, we compare the classical and quantum models on the two-dimensional lattice and depict a road where a walker goes from the central point to the endpoint on one diagonal of the lattice by the numerical simulation. It's seen that the walker can only select one classical road while another is the coherence of all possible roads in the quantum case where the coin operator is a $4\times 4$ large matrix of formalism Eq. (\ref{eq-7}) and the shift operator is analogous to Eq. (\ref{eq-5}) but involves the four directions, up, down, left and right.

\textbf{(b) Szegedy quantum walk}

Inspired by the discrete-time Markov chain (time-homogeneous), Szegedy~\cite{szegedy2004quantum} proposed its quantum version with the result of faster searching for the marked vertex of one graph in 2004. After that, this quantum walk model, called Szegedy quantum walk (SQW), spreads widely its application around the problems about the certain graphs, such as the complete graph~\cite{loke2017efficient}, the directed graph~\cite{santha2008quantum,qiang2016efficient} and so on.

Considering a time-homogeneous Markov chain during a period of discrete time $n$, Eq. (\ref{eq-4}) can be modified by $p_j(n)=\sum_iP_{ij}(n)p_i(0)$. Originating from this, we can define the unitary evolution operator as a one-step coinless discrete-time quantum walk, which determines a rule that one walker moves on an enlarged symmetric bipartite graph.

This enlarged graph should have two equivalent parts $A$ and $B$, both of which consist of the all vertices of the original graph. Let 
$i\in A$ and $j\in B$, then the $p_{ij}$ is the transition probability of which walker goes from vertex $i$ to vertex $j$, and the inverse process is expressed as $q_{ij}$. The evolution operator is defined as follows:
\begin{equation}
    U=W_BW_A,
\end{equation}
where 
$$\left\{ {\begin{array}{*{20}{c}}
{W_A=2\sum_{i\in A}|\phi_i\rangle\langle\phi_i|-I}\\
{W_B=2\sum_{j\in B}|\psi_j\rangle\langle\psi_j|-I}
\end{array}} \right.$$
 with $\phi_i=|i\rangle\otimes\sum_j\sqrt{p_{ij}}|j\rangle$ and $\psi_j=\sum_i\sqrt{q_{ji}}|i\rangle\otimes|j\rangle$.

 To clearly see this model's diffusion effect on single-node state, assuming that the initial state is $|i_0, j_0\rangle$, then after $W_A$ evolution, it is straightforward to get that
 \begin{equation}
    2\sqrt{p_{i_0j_0}}|i_0\rangle\otimes\left(\sum_{j\neq j_0}\sqrt{p_{i,j}}|j\rangle\right)+(2p_{i_0j_0}-1)|i_0\rangle\otimes|j_0\rangle,
 \end{equation}
where the first part means the walker moves to the other position with probability $p_{i_0j_0}p_{ij}$, which is the same as the classical Markov process despite a global factor $p_{i_0j_0}$. This also implies that Szegedy quantum walk is a quantization of classical random walk.
Of course, if the superposition state were selected as the initialization, interference would arise between different paths, which contributes to the main reason that this model performs quadratically faster than classical models in some scenarios~\cite{magniez2007search,santha2008quantum}. A further introduction to the properties of Szegedy quantum walk can be found in Ref~\cite{portugal2016staggered-0,portugal2013quantum}.

\textbf{(c) Staggered quantum walk}

The staggered quantum walk is fully introduced by Protugal in Refs~\cite{portugal2016staggered-0,portugal2016staggered-1}. In this model, a graph is split into several tessellations, where any tessellation is made up of some polygons and the union of these polygons in one tessellation surrounds all vertices of the graph. By constructing a reflection operator associated with one of these tessellations, the walker is restricted to a one-step motion in polygons of one tessellation, so a staggered movement, defined by these reflection operators of tessellations in several steps, can spread all paths in one graph~\cite{portugal2016staggered-0}. The staggered model can be used for the search problem if a separate reflection operator, which is related to a partial tessellation with the marked nodes, is interleaved with the model. The property of the staggered model can be studied by spectral analysis in the same way as the Szegedy model, and especially, the two-tessellate case has been discussed in Refs~\cite{konno2018spectral,konno2018partition}.
Except for using the reflection operators directly, taking the reflection operator as a Hamiltonian provides an alternative idea for staggered models~\cite{portugal2017staggered,portugal2013quantum,coutinho2019discretization}, which is more friendly for the physical implementation.
More properties about the staggered model have been studied in recent works, such as the spectral characteristic of the evolution operator in two-tessellate case~\cite{konno2018spectral}, model robustness for noise like decoherence phenomenon~\cite{santos2022decoherence}.

\subsubsection{Continuous-time quantum walk}\label{sec-2.1}

The idea of continuous-time quantum walk (CTQW) also originates from its classical counterpart, which is called the Markov chain (or one-dimensional random walk) of discrete-time in countable state space~\cite{childs2003exponential,chandrashekar2010relationship}. According to Kolmogorov backward equations, $\frac{d P_{ij}(t)}{dt}=\sum_{j=1}^N H_{ik}P_{kj}(t)$ holds for the transition probability $P_{ij}(t)$ and transition rate $H_{ij}$ to describe the jumping process instead of diffusion process when a state $i$ changes to the state $j$. For an undirected graph $G$ of $N$ vertices, we can define 
\begin{equation}
    H_{ij}=\left\{ {\begin{array}{*{20}{ll}}
\gamma &\text{if vertices $i$ and $j$ are connected}, \\
0 &\text{if vertices $i$ and $j$ are not connected},\\
-\gamma d_i &\text{if $i=j$}
\end{array}} \right.
\end{equation}
where $\gamma$ represents the hopping rate, and $d_i$ is the degree of the vertex $i$, namely a summation of connected edges.

Let $p_i(t)$ be the probability at the position of vertex $i$ at time $t$, then, using $p_i(t)=P_{ij}(t)p_j(0)$ where $\sum_{i=1}^Np_i(t)=1$ preserves, it follows that

\begin{equation}\label{eq-4}
    \frac{dp_i(t)}{dt}=\sum_{j=1}^N H_{ij} p_j(0).
\end{equation}

For a quantum particle (or quantum walker) placed on $G$, a superposition state $|\psi(t)\rangle$ is used to describe its position with probability
$p_j(t)=|\langle\psi(t)|j\rangle|^2$ where the orthonormal basis $\left\{ {\left| 1 \right\rangle ,\left| 2 \right\rangle , \dots ,\left| N \right\rangle } \right\}$ corresponds to different vertices of $G$. Here the imaginary unit $i$ is introduced into the left part of Eq. (\ref{eq-4}), so it can be modified as $i{d|\psi(t)\rangle}/{dt}=H|\psi(0)\rangle$.
The CTQW evolution complying with quantum mechanics are further formulated by
\begin{equation}\label{eq-10}
    |\psi(t)\rangle=e^{-iHt}|\psi(0)\rangle
\end{equation}
where $H=\gamma(D-A)$ with degree matrix $D={\rm diag}(d_1,d_2,\dots,d_N)$ and adjacent matrix $A$ whose entries ${A_{jk}} = 1$ if vertices $j$ and $k$ are connected by an edge in $G$, and ${A_{jk}} = 0$ otherwise. The setting of the Hamiltonian $H$ depends on what it is used for, for example, we set $H=A$ or $H=D-A$ for graph isomorphism problems and use $H=|\omega\rangle\langle\omega|-A$ as a search Hamiltonian when searching a marked vertice expressed as an oracle Hamiltonian $|\omega\rangle\langle\omega|$~\cite{qiang_implementing_2021}. 

It is found that CTQW has computational speedup power in some particular graphs over classical algorithms. The first example is regarding the graph traversal problems known as glued-tree graphs, where Childs et al. proposed a black-box quantum scheme to solve it in polynomial queries and no classical case can give a solution in subexponential time~\cite{childs2003exponential}. Other examples like spatial search have also been proven a quadratic advantage~\cite{Chakraborty2020Finding,Apers2022Quadratic} if we use CTQW methods, not only for searching a marked node but also for finding several vertices in graphs.

\subsubsection{Discontinuous quantum walk}

Discontinuous means a fusion of both discrete-time and continuous-time, which infers a model that exhibits a discrete transition from node to node via continuous-time evolution. The first introduction of the discontinuous model was used for devising a scheme of implementing any universal gate~\cite{Underwood_universal_2010} by way of chosen perfect state transfers (PSTs)~\cite{christandl2004perfect}.

Every step of a discontinuous quantum walk can be seen as a continuous-time quantum walk as defined in Eq. (\ref{eq-10}), namely  
\begin{equation}
    |v_2\rangle = e^{-iAt}|v_1\rangle,
\end{equation}
which implements a CTQW from $|v_1\rangle$ to $|v_2\rangle$.
A PST is achieved if $v_1$ and $v_2$ mean a pair of vertices in a graph. The discontinuous model makes it much easier to employ a PST between non-adjacent vertices by splitting a CTQW of an entire graph into several small-scale CTQWs on subgraphs, and further, several PSTs on vertices of well-designed subgraphs can perform a universal quantum operation~\cite{Underwood_universal_2010}.

\subsubsection{Non-unitary quantum walks}

Different from the above unitary models, the following two models are non-unitary and act on an open quantum system. 
The first formalism of the non-unitary quantum walk is a continuous-time model, namely stochastic quantum walk. This model defines a quantum stochastic process described by density operators instead of the probability amplitudes of quantum states~\cite{whitfield2010quantum}, where the Lindblad master equation is used as a general type of evolution that involves a set of axioms about all possible transition processes of connected vertices, so it can express more random walks besides classical Markov process and quantum walk of the continuous time. Nevertheless, this model may have difficulty in experimental realization in near-quantum devices, so a relevant approach to circumvent this limitation has been proposed in Ref~\cite{govia2017quantum}. Due to its advantages to describe the evolution of the open system, both the decoherence and coherence process, multiple applications for this model can be found in near reports, like analyzing the transport process in Photosynthesis~\cite{dudhe2022testing}, Pagerank~\cite{dudhe2022resolving}, and function approximation, data classification and otherwise~\cite{wang2022implementation}. 

Analogously, there is also one kind of discrete-time model for the open quantum system, namely the open quantum walk model~\cite{attal2012open}, which is derived directly from the formalism of the operator-sum representation. This model sparked some discussions about quantum Markov process~\cite{dhahri2019open,dhahri2019quantum,souissi2023structure,kang2023markov}, and studies in its properties, such as limit distribution~\cite{konno2013limit,attal2015central}, hitting time~\cite{lardizabal2016open}, and so on.

\subsection{Quantum walk model comparisons}

There clearly exist differences and relationships among different random walks. Here we make a detailed analysis of the difference and relation between original discrete-time and continuous-time quantum walk, the interchangeability of various discrete quantum walk models, including coin-based quantum walk, Szegedy quantum walk and staggered quantum walk, and how non-unitary models unify both the classical random walk and parts of quantum walk models.

 Quantum walk models have been studied rapidly in recent years, but two original branches of discrete-time and continuous-time quantum walk models still matter in this field. No doubt there exists an obvious difference because the coin-based discrete model (expressed as DTQW later) has an extra space while the continuous-time model does not. Some other interesting differences should not be ignored. First, the DTQW needs different numbers of quantum states when performing a quantum walk on a graph. For an arbitrary graph, DTQW needs to introduce quantum states for each edge of every vertex, while CTQW needs only to introduce quantum states for each vertex. Second, it is feasible for DTQW to regulate the coin of each step while the CTQW cannot but control evolution by modifying Hamiltonian. So, in the design of algorithms, we should select an appropriate model according to the specific situations.

 \begin{figure}[!t]
    \centering
    \includegraphics[width=1\textwidth]{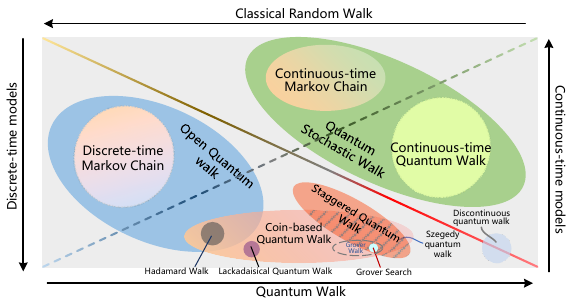}
    \caption{\textbf{Relationship between random walk models.} Arrows are used to distinguish the models of different classes such as discrete and continuous, classical and quantum, as well as their combinations. Every arrow points to the right-angle vertex of the triangle area which represents the covering range of the corresponding model.}
    \label{fig-2}
\end{figure}

 Despite the significant distinctions listed above, finding the underlying relation between the two models is a fundamental problem. It has been shown that their spreading properties are quite similar like faster diffusion than classical models~\cite{childs2003exponential,ambainis2004coins}, and the long-time limit of the DTQW with a certain coin operator is approximately equivalent to the CTQW case in terms of the probability density~\cite{konno2002quantum,grimmett2004weak,konno2005limit,konno2005new}. Moreover, in various applications, both DTQW and CTQW algorithms work, holding the same performance for many algorithms, such as search~\cite{childs2004spatial, lovett2012spatial}, graph isomorphism~\cite{douglas2008classical,rudinger2013comparing}, and they both can be used for universal computation~\cite{childs2009universal, lovett2010universal}. Due to these similarities, some studies are more concerned about how to convert between them, such as in Ref~\cite{strauch2006connecting}, Strauch first investigated which case of coin-based or coinless discrete QW models can be transformed into the CTQW by taking the limit of certain parameters, and blazed a way of how to generalize such a transformation to the higher dimensional walks. A more specific work can be seen in Ref~\cite{childs2010relationship}, where Childs first showed a procedure for constructing a Szegedy quantum walk to simulate the CTQW with any Hamiltonian. There is a similar train of thought that the Hamiltonian of CTQW can be discretized by the commuting adjacent matrices under one graph's partition, where the evolution based on one adjacent matrix constitutes one step of the staggered model with Hamiltonians, which is applied to numerous graphs, such as Cayley graph, as stated in Ref~\cite{coutinho2019discretization}.
These demonstrate that the two models are interchangeable to some extent, but note that this interchangeability only means a strict approximation rather than a direct equivalence.

However, direct interchangeability exists among various discrete models, and their relationship has been depicted in Figure~\ref{fig-2}. For the discrete quantum walks, any Szegedy's walk is equivalent to the two-tessellable staggered quantum walk on the line graph of one bipartite graph~\cite{portugal2016staggered-0}, while both of them can also be converted to coin-based quantum walk models on the multigraphs satisfying the regularity and other conditions listed in~\cite{portugal2016establishing}. Conversely, coin-based quantum walk will be regarded as the Szegedy quantum walk on the bipartite graph if the degree of vertices on one side is two, or staggered quantum walk on the line graph if the tesselation covers the perfect matching, which is involved in Ref~\cite{portugal2016establishing} and Ref~\cite{portugal2016staggered-1}, respectively. More generally, any coin-based model using the flip-flop shift operator can be cast into Szegedy's model or staggered model only if the coin operator is an orthogonal reflection despite the graph being non-regular~\cite{portugal2016establishing}, and a special example is the Grover search, which can be seen as two steps of Grover walk~\cite{ambainis2004coins}, one kind of coined-based quantum walk with the Grover coin. Not considering strict conditions about graphs or models, a family of discrete walks regarding two local operators, called partition-based quantum walks~\cite{konno2018partition}, are unitary equivalent, that is, the evolution operators of these models including any Szegedy's model, two-tessellate staggered model and two-step coined model, can be interchangeable except for a unitary multiplier~\cite{portugal2017connecting,konno2018partition}. 

Note that two models for describing walks in an open environment can cover both classical and quantum models. One is the open quantum random walk, which is a new model but also embodies existing well-known models using unitary operators like Hadamard walk and can be changed to the classical models by adding measurements after each step, like homogeneous discrete Markov chain if using well-designed Kraus operators, or non-homogeneous case for general Kraus operators. For the continuous-time case, the stochastic quantum walk introduces an axiomatic formalism by means of a master equation, which can not only recover both classical random walk like continuous-time Markov chain and continuous-time quantum walk as described in Section~\ref{sec-2.1}, but also enlarge the region of continuous-time random walk models, which is shown in Figure~\ref{fig-2}.

\subsection{Quantum walk characteristics}

Quantum mechanics effects cause quantum walks to behave different from classical walks on certain properties, such as mixing time, hitting time, and so on, and allow quantum walks to handle complex problems, like quantum sampling problems, which are difficult for classical simulation. So, in this section, 
two main classes of these characteristics, including evolution and sampling, are concluded as follows.

\subsubsection{Evolution Characteristics}

Here we will introduce some available characteristics during the evolution process of a quantum walk, which have a close connection with the evolution time or probability distribution of states and are useful for demonstrating the merits of quantum models compared with classical random walks. As is shown in Figure \ref{fig-1}, quantum walk reveals a faster propagation and different localization compared with one of the famous classical random walk models.

\begin{itemize}
\small
    \item \textbf{Mixing time}
    
 Mixing time is defined as a quantity for describing the shortest time when the probabilities $P(x,t)$, for different vertices that the walker stays at after time $t$, get close enough to a \emph{stationary distribution} $\pi(x)$.
Define the time-averaged distribution by  $\bar{P}(x,T) \equiv \frac{1}{T}\sum_{t=1}^{T} P(x,t)$, then the stationary distribution is $\pi(x)=\lim_{T\to\infty} \bar{P}(x,T)$.
    So, for a certain small constant $\epsilon>0$, mixing time~\cite{marquezino2008mixing, Chakraborty2020Finding} is
\begin{equation}\label{eq-12}
    T_{M}^\epsilon = \min\left\{t\ \big|\ \lvert \bar{P}(x,T)-\pi(x)\rvert_1<\epsilon\right\},
\end{equation}
    where  $\lvert\cdot \rvert_1$ is the $l_1\ norm$.

    Note that the mixing time for the continuous-time walk~\cite{Chakraborty2020Finding} shares the same definition as Eq. (\ref{eq-12}) after replacing $\bar{P}(X,t)$ with $ \bar{P}(X,t) = \frac{1}{T}\int_{0}^{T}P(X,t) dt $.

    \item \textbf{Hitting time}
    
 Hitting time is used to evaluate the time from a starting vertex to a marked vertex when detecting or searching a target on a graph, which is a vital quantity for comparing the performance of different random walks even under the condition of no idea of the exact time complexity. There are various ways of definition. Generally, classical hitting time is the expected time~\cite{boito2023quantum} (i.e. first hitting time) at first reach of the end vertex, which is directly defined by
    \begin{equation}\label{eq-13}
        T_H^{ij} = \sum_{t=0}^\infty tP_{ij}(t),
    \end{equation}
where $P_{ij}(t)$ is the probability that the walker starts from the vertex $i$ to the $j$'s vertex after time $t$. Analogously, the continuous version of Eq. (\ref{eq-13}) is $\lim_{T\to\infty}\int_{0}^{T}tP_{ij}(t)\ dt$.

For quantum walks, we care more about the time when the probability of hitting the target is greater than a threshold $\epsilon$, and it has been covered that quantum walks propagate exponentially faster than classical walks on some graphs~\cite{childs2002example,childs2003exponential}. Thus the hitting time can also be defined by 
\begin{equation}
    \tilde{T}_H^{ij} = \min\{t:\bar{P}_{ij}(t)>\epsilon\}
\end{equation}
where $\bar{P}_{ij}(t)$ is the probability of finding the marked vertex all over the time $t$.
Another definition means an optimal hitting, namely the time related to the maximal arriving probability, which was shown exponentially faster than the classical model~\cite{tang_experimental_2018,wang_large-scale_2022}. 

Different from the above notions, quantum hitting time for Szegedy quantum walk was aimed to quantify the running time of searching a marked vertex on the complete graph, and by this, a quadratic speedup than classical algorithms is proven~\cite{boito2023quantum, santos2010quantum}.

    \item \textbf{Commute time}

  The term of commute time means an expected time spent on the process adding a return to the starting vertex except arriving at the end vertex from the start. Accordingly, the commute time from vertex $i$ to $j$ is defined as
    \begin{equation}
        T_C^{ij} = T_H^{ij}+T_H^{ji},
    \end{equation}
    from which it's found that $T_C^{ij} = T_C^{ji}$ all the time, so the commute time is a symmetric quantity. 
    
    For the CTQW, commute time can be directly computed by the Laplacian's spectrums, and help to analyze the graph-embedding property and distinguish clusters in a set of graphs~\cite{emms2007graph}.
    
    \item \textbf{Cover time}

     Cover time~\cite{jonasson1998cover} for the graph $G$, denoted as ${\rm Cover}(x,G)$ if the starting vertex is $x$, is the expected time to traverse every vertex. If the starting vertex is not assigned, then the cover time for an undirected graph is expressed as
     \begin{equation}
        {\rm Cover}(G) = {\max}_x \ {\rm Cover}(x,G). 
     \end{equation}
     This term reflects on the searching efficiency on a region of a graph rather than some or other vertex, which is the key feature different from hitting time~\cite{chupeau2015cover}.
     
    \item \textbf{Anderson Localization}

   Anderson localization is a phenomenon of diffusion's absence due to the interference of waves from various paths in disordered circumstances, in which a tight-binding model~\cite{anderson1958absence} was first proposed by Anderson to describe the wave function (means a time-independent probability amplitude) of one particle on an impure lattice. Taking the example of an electron trapped in a potential field, this phenomenon also provides a credible explanation for the Metal-insulators Transition (MIT)~\cite{ying2016anderson} because the electron would be localized at the neighbouring of the starting position, leading to a loss of conductivity properties of the systems, if we set the energy at each lattice site using a random way~\cite{dominguez2004simple,hundertmark2008short}.

  Anderson location is also a significant characteristic for quantum walk models~\cite{ortuno2009random}, where the analogous propagation impediment will arise with increasing the randomization condition like setting each-step phase-shift operator~\cite{crespi_anderson_2013} or randomly removing lattice sites~\cite{duda2023quantum}. As one of the most popular propagation models, the quantum walk also has been selected to experimentally demonstrate Anderson localization in various systems~\cite{crespi_anderson_2013,duda2023quantum,karamlou2022quantum,ghosh2014simulating}. Different from the direct observation of distribution on all sites, in these experiments, more detailed qualification also plays a vital part in examining the degree of the localization, for example, the average distance and particle numbers on initial sites are welcome in recent related experimental works~\cite{crespi_anderson_2013,karamlou2022quantum}.
    
\end{itemize}

\subsubsection{Quantum Dynamics and Sampling Complexity}

Besides the evolution merits compared with the classical random models,
quantum walks also promise quantum supremacy in terms of computing power in various scenarios, especially for sampling problems, which are shown to be classically hard to simulate for both single- and multi-particle cases~\cite{qiang2016efficient,muraleedharan2019quantum}. 

Here we first introduce the multi-particle quantum walk,
and as an enhanced version of the single-particle model, we further set forth its capability for more complex tasks, like simulation on a larger graph using relatively few resources, by harnessing the multi-particle properties.
Then the quantum sampling problems are briefly elaborated, and two examples of the single-particle quantum walk on the circulant graph and Boson sampling are presented to demonstrate the computational advantages beyond classical algorithms.

\begin{itemize}

    \small
    \item \textbf{Quantum dynamics of multiple particles}

    Different from distinguishable classical particles, microscopic particles at the quantum scale, like elementary particles, atoms as well as molecules, are indistinguishable in principle.
    Indistinguishable quantum particles—whether bosons, fermions, or anyons—exhibit distinct behaviors in quantum statistics.  In a multi-particle system, bosons demonstrate symmetric exchange, fermions antisymmetric, while anyons, unique to two-dimensional systems, display fractional statistics with intermediate exchange properties. So, the fact is straightforward:
    \begin{equation}
        \left\{ {\begin{array}{*{20}{c}}
            {a_k^\dagger a_j^\dagger-\zeta a_j^\dagger a_k^\dagger=0},\\
            {a_ka_j-\zeta a_ja_k=0},\\
            {a_ka_j^\dagger-\zeta a_j^\dagger a_k=\delta_{k,j}},
            \end{array}} \right.
    \end{equation}
    where a creation (annihilation) operator $a_i^\dagger(a_i)$ means an increase (or decrease) of particle number on mode $i$, and $\zeta$ is set to 1, -1 or any fraction for bosons, fermions or anyons, respectively.    
    
    Consider any $M$-particle quantum walk model in Fock space, which is denoted by a unitary $U$ without loss of generality, its initial state can be characterized by creation or annihilation operators, given by
    \begin{equation}\label{eq-16}
        |\phi_{in}\rangle=\prod\limits_{i} (a_i^\dagger)^{m_i}|0^v\rangle=(\prod_i{m_i!})^{\frac{1}{2}}|M\rangle,
    \end{equation}
    where $|0\rangle$ is the vacuum state and $|M\rangle=|m_1,m_2,\cdots\rangle$ is the Fock state in occupation number representation and $M=\sum_im_i$. Specially, for $m$ particles with a single-mode, Eq. (\ref{eq-16}) can be converted to $(a^\dagger)^m|0^v\rangle$.

    Then, after a time $t$ of evolution, the final state becomes
    \begin{equation}
        |\phi_{out}\rangle=\prod\limits_{i}\sum_j(U_{ij}^ta_j^\dagger)^{m_i}|0^v\rangle.
    \end{equation}

    Note that most problems are related to the detecting probability $P_{M\rightarrow T}$ of any output state $|T\rangle=|t_1,t_2,\cdots\rangle$ with initial state $|M\rangle$, so we formulate it by
    \begin{equation}\label{eq-19}
        P_{M\rightarrow T}=\left|\langle T|\phi_{out}\rangle\right|=\frac{\left|\sum_{\hat\sigma\in S_M}\zeta^{{\rm inv}(\hat\sigma)}\prod_{i=1}^n(U_{T,M}^t)_{i,\hat\sigma(i)}\right|^2}{\prod_{i}t_i!\prod_{j}m_j!}
    \end{equation}
    where $S_M$ is the symmetric group defined over all permutations of $M$ elements, $U_{T,M}$ is constructed from the original unitary $U$ by repeating the $i$th row $t_i$ times and the $j$th column $m_j$ times and the function ${\rm inv}(\cdot)$ represents the inversion number of one permutation.

    From the fact that the dimension of Fock space grows exponentially with the number of particles, it becomes feasible that quantum walks on exponentially large graphs are performed using the same time as the one-particle case on a small graph~\cite{chandrashekar2012quantum,rudinger2012noninteracting,sansoni2012two, qiang2016efficient}. 
    To see this, consider a multi-particle CTQW on the $N$-dimensional Hilbert space, 
    its simulation scale of one-particle CTQW will be enlarged to $N^M$-dimensional if $M$ particles begin walking simultaneously, which is seen from the corresponding non-interacting Hamiltonian $H(M)$
    that is expressed as a direct sum of $M$ one-particle Hamiltonian $H$:
    \begin{equation}
        H(M)=H^{\oplus M},\qquad H=\sum_{ij} h_{ij}a_i a_j^\dagger.
    \end{equation}
    It directly concluded that the multi-particle model can simulate the Cartesian product of graphs, one kind of $N^M$-dimensional graph if all particles are distinguishable~\cite{qiang_implementing_2021}. 
        
    As for identical particles, the exchange symmetry between them results in a lower space dimension than the distinguishable case but notably bigger than the one-particle. Here are two particular examples of bosons and fermions, both of which need to introduce a symmetrization operator $\hat P$ such as to map the Hamiltonian from Fock space to Hilbert space. Define the basis state $|\Phi\rangle=|k_1\rangle\otimes|k_2\otimes\cdots\otimes|k_M\rangle$ on Hilbert space $\mathcal{H}^{\otimes M}$, the symmetrization operators is described as
    \begin{equation}
        \hat{P}_{B(F)}|\Phi\rangle=\sqrt{\frac{\prod_{i}m_i!}{M!}}\sum\nolimits_{\hat\sigma\in S_M} (\zeta)^{{\rm inv}(\hat\sigma)}U_\sigma |\Phi\rangle,
    \end{equation}
    in which capitals $B$ and $F$ means bosons $(\zeta=1)$ and fermions ($\zeta=-1$) separately and $U_{\hat\sigma}$ is the permutation matrix associated with $\hat\sigma$. For fermions, no two particles occupy the same vertex so that $k_1<k_2<\cdots<k_M$, then there are $N\choose2$ basis states which means a reduction to the one-particle CTQW on $N\choose2$-dimensional graph; Analogously for bosons, let $k_1\leq k_2\leq \cdots\leq k_M$, then it is concluded that CTQW using multiple bosons can simulate a larger graph with $N+1\choose2$ vertices~\cite{rudinger2012noninteracting,qiang_implementing_2021}. 

    \item \textbf{The supremacy of quantum walks in sampling problems}

    A sampling problem is a type of problem that collects samples from a given probability distribution
    and acts as a solid foundation for lots of disciplines. And quantum sampling problems refer to the ones being solved by sampling (or measuring) outputs of quantum circuits, 
    some of which are simulated difficultly by any classical algorithm, thus promising a demonstration of quantum supremacy in recent quantum devices~\cite{bremner2011classical,aaronson2011computational,lund2017quantum,brod2021bosons}. So does sampling the quantum walk evolution, which shows hardness for classical algorithms, while it could be achievable in quantum way. Here are two examples.

    As an important class of quantum walk models, one-particle CTQWs are allowed to be efficiently performed on  a circulant class of graphs but the classical simulation algorithms are not~\cite{qiang2016efficient}, 
    when the initial state is $|0\rangle$ and a projective measurement is applied under the basis $|0\rangle\langle0|$.
    This process can be described by the probability formulation, which is
    \begin{equation}
        \begin{array}{rl}
            {\left|{\langle 0|e^{-itH_c}|0\rangle}\right|^2}&{=\left|\langle0|F^\dagger DF|0\rangle\right|^2}\\
            {}&{ = \left|\langle0|H^{\otimes n} DH^{\otimes n}|0\rangle\right|^2.}
            \end{array}
    \end{equation}
    This equation is easily seen to hold since the adjacency matrix $H_c$ of any circulant graph can be factorized as quantum Fourier transformation $F$ and a diagonal matrix $D'$ such that $D=e^{-itD'}$ is a diagonal matrix too. From this, a connection has been established between the CTQW and the IQP circuits~\cite{takeuchi2016ancilla} that were confirmed hard sampling classically\cite{bremner2011classical}.

    The same PH collapse occurs in the multi-particle quantum walks, which can be seen from the fact that the quantum walks of multiple bosons are equivalent to Boson Sampling problems in sense of the behaviors of quantum particles.
    A vital step towards classical sampling of one quantum circuit is 
    computing probability $P_{M\rightarrow T}$ by equation (\ref{eq-19}), consider that all particles are bosons, it directly follows that
    \begin{equation}
        P_{M\rightarrow T}=\frac{{\rm Perm}(\left|U_{M,T}\right|^2)}{\prod_{i}t_i!\prod_{j}m_j!},
    \end{equation}
     where ${\rm Perm}|\cdot|$ means the permanent of any matrix and computing it is a \#P-hard problem theoretically. Therefore, there doesn't exist any polynomial-time classical algorithm to simulate Boson Sampling~\cite{bremner2016average}.

    Note that both of the above two cases still hold under the condition that the classical simulation or quantum computer is noisy and imperfect, which has been confirmed based on the unproven worst-case or average-case hardness conjectures~\cite{aaronson2011computational,bremner2016average}. 

\end{itemize}

\begin{table}[htbp]
    \centering
    \tiny
    \renewcommand\arraystretch{1.4}
    \caption{Summary of physical implementations of quantum walks.}\label{tab-2}
    \begin{tabular}{m{0.5in}<{\centering}m{0.4in}<{\centering}m{0.4in}<{\centering}m{0.3in}<{\centering}m{0.9in}<{\centering}m{0.9in}<{\centering}m{0.8in}<{\centering}m{0.5in}<{\centering}}
        \toprule
        & \multicolumn{2}{m{0.8in}<{\centering}}{Physical systems}	& QW models	& Encoding ways	& Programmability	& Scalability	& References \\
        \midrule
        {Digital physical implementation approach}	& \multicolumn{2}{m{1in}<\centering}{Quantum circuit implementations}	& CTQW, DTQW, SQW	& Qubits encoded, and quantum walk evolution via quantum circuits	& Full programmability through quantum circuits	&Efficient circuit required for achieving speedup over classical computer	& \cite{douglas2009efficient, loke2012efficient, loke_efficient_2017, qiang2018large, childs2010relationship, qiang2016efficient, du2003experimental,ryan2005experimental}\\
        \midrule
        \multirow{15}{0.6 in}[-0.3\textheight]{Analogue physical implementation approaches} & \multirow{4}{0.5in}[-10.5em]{Solid-state quantum systems} & \multirow{2}{0.5in}[-4em]{Supercond-ucting system} & DTQW &  Each lattice site encoded into a coherent state of the cavity, and coin space into the transmon qubit & Bloch-oscillating QWs from an arbitrary split- or single-step QWs, with controllable coin operations & Scaling by increasing superconducting qubit array & \cite{ramasesh_direct_2017}\\
        \Xcline{4-8}{0.5pt}
        & & & CTQW & Each site encoded into a superconducting transmon qubit &	Programmable propagation paths, tunable disorders on the evolution paths and interactions between neighbouring sites, depending on the superconducting qubit array	& Scaling by increasing superconducting qubit array &	\cite{yan_strongly_2019, gong_quantum_2021}\\
        \Xcline{3-8}{0.5pt}
        & & Trapped ions system	& DTQW &	Position space encoded into the phase space and coin space encoded into two electronic (hyperfine) states of the ion &	Controllable coin operation	& Scaling up one-dimensional DTQW step number and possibly QW in higher dimension	& \cite{schmitz2009quantum,zahringer2010realization,matjeschk_experimental_2012}\\
        \Xcline{3-8}{0.5pt}
        & & Trapped atoms system &	DTQW	& Coin space encoded into a two-level particle with internal states, and position space encoded into one-dimensional spin-dependent optical lattice	& Controllable coin operation	& Scaling up one-dimensional DTQW step number &	\cite{karski2009quantum}\\
        \Xcline{2-8}{0.5pt}
        & \multirow{8}{0.4in}[-13.5em]{Optical quantum systems} & \multirow{5}{0.4 in}[-7.5em]{Bulk optics system} & \multirow{4}{*}[-5em]{DTQW} &  Polarization for coin space, time-bin for position space	& Controllable coin operation and initial state	& Scaling up one-dimensional DTQW step number	& \cite{barkhofen_supersymmetric_2018,lorz_photonic_2019, xu_measuring_2018}\\
        \Xcline{5-8}{0.5pt}
        & & & & Polarization for coin space, spatial mode for position space	& Controllable coin operation and initial state	& Scaling up one-dimensional DTQW step number	& \cite{wang_simulating_2019,zhan_detecting_2017,xue_experimental_2015,xiao_observation_2017,wang_generalized_2023}\\
        \Xcline{5-8}{0.5pt}
        & & & & SAM for coin space, OAM for position space	& Controllable coin operation and initial state	& Scaling up one-dimensional DTQW step number	& \cite{zhang_implementation_2010}\\
        \Xcline{5-8}{0.5pt}
        & & & & Polarization for coin space, OAM for position space	& Controllable coin operation	& Scaling up one-dimensional DTQW step number	& \cite{goyal_implementing_2013, giordani_experimental_2019}\\
        \Xcline{4-8}{0.5pt}
        & & &CTQW	& Quantum states are encoded in both polarization and spatial degrees of freedom	& Controllable diffusion operation and initial state	& Designed setup corresponding to the target quantum circuits &	\cite{qu2022deterministic}\\
        \Xcline{3-8}{0.5pt}
        & & \multirow{3}{0.4 in}[-4.8em]{Fiber system}	& DTQW	& Polarization for coin space, time for position space	& Controllable coin operation and initial state	& Scaling up one-dimensional DTQW step number	& \cite{schreiber2010photons}\\
        \Xcline{4-8}{0.5pt}
        & & & \multirow{2}{*}[-2em]{CTQW} & Site encoded in the core of multicore fiber	& Quasiperiodic photonics lattices possessing both on- and off-diagonal deterministic disorder	& Scaling by increasing photon and waveguide numbers & 	\cite{nguyen_quantum_2020}\\
        \Xcline{5-8}{0.5pt}
        & & & & Site encoded in supporting modes of multimode fiber	& Controllable Hamiltonian configurations and initial states	& Scaling by increasing the number of modes and photons &	\cite{defienne_two-photon_2016}\\
        \Xcline{2-8}{0.5pt}
        & \multirow{3}{0.4 in}[-7.5em]{Integrated photonics systems} & \multirow{2}{0.5in}[-2em]{Directly laser waveguide writing chips} & CTQW &	Spatial mode	& Fixed Hamiltonian and evolution time for a given device& Scaling by increasing photon number and waveguide array size	& \cite{peruzzo2010quantum, poulios2014quantum, benedetti_quantum_2021, crespi_suppression_2016, tang_experimental_2018, xu_quantum_2021, jiao_two-dimensional_2020,tang_experimental_2018-1}\\
        \Xcline{4-8}{0.5pt}
        & & & DTQW	& Spatial mode	& Fixed steps for QW for a given device, with possible particle features control	& Scaling by increasing photon number, step number and waveguide array structure complexity	& \cite{crespi_anderson_2013, sansoni2012two}\\
        \Xcline{3-8}{0.5pt}
        &  & Silicon photonic chips& 	CTQW	& Spatial mode	& Fully programmable Hamiltonian/Evolution time/Particle features (using proposed entanglement scheme) on a same device	& Scaling by increasing photon number and network size	& \cite{qiang_implementing_2021,wang_large-scale_2022,wang_experimental_2022}\\
        \bottomrule
    \end{tabular}
\end{table}

\section{Quantum Walk Implementations}

A variety of physical quantum systems have been used for implementing quantum walks, including superconducting qubits~\cite{gong_quantum_2021,yan_strongly_2019,ramasesh_direct_2017}, trapped ions~\cite{schmitz2009quantum,zahringer2010realization,matjeschk_experimental_2012}, nuclear magnetic resonance~\cite{du2003experimental,ryan2005experimental}, trapped atoms~\cite{karski2009quantum}, laser beam and bulk optics~\cite{barkhofen_supersymmetric_2018,lorz_photonic_2019, xu_measuring_2018,wang_simulating_2019,zhan_detecting_2017,xue_experimental_2015,xiao_observation_2017,wang_generalized_2023,zhang_implementation_2010,goyal_implementing_2013, giordani_experimental_2019,qu2022deterministic,schreiber2010photons,nguyen_quantum_2020,defienne_two-photon_2016}, and integrated photonics systems~\cite{peruzzo2010quantum, poulios2014quantum, benedetti_quantum_2021, crespi_suppression_2016, tang_experimental_2018, xu_quantum_2021, jiao_two-dimensional_2020,tang_experimental_2018-1,crespi_anderson_2013, sansoni2012two,qiang_implementing_2021,wang_large-scale_2022,wang_experimental_2022}. Different quantum walk models such as discrete-time and continuous-time quantum walks and quantum walk based algorithms and applications have been experimentally demonstrated. 

There are two different approaches for physically implementing quantum walks: analogue physical simulation of quantum walks and digital physical simulation of quantum walks. In the analogue physical simulations of quantum walks, the apparatus is dedicated to implementing specific instances of Hamiltonian without translation onto quantum logic, where most of the existing experimental demonstrations belong to this kind. The analogue physical simulation of quantum walks is not limited to the graph structure, though for single-particle quantum walks it does not scale efficiently in resources when simulating broad classes of large graphs. With the evolving particles increasing, the simulated multi-particle quantum walks become classically intractable because of the exponentially large Hilbert space. The analogue approach can scale by increasing the particle numbers and the dimension of the Hamiltonian evolving, on the other hand, few existing methods exist for realizing error correction or providing fault tolerance in implementing analogue physical simulations of quantum walks.

In the digital physical simulations of quantum walks, it is required to construct quantum circuits for implementing quantum walk evolutions. If an efficient quantum circuit can be constructed, quantum speed-up can be achieved for implementing quantum walks. With such quantum circuit implementations of quantum walks, one can either implement further quantum circuit operations or perform direct measurements of the output evolution states~\cite{qiang2016efficient}. It is in general difficult to design efficient quantum circuits for simulating quantum walks, though many of them have been found for families of graphs for different quantum walk models. However, with the digital simulation of quantum walks, i.e., with quantum circuit implementations, fault tolerance and error corrections in simulating quantum walks would be available by using the existing technologies in general quantum circuit computing.

We summarize in Table~\ref{tab-2} different kinds of physical implementations of quantum walks, including both digital and analogue physical implementation approaches, and compare these approaches in quantum walk models, encoding ways, programmability, and scalability. The different physical implementations of quantum walks and the recent progress are further described in detail in the following sections. 

\subsection{Digital physical implementation of quantum walks}

\begin{figure}[!htb]
    \centering
    \includegraphics[width=1\textwidth]{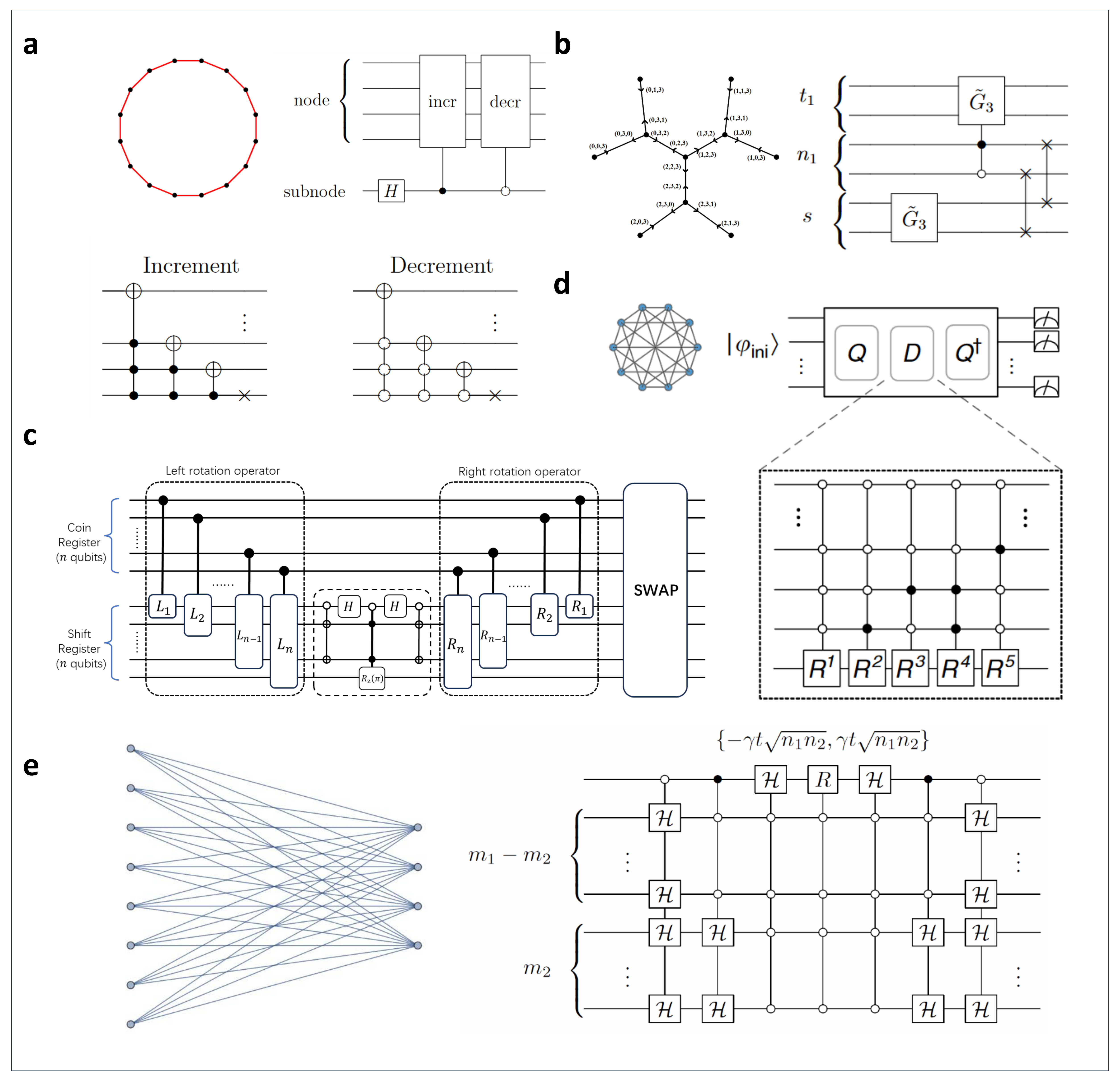}
    \caption{\textbf{Quantum circuit implementation of quantum walks.} \textbf{a}, Quantum circuit for implementing DTQW on a 16-vertex circle graph. Reprinted figure with permission from Ref~\cite{douglas2009efficient} Copyright \copyright2009 by the American Physical Society. \textbf{b}, Quantum circuit for implementing DTQW on a 3CT-2 graph. Reprinted figure with permission from Ref~\cite{loke2012efficient} Copyright \copyright2012 by the American Physical Society. \textbf{c}, Quantum circuit for implementing Szegedy quantum walks on a cycle graph~\cite{loke2017efficient}. \textbf{d}, Quantum circuit for implementing CTQW on a circulant graph, reproduced from Ref~\cite{qiang2016efficient}. \textbf{e}, Quantum circuit for implementing CTQW on the complete bipartite graph~\cite{loke_efficient_2017}\copyright~2017 IOP Publishing Ltd. Reproduced with permission. All rights reserved.}
    \label{fig:exp-digital}
\end{figure}

Quantum walk simulation in general can be implemented on a digital quantum computer, but only efficient quantum circuits promise quantum speed-up compared to classical computation. Consider that if a quantum circuit which uses $O(\log(N))$ qubits and $O(\log(N)$ elementary gates can implement quantum walks on a graph of $N$ vertices, the quantum circuit is then efficient. 
Efficient quantum circuit implementation for quantum walks on various kinds of graphs has been studied, as shown in Figure~\ref{fig:exp-digital}. For DTQW models, efficient quantum circuits have been proposed for some highly symmetric graphs, including cycle, hypercycle, complete graphs and ``twisted" toroidal lattice graphs~\cite{douglas2009efficient}. For non-degree-regular graphs, such as star graphs and Cayley trees with an arbitrary number of layers, quantum circuits for simulating DTQW on these graphs have also been proposed, and a quantum walk based search algorithm on these graphs was shown to be implemented using these circuits~\cite{loke2012efficient}. Lokes et al. also presented a general scheme to construct efficient quantum circuits for Szegedy quantum walks that correspond to classical Markov chains possessing transformational symmetry in the columns of the transition matrix~\cite{loke2017efficient}, and they further applied this scheme to construct efficient quantum circuits simulating the Szegedy walks used in the quantum Pagerank algorithm for some classes of non-trivial graphs, providing a necessary tool for experimental demonstration of the quantum Pagerank algorithm. This quantum circuit implementation of Szegedy quantum walk has been demonstrated on a programmable silicon photonic quantum processor~\cite{qiang2018large}. 

For the continuous-time quantum walk model, efficient quantum circuit implementation has been shown on several classes of graphs, including commuting graphs and Cartesian products of graphs~\cite{loke_efficient_2017}. There are a select few classes of graphs for which their CTQWs can be fast-forwarded to obtain an efficient quantum circuit implementation. It has been pointed out that the glued tree, complete graph, complete bipartite graph, and star graph can be simulated efficiently using the diagonalization approach~\cite{childs2010relationship}. A certain class of circulant graphs have also been identified as being efficiently implementable~\cite{qiang2016efficient}. The circulant graphs can be diagonalized using the quantum Fourier transform, and by using the quantum Fourier transform circuit and inverse quantum Fourier transform circuit, a whole efficient quantum circuit for CTQW on circulant graphs can be constructed if the given circulant graph of $2^n$ vertices has $O(\poly(n))$ non-zero eigenvalues, or it has more distinct eigenvalues but can be characterized efficiently (such as the cycle graphs and M\"{o}bius ladder graphs)~\cite{qiang2016efficient}. It is noteworthy that efficient quantum circuits for CTQWs on circulant graphs, on the one hand, enable simulation of circulant molecular dynamics (for example, the DNA M\"{o}bius strips) via quantum walks. On the other hand, it has been proven from the computational complexity theory that sampling the output probability distribution of even single-particle quantum walks on exponentially large circulant graphs could be classically intractable and thus demonstrate quantum supremacy over classical computers --- this shows a new link between continuous-time quantum walks and computational complexity theory. The experimental demonstration has been realized using a bulk optics quantum processor setup~\cite{qiang2016efficient}. 
Also, in nuclear magnetic resonance (NMR) systems, both CTQW and DTQW using qubit models have been demonstrated~\cite{du2003experimental,ryan2005experimental}, where using a two-qubit NMR quantum processor, CTQW on a circle with four vertices have been experimentally demonstrated~\cite{du2003experimental}, and using a three-qubit liquid state NMR quantum processor DTQW on a square was experimentally demonstrated~\cite{ryan2005experimental}. Among the three qubits, one qubit describes the coin state and two for the position state.

The digital physical simulation approach of quantum walks, i.e., quantum circuit implementation for quantum walks can have the following benefits: first, quantum walks are used as a tool for designing quantum algorithms. Quantum circuit implementation can be used to efficiently implement quantum walk oracle, and further implement the quantum walk algorithms; second, quantum circuit implementation can provide a better estimate of the complexity and resources needed to implement quantum walk on a given graph, as compared to a “black-box” oracle; third, to realize quantum walk based algorithms on a universal quantum computer, an efficient quantum circuit would be needed to implement the required quantum walk evolutions.

\subsection{Analogue physical implementation of quantum walks}

The analogue physical implementation of quantum walks has been demonstrated by using different physical systems and has also shown great potential for accelerating useful NISQ quantum computing systems. 
Experimental and technical advancements have been achieved recently in various physical systems for simulating quantum walks. Through the analogue physical implementation approaches of quantum walks, a quantum walk based application-oriented quantum computing system is being pursued, which aims to implement different quantum walk simulations and quantum walk based applications. It usually requires that the quantum walk system has full programmability for quantum walk evolutions. Such programmability can include: i) Hamiltonian programmability, which requires configuring the corresponding Hamiltonians of quantum walks for different instances of the problem; ii) Evolution time programmability, which requires controlling the evolution time of quantum walks; iii) Initial states programmability, which requires configuring the initial state of quantum walks, e.g., uniform superposition state or single basis state; iv) Particle features programmability, given that particle distinguishability, particle exchange symmetry, and particle interactions can affect the output results of multi-particle quantum walks. The current analogue implementation approaches of quantum walks, including solid-state, bulk optics and integrated photonics systems, have shown the capability of these programmable abilities for various quantum walk simulations and quantum walk based applications. 

\begin{figure}[!htbp]
    \centering
    \includegraphics[width=\textwidth]{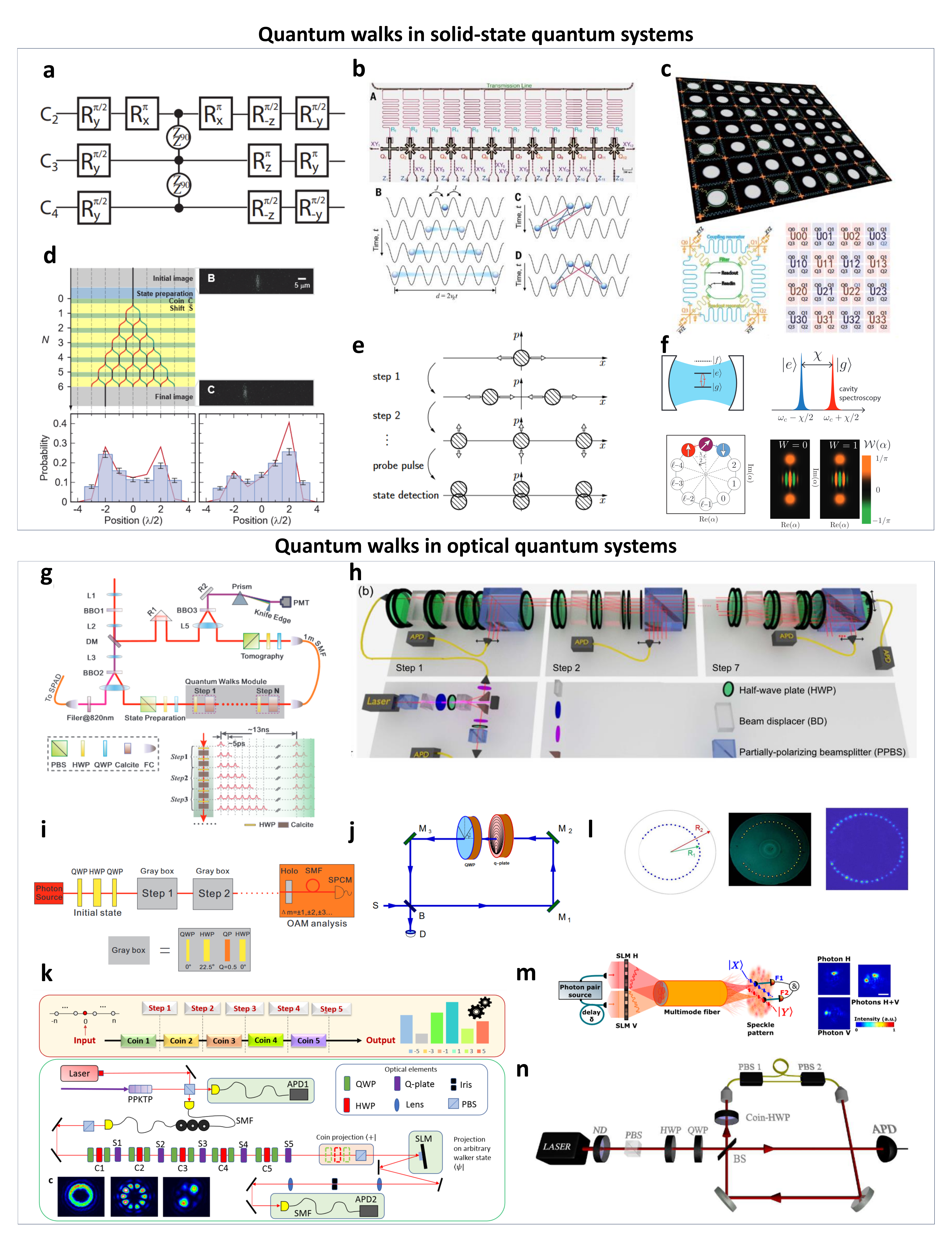}
    \caption{\small{\textbf{Physical implementations of quantum walks in solid-state and optical quantum systems.} \textbf{a}, NMR pulse sequence representing a single step of DTQW. Reprinted figure with permission from Ref~\cite{ryan2005experimental} Copyright \copyright2005 by the American Physical Society. \textbf{b}, Simulating QWs of one and two photons}}
    \label{exp-1}
\end{figure}
\begin{figure}[t]
\caption*{\small{in a one-dimensional lattice of a superconducting processor. From Ref~\cite{yan_strongly_2019}. Reprinted with permission from AAAS. \textbf{c}, The schematic diagram of a 2D superconducting quantum processor, on which QWs on a 2D superconducting qubits array were implemented. From Ref~\cite{ gong_quantum_2021}. Reprinted with permission from AAAS. \textbf{d}, 1D DTQW was implemented in position space with single optically trapped atoms. From Ref~\cite{karski2009quantum}. Reprinted with permission from AAAS. \textbf{e}, Schematic demonstration of DTQW on a line in phase space using ions. Reprinted figure with permission from Ref~\cite{zahringer2010realization} Copyright \copyright2010 by the American Physical Society. \textbf{f}, Schematic of a proposed cQED setup for realizing QWs using a superconducting cavity mode coupled to a transmon qubit. Reprinted figure with permission from Ref~\cite{ramasesh_direct_2017} Copyright \copyright2017 by the American Physical Society. \textbf{g}, 1D DTQW implemented in an optical system where coin space is encoded in polarization and position space is encoded in time-bin. Reprinted figure with permission from Ref~\cite{xu_measuring_2018} Copyright ©2018 by the American Physical Society. \textbf{h}, 1D DTQW implemented in an optical system where coin space is encoded in polarization and position space is encoded in spatial mode. Reprinted figure with permission from Ref~\cite{zhan_detecting_2017} Copyright \copyright2017 by the American Physical Society. \textbf{i}, 1D DTQW implemented in an optical system where coin space is encoded in SAM and position space is encoded in OAM. Reprinted figure with permission from Ref~\cite{zhang_implementation_2010} Copyright \copyright2010 by the American Physical Society. \textbf{j}, 1D DTQW implemented in the OAM space of a laser beam, where coin space is encoded in polarization and position space is encoded in OAM. Reprinted figure with permission from Ref~\cite{goyal_implementing_2013} Copyright \copyright2013 by the American Physical Society. \textbf{k}, 1D DTQW implemented in an optical system where coin space is encoded in polarization and position space is encoded in OAM. Reprinted figure with permission from Ref~\cite{giordani_experimental_2019} Copyright ©2019 by the American Physical Society. \textbf{l}, CTQW implemented in periodic and quasiperiodic Fibonacci multicore fibres, reproduced from Ref~\cite{nguyen_quantum_2020}. \textbf{m}, CTQW implemented in a multimode fibre system where each site is encoded in the mode of the multimode fibre. From~\cite{defienne_two-photon_2016}. \copyright The Authors, some rights reserved; exclusive licensee AAAS. Distributed under a CC BY-NC 4.0 license http://creativecommons.org/licenses/by-nc/4.0/”.  Reprinted with permission from AAAS. \textbf{n}, DTQW implemented in a fibre optics system where coin space is encoded in polarization and position space is encoded in time-bin. Reprinted figure with permission from Ref~\cite{schreiber2010photons} Copyright \copyright2009 by the American Physical Society.}}
\end{figure}

\subsubsection{Quantum walks in solid-state quantum systems}
Quantum walks have been demonstrated by using solid-state quantum systems, such as NMR systems, superconducting systems, trapped ion systems and trapped atom systems. In NMR systems, the demonstrations of quantum walks were mainly using qubit models, i.e., in the digital physical implementation approach~\cite{du2003experimental,ryan2005experimental}, while the other systems use analogue physical implementation approaches as follows.

In superconducting quantum systems, superconducting qubits can be used as artificial atoms with high-fidelity manipulation and tomographic readout, and thus they can easily simulate interacting quantum walks and explore even the hard-core boson interference in quantum walk evolutions.
In a circuit of quantum electrodynamics architecture where a superconducting transmon qubit is coupled to a high-quality-factor electromagnetic cavity, considering that the quantum walk takes place in the phase space of the cavity mode, one can realize a quantum walk on a circular lattice in cavity phase space. Ramasesh et al. demonstrated that such a simulation platform associated with discrete-time quantum walks is naturally suited for the direct extraction of topological invariants~\cite{ramasesh_direct_2017}. One and two strongly correlated microwave photons quantum walks in a one-dimensional array of 12 superconducting qubits in the presence of strong attractive interactions have afterward been demonstrated~\cite{yan_strongly_2019}. Furthermore, with a 62 superconducting qubit processor, the easy programmability of the circuit enables the realization of a fully configurable two-dimensional quantum walks~\cite{gong_quantum_2021}, in which the experimental demonstrations show the realization of programmable quantum walks, enabling to define propagation paths for the quantum walkers, with the remarkable control of not only the qubits frequencies but also the tunnelling amplitude and phase between neighbouring sites.

In trapped-ion systems, discrete-time quantum walks have been demonstrated, where the simulated walker evolves along a line in a phase space~\cite{schmitz2009quantum,zahringer2010realization,matjeschk_experimental_2012}. By extending the quantum walks using two ions, the walker achieves the additional possibility to stay instead of taking a step~\cite{zahringer2010realization} --- the coin operation has three options: move left, move right, or stay. For increasing the number of steps of DTQW using a trapped-ion system, it requires to take account of the high-order terms of the quantum evolution, and a novel protocol for quantum walks has been proposed, based on a scheme from the field of quantum information processing using photon kicks~\cite{garcia2003speed,garcia-ripoll_coherent_2005}, to overcome these restrictions and to allow in principle for hundreds of steps, and can be extendable to quantum walks in higher dimensions~\cite{matjeschk_experimental_2012}. 

In trapped atom systems, a discrete-time quantum walk of up to 24 steps on a line with single neutral atoms has been demonstrated, by deterministically delocalizing the atoms over the sites of a one-dimensional spin-dependent optical lattice~\cite{karski2009quantum}. To achieve many steps, atomic systems require formidable control and isolation from their environment, which should be considered to overcome. 

\subsubsection{Quantum walks in optics systems}
Optics or photon systems are currently well-developed for implementing quantum walks, among the analogue implementation approaches of quantum walks. With various degrees of freedom (DOFs) encoding ways, such as spatial modes (or path), polarization, time-bin, orbital angular momentum (OAM), and frequency, different photonic platforms have been used for implementing quantum walks. These photon platforms include bulk optics systems, fibre systems, and integrated photonics systems described in the next subsection, showing extensive techniques and promising progress in the photonic implementations of quantum walks. 

In bulk optics systems, several different ways have been used for implementing DTQWs. With laser pulses as the simulated walkers, by encoding coin states in the polarizations of photons and encoding position states in the time-bin of photons, Barkhofen et al.~\cite{barkhofen_supersymmetric_2018} and Lorz et al.~\cite{lorz_photonic_2019} successively demonstrated both one-dimensional and two-dimensional DTQWs with controllable coin operations. To overcome the extra loss in a conventional loop structure, Xu et al. introduce birefringent crystals to implement the conditional shift operations in DTQWs, and with polarization encoded coin space and time-bin encoded position space~\cite{xu_measuring_2018}. Alternatively, Xue et al. used photon's polarization for coin state and spatial mode for position state, where the wave plates are used for coin operations and beam displacers (BDs) is used for conditional shift operations, and a variety of DTQWs and DTQW-based application experiments have been demonstrated~\cite{wang_simulating_2019,zhan_detecting_2017,xue_experimental_2015,xiao_observation_2017,wang_generalized_2023}. 

By using the DOFs of photons: spin momentum (SAM) and OAM, a qubit system and a high-dimension system can be encoded and realized respectively, which provides a feasible way for implementing discrete-time quantum walks. By encoding coin state space in SAM and position state space in OAM, Zhang et al. used a spin-orbital system for implementing one-dimensional DTQW~\cite{zhang_implementation_2010}. Goyal et al. alternatively used OAM modes as the lattice sites and polarizations as the coin state to implement DTQW with a laser beam, in a proposed ring interferometer~\cite{goyal_implementing_2013}. Furthermore, Giordani et al. further used a similar linear-optics platform for DTQW with OAM of photons to investigate quantum walk dynamics, showing a quantum state-engineering toolbox via quantum walks~\cite{giordani_experimental_2019}. In their experiments, they used the OAM of light as the physical embodiment of the walker, while the logical states of the coin are encoded in circular polarization states.

Fibre systems are also used for implementing quantum walks, considering that bulk optical systems could have larger sizes and challenging scalability and stability. Fiber systems on the one hand can easily use fibre loops to implement time-bin encoded quantum states, and on the other hand, recently some kinds of specialized fibers allowed new ways for encoding quantum walks. In 2010, by encoding coin states in the polarization of photons and position state in time zone, Schreiber et al. used a fibre-loop setup to implement DTQW of five steps on a line with adjustable coin operations~\cite{schreiber2010photons}. With the customized multicore ring fibres, Nguyen et al. demonstrated CTQWs and localized CTQWs in a new optical fibre that has a ring of cores constructed with both periodic and quasiperiodic Fibonacci sequences, respectively~\cite{nguyen_quantum_2020}. In this case, each core represents a site of CTQW and photons can transmit between different cores during the CTQW evolution. Furthermore, Defienne et al. used a multimode fibre (MMF) to implement a quantum walk of indistinguishable photon pairs in a MMF fibre supporting 380 modes where the transmission matrix knowledge of an MMF combined with wavefront shaping methods are used, which establishes MMFs as a reconfigurable, high-dimensional platform for multi-photon quantum walks~\cite{defienne_two-photon_2016}. 

\subsubsection{Quantum walks in integrated photonics systems}
Integrated quantum photonics is an engineering solution proposed for robust and exquisite control of photonic quantum information by allowing the generation, manipulation, and detection of photons on a single chip. It enables the miniaturization of quantum-photonic experiments into chip-scale waveguide circuits, and thus provides an ideal platform for realizing devices with high sub-wavelength stability for generalized quantum interference experiments, such as multi-photon quantum walks. By employing photons as quantum walkers on intrinsically stable integrated waveguide arrays that are used as the quantum walk evolving network, integrated photonics devices have been extensively shown for implementing multi-particle quantum walks on large networks with even more complicated structures. In some of the integrated photonics devices, each waveguide is in general coupled to several other waveguides, enabling the fabrication of structures that are directly map to graphs with a high degree of connectivity, with different coupling strengths for nearest and non-nearest neighbouring waveguides. In this way, it is equivalent to the adjacency matrix of the graph representing the waveguide structure on which the CTQW evolves. In addition, a reconfigurable on-chip optical network, consisting of Mach-Zehnder interferometers (MZIs) and phase shifters, can be programmed into an arbitrary optical unitary transformation~\cite{carolan2015universal} and thus is used for implementing the quantum walk evolution with different Hamiltonians by reconfiguring the on-chip network easily. 

In 2010, Peruzzo et al. first demonstrated a one-dimensional continuous-time quantum walk in an array of 21 continuously evanescently coupled waveguides in a SiOxNy (silicon oxynitride) chip, and they simulated one-particle walking on the two-dimensional lattice by injecting two indistinguishable photons into the chip~\cite{peruzzo2010quantum}. With the technology of direct laser waveguide writing, waveguide chips were fabricated for one-dimensional to two-dimensional quantum walks, and even for quantum walks with complicated structures. In 2014, Poulios et al. demonstrated the quantum walks on an integrated photonic waveguide chip with a two-dimensional “swiss cross” structure, and they further implemented quantum walks on a more complex graph structure of 45 vertices and 126 links by performing quantum walks of two indistinguishable photons in the “swiss cross” waveguides~\cite{poulios2014quantum}. In 2021, Benedetti et al. demonstrated the CTQW evolutions in a planar triangular lattice structure waveguide chip in fused-silica substrates which is fabricated by the femtosecond laser writing technique, and they implemented CTQW-based spatial search on the two-dimensional triangular graphs~\cite{benedetti_quantum_2021}. With a maze of tens of waveguide modes fabricated by femtosecond laser waveguide writing, Caruso et al. demonstrated a quantum walker travelling in a maze could lead to extremely efficient and fast transmission~\cite{crespi_suppression_2016}. 

\begin{figure}[p]
    \centering
    \includegraphics[width=0.9\textwidth]{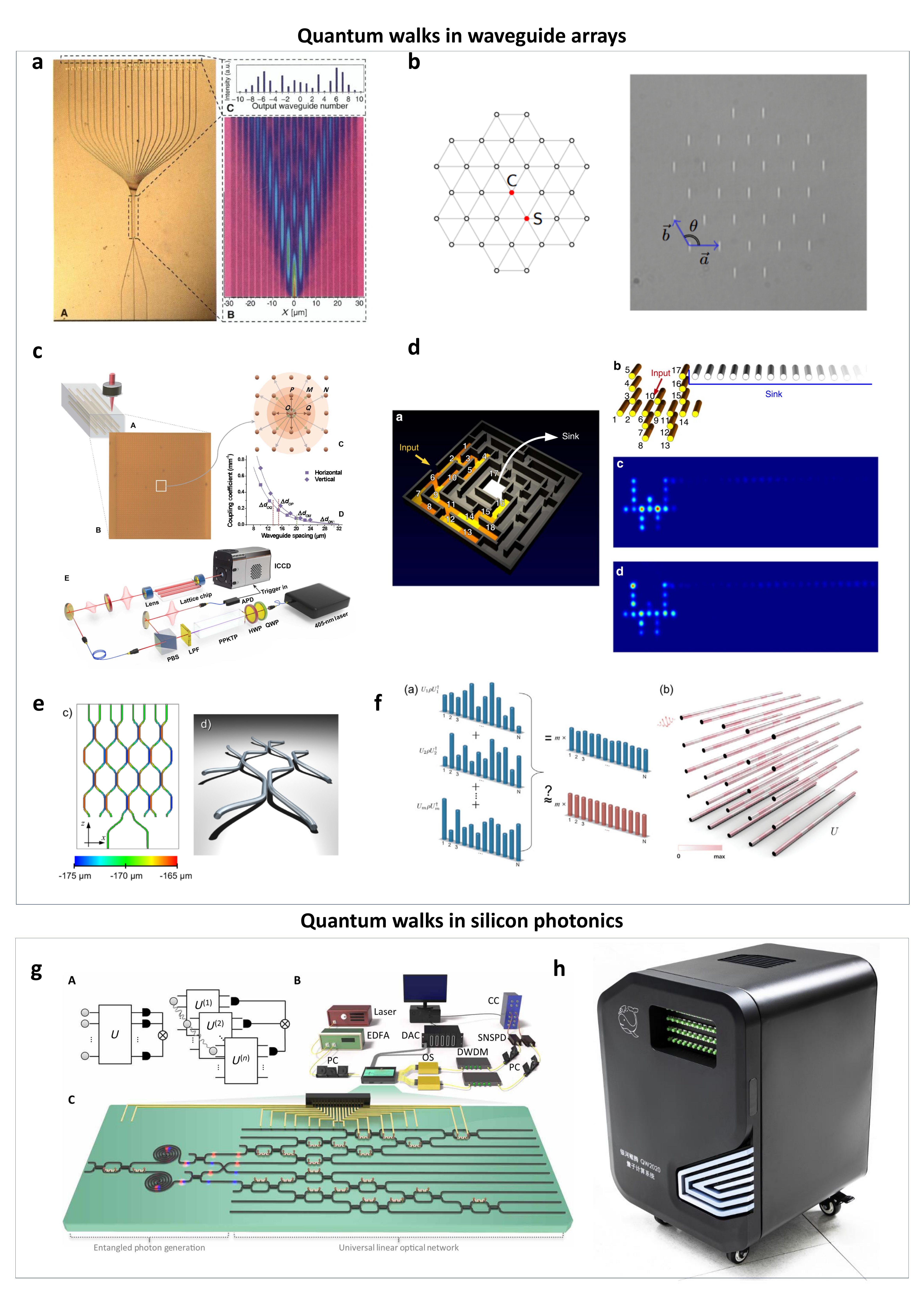}
    \caption{\small{\textbf{Physical implementations of quantum walks in integrated quantum photonics devices.} \textbf{a}, A continuously coupled waveguide array in a SiOxNy chip for implementing CTQWs of correlated photons.}}
    \label{exp-2}
\end{figure}
\begin{figure}
    \caption*{From Ref~\cite{peruzzo2010quantum}. Reprinted with permission from AAAS. \textbf{b}, CTQW-based search demonstrated in a two-dimensional array of optical waveguides fabricated by the femtosecond laser direct writing techniques.
    Reprinted figure with permission from Ref~\cite{benedetti_quantum_2021} Copyright \copyright2021 by the American Physical Society. \textbf{c}, A two-dimensional waveguide array for CTQWs on lattices up to $49\times49$ nodes fabricated by femtosecond-laser direct writing techniques. From ~\cite{tang_experimental_2018-1}. \copyright The Authors, some rights reserved; exclusive licensee AAAS. Distributed under a CC BY-NC 4.0 license http://creativecommons.org/licenses/by-nc/4.0/”.  Reprinted with permission from AAAS. \textbf{d}, A two-dimensional integrated waveguide array for stochastic quantum walk on the maze, reproduced from Ref~\cite{caruso2016fast}. \textbf{e}, An integrated optical circuit consisting of waveguides and directional couplers fabricated by the femtosecond laser direct writing techniques and it implements a four step of DTQW. Reprinted figure with permission from Ref~\cite{sansoni2012two} Copyright \copyright2012 by the American Physical Society. \textbf{f}, Illustration of two-dimensional stochastic quantum walks on the integrated photonic chip used for generating Haar-uniform randomness. Reprinted figure with permission from Ref~\cite{tang2022generating} Copyright \copyright2022 by the American Physical Society. \textbf{g}, A programmable silicon quantum photonic processor for implementing CTQW with full control over various QW evolution parameters in one device. From~\cite{qiang_implementing_2021}. \copyright The Authors, some rights reserved; exclusive licensee AAAS. Distributed under a CC BY-NC 4.0 license http://creativecommons.org/licenses/by-nc/4.0/”.  Reprinted with permission from AAAS.  \textbf{h}, A system with the core of fully-programmable silicon quantum photonics chip for implementing large-scale quantum walks, reproduced from Ref~\cite{wang_large-scale_2022}.}
\end{figure}

Both the scalability and complexity of the waveguide structure fabricated by femtosecond-laser writing techniques have achieved rapid developments in recent years, and different quantum walk evolution dynamics have been demonstrated. With a two-dimensional structure of waveguides up to 160 modes, Jin et al. experimentally demonstrated quantum fast hitting on hexagonal graphs, achieving a linear relationship between the optimal hitting time of CTQW and the network depth~\cite{tang_experimental_2018}. By using a waveguide chip containing nearly 700 sites, they further implemented CTQWs in fractal photonic lattices and experimentally investigated quantum transport in fractal networks, which further reveals the transport process through either the direct observation of photon evolution patterns or the quantitative characterization~\cite{xu_quantum_2021}. By using femtosecond laser direct writing techniques, various large-scale structures that form a two-dimensional lattice with up to $37\times 37$ and $49 \times 49$ sites have been implemented on photonic chips~\cite{jiao_two-dimensional_2020,tang_experimental_2018-1}. These chips can be used for simulated much larger graph structures by using an injected two-photon state, for example, with the chip of $37\times 37$ sites, Jiao et al. simulated a single-photon CTQWs in a higher dimensional graph structure that contains 1369 sites and 6600 edges~\cite{jiao_two-dimensional_2020}. 

Though the femtosecond-laser writing technique mostly is used for fabricating waveguide chips for implementing continuous-time quantum walks, it also provides a good platform for implementing discrete-time quantum walks. With a disordered lattice that is realized by an integrated array of interferometers fabricated in glass by femtosecond laser writing, Crespi et al. experimentally implemented the DTQW of eight steps and observed Anderson localization of entangled photons in a DTQW affected by position-dependent disorder~\cite{crespi_anderson_2013}. With a femtosecond laser-written waveguide chip, Sansoni et al. implemented an array of beam splitters for a four-step quantum walk, and they experimentally demonstrated how the particle statistics, either bosonic or fermionic, influence a two-particle discrete-time quantum walk~\cite{sansoni2012two}.

Silicon quantum photonics, taking advantage of its high component density, full programmability, and CMOS compatibility, becomes a very promising approach for implementing large-scale photonic quantum computing~\cite{bao_very-large-scale_2023,qiang2018large,wang2018multidimensional,qiang_implementing_2021,chi_programmable_2022} and universal quantum computing~\cite{bartolucci_fusion-based_2023}. With its well-developed on-chip photonic quantum state generation and manipulation, silicon quantum photonics also presents an ideal platform for implementing quantum walk simulations and quantum walk based algorithms. 
Silicon photonics can implement on-chip reconfigurable linear optical networks consisting of MZIs and phase shifters that are universal for arbitrary unitary operations. With such reconfigurable on-chip networks, the quantum walk evolutions on arbitrary graph structures can be implemented, and with multiple photons injected into the network, multi-photon quantum walks can be performed. In addition to that, with the control over entangled photons and reconfigurable optical networks on chip, the full programmability over quantum walks, such as Hamiltonian programmability, evolution time programmability, initial state programmability and particle features programmability, have been studied~\cite{sansoni2012two,matthews2013observing} and experimentally demonstrated in silicon quantum photonics~\cite{qiang_implementing_2021,wang_large-scale_2022,wang_experimental_2022}. 

In 2021, Qiang et al. first implemented a universal multi-particle quantum walk processor in silicon quantum photonics by exploiting an entanglement-driven scheme~\cite{qiang_implementing_2021}, and they realized quantum walks with full control over various parameters in one device, including evolution time, initial state, underlying graph structures, particle exchange symmetry and particle indistinguishability. The device can implement two-photon quantum walks on any five-vertex graph, with continuously tunable particle exchange symmetry and indistinguishability. With the device, they showed how this simulates single-particle walks on larger graphs, with size and geometry controlled by tuning the properties of the composite quantum walks. By employing such full programmability, different quantum walk based algorithms were experimentally demonstrated using the device, such as searching vertices in graphs and testing for graph isomorphism. The increasing integration scale of silicon quantum photonics enables a larger scale of quantum walk simulations. Using the same protocol of universal multi-particle quantum walk simulation, Wang et al. further implemented larger-scale quantum walks using a fully programmable large-scale silicon photonic quantum walk processor~\cite{wang_large-scale_2022}, which enables the simulation of quantum walk dynamics on graphs with up to 400 vertices. 
With this system, they demonstrated the exponentially faster hitting and quadratically faster mixing performance of quantum walks over classical random walks, achieving more than two orders of magnitude of enhancement in the experimental hitting efficiency and almost half of the reduction in the experimental evolution time for mixing, and implemented a bunch of large-scale quantum walk based applications, including measuring the vertex centrality and searching marked vertices in networks of up to hundreds of nodes. Moreover, such full programmability over quantum walks by using silicon quantum photonics provides ways for simulating other physics via dynamics behaviours of quantum walk evolution, for example, the topological phase of higher-order topological insulators can be simulated in the same quantum walk processor~\cite{wang_large-scale_2022}, and by implementing a time-reversal symmetry-breaking quantum walks on a reconfigurable silicon photonic device, the enhancement introduced by breaking time-reversal symmetry was experimentally demonstrated~\cite{wang_experimental_2022}.

\section{Quantum Walk Applications}

\begin{figure}[!htb]
    \centering
    \includegraphics[width=1.0\textwidth]{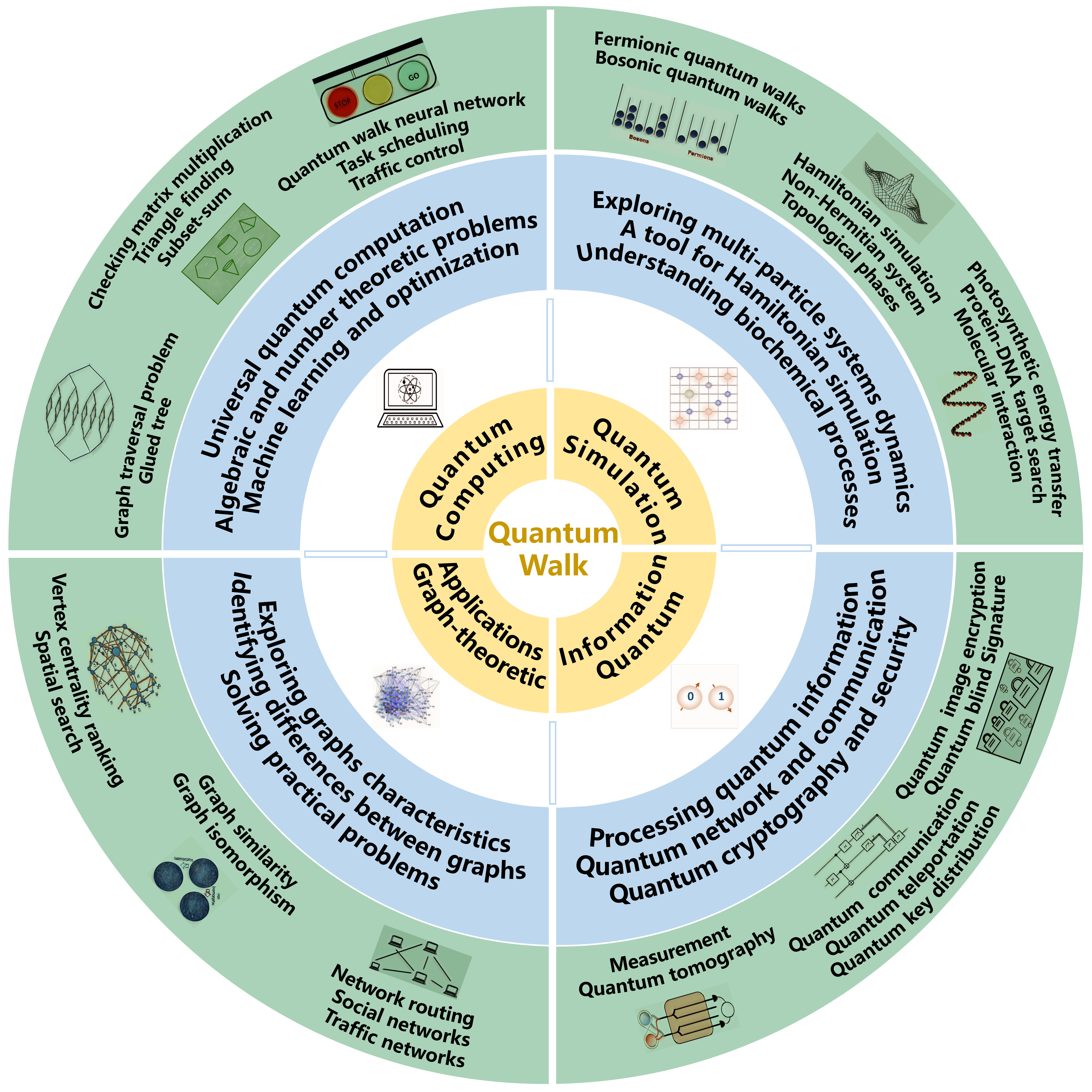}
    \caption{\textbf{Summary of quantum walk applications.} They are mainly divided into four categories: quantum computing, quantum simulation, quantum information processing, and graph-theoretic applications.}
    \label{fig:3}
\end{figure}

Along with the abundant development of quantum walk theories and rapid progress of quantum walk implementation techniques, the exploration of quantum walk applications shows flourishing development. Here we summarize these applications into four main categories: quantum computing, quantum simulation, quantum information processing and graph-theoretic applications, as shown in Figure~\ref{fig:3}. These four categories in fact can represent different aspects of exploiting the potentials of quantum walks. Firstly, as the quantum counterpart of the well-known classical random walk, quantum walks are naturally assumed to be capable of designing quantum walk based algorithms that may outperform their classical counterparts in a broad class of applications, and considering their capability for universal quantum computation, quantum walk models provide sufficient possibility for designing quantum algorithms. Secondly, quantum walk evolution describes a physical process of quantum particles evolving, and it presents a way for solving quantum simulation applications via quantum walks. On the one hand, quantum walk on its own provides a framework for studying multi-particle properties, e.g., bosonic, fermionic, and anyonic particles characteristics, and on the other hand, quantum walk can act as a general approach for simulating Hamiltonian dynamics and further in applications of molecular analysis and biochemical processes. Thirdly, quantum walks can be served as a comprehensive toolbox in the field of quantum information processing for the preparation, manipulation, characterization, and transmission of quantum states, and this paves a way for developing new techniques for quantum information processing applications, for example, based on quantum walks, novel protocols for quantum communications and new algorithms for quantum cryptography have been proposed. Lastly, since the underlying Hamiltonian that determines the evolution of quantum walks is directly associated with the structure of a graph, quantum walks become a promising tool for solving graph-theoretic applications, such as by exploring the structure of graph more efficiently or distinguishing the difference between graph structures more precisely. A variety of network applications can thus be addressed via quantum walks, by mapping them into related graph-theoretic problems.

\subsection{Quantum computing}

The critical part of quantum computation is to design appropriate quantum algorithms for different problems, which is quite a challenging task. Similar to the commonly used quantum circuit model in which a quantum circuit acts on a fixed number of qubits and terminates with a measurement, quantum walk models provide a new approach to design quantum computing algorithms, where a particular quantum walk evolution with input initial state is performed and the probability distribution of the output evolution state is usually sampled for extracting results. Quantum walk models become quite a versatile tool in quantum computation, on the one hand, they are proven to be universal for quantum computation, and thus any other quantum algorithms in theory can be implemented via quantum walks, on the other hand, quantum walks are the quantum counterparts of the well-known classical random walks, and thus it opens up a new avenue for designing quantum walk algorithms in various applications that can outperform over the classical random walk algorithms.

Quantum walk has been shown to be capable of performing universal quantum computation. The power of quantum walk as a universal model of quantum computation was first explored by Childs in 2009, where he demonstrated that CTQW can be used for universal computation in the sense that any quantum computation can be simulated by a quantum walk on a sparse, unweighted graph of low degree~\cite{childs2009universal}. Afterward, Childs, Gosset, and Webb further considered the multi-particle quantum walks with interactions and showed that any $n$-qubit circuit with $g$ gates can be simulated by the dynamics of $O(n)$ particles that interact for a time $\poly(n,g)$ on a weighted planar graph of maximum degree four with $\poly(n,g)$ vertices, which means a multi-particle quantum walk is capable of performing efficient universal quantum computation~\cite{childs2013universal}. In 2010, Lovett et al. showed that DTQW is also capable of universal quantum computation, and thus can be regarded as a computational primitive as CTQW~\cite{lovett2010universal}. Furthermore, Underwood and Feder extended this concept to discontinuous quantum walk in which a quantum walker takes discrete steps of continuous-time quantum walk evolution~\cite{Underwood_universal_2010}, and they showed that it is also a universal model of quantum computation. These studies have collectively shown that quantum walk models are capable of universal quantum computation, which on the one hand may inspire new architecture of a quantum computer that does not require complex control, and on the other hand, provide tools for designing quantum algorithms and analyzing quantum computation theory.

Quantum walks have been widely used in designing quantum algorithms that can achieve computational speedup over classical algorithms in various problems, particularly for algebraic and number theoretic problems. Typically for some black-box problems, quantum walk algorithms can obtain exponential speedups. A well-known example is the glued-tree problem~\cite{childs2003exponential}. A glued tree is a graph obtained by taking two binary trees of equal height and connecting each leaf of one tree to exactly two leaves of the other tree, so that the degree of each node that was a leaf in one of the original trees is now exactly $3$. It has been shown that the time of CTQW on this glued tree can reach the right root from the left one is exponentially faster than classical random walk~\cite{childs2003exponential}, and a proof-of-principle experimental demonstration was also presented~\cite{shi2020quantum}. 
 Another example is the quantum algorithm for finding hidden nonlinear structures over finite fields~\cite{childs2007quantum}. The algorithm is based on quantum walk on a graph that encodes the structure of the black box function, and an exponential speedup arises from the quantum interference effect that allows the quantum walker to explore the search space more efficiently than classical algorithms. 

 For quite some algebraic problems, quantum walks show the potential of achieving polynomial computational speedup over classical algorithms, leading to faster and more efficient solutions. For example, Ambainis showed an optimal quantum algorithm for the element distinctness problem~\cite{ambainis2007quantum}, where for determining whether all elements of a list of size $N$ are distinct, only $\Theta(N^{2/3})$ queries are required using a quantum walk algorithm, while classically $\Omega (N)$ queries are required. In the problem of formula evaluation, a formula is encoded into a tree with a gate at each internal node and an input bit at each leaf node, and then the task is to evaluate the output of the root node when given oracle access to the input. With a quantum walk based algorithm, the formula evaluation problem can be solved using much fewer queries, achieving a polynomial speedup compared to classical algorithms~\cite{Reichardt_span_2008}. In addition, quantum walk has been applied to other problems in algebraic and number theory, such as checking matrix multiplication~\cite{Buhrman_quantum_2006}, testing group commutativity~\cite{Magniez_quantum_2007}, subset-sum~\cite{Becker_improved_2011}, and triangle finding~\cite{magniez2007quantum}. These quantum walk algebraic algorithms can have significant implications for applications in various fields, including cryptography, data analysis, drug discovery, and optimization problems, and become the basis for designing other quantum computing algorithms or protocols.

Quantum walk models have also been investigated in the fields of machine learning and optimizations, showing promising potential in designing quantum neural networks of new structures and algorithms of quantum machine learning and optimizations. For example, Dernbach et al. proposed a quantum walk neural network~\cite{dernbach_quantum_2019}, where the neural network architecture learns a quantum walk on a graph through learning the coin operator, and the network then uses this learned quantum walk to form a diffusion operator to act on the input data. They obtained competitive results when compared to other graph neural network approaches. Schuld et al. studied to use quantum walks as an approach to quantum neural networks (QNNs), where they use the formalism of quantum walks in order to find the dynamic evolution of a QNN that optimizes the computational properties of classical neural networks~\cite{schuld_quantum_2014}. On the other hand, quantum walks are explored to enhance the classical neural networks, such as in training a classical artificial neural network where a quantum walk algorithm is applied to search the weight set values to train the neural network~\cite{de_souza_classical_2022}. 
For addressing the usually NP-hard optimization problems, a quantum walk optimization algorithm (QWOA) was developed by Marsh and Wang~\cite{marsh_quantum_2019,marsh_combinatorial_2020}, which utilizes an efficient indexing algorithm in conjunction with a generalization of the QAOA mixing operator to a continuous-time quantum walk over a circulant graph that connects all feasible solutions. 
The algorithm was further applied to the NP-hard problem of portfolio optimization with discrete asset constraints, in finding high-quality solutions to the portfolio optimization problem~\cite{slate_quantum_2021}, and was also demonstrated the applicability of QWOA to the capacitated vehicle routing problem~\cite{bennett_quantum_2021}, where the algorithm achieves expected qualities with the less computational effort that required by classical random sampling. Other optimization algorithms based on quantum walks are also proposed, such as the clustering algorithm based on one-dimensional quantum random walks which were applied in network cluster server traffic control and task scheduling~\cite{yumin_novel_2014}, and the quantum backtracking algorithm proposed by Montanaro which consists in solving a constraint satisfaction problem by constructing a backtracking tree and exploring it using a quantum walk~\cite{montanaro_quantum-walk_2018}.

\subsection{Quantum simulation}

Quantum simulation is using a controllable quantum system to simulate the behaviour of other quantum systems. It aims to gain insights into complex quantum phenomena that are difficult or impossible to analyze using classical computers. 
It is a rapidly growing field ranging from molecules, materials to quantum field theories. Quantum walks can be used as a tool for quantum simulations, which has potential applications in simulating multi-particle systems, solving complex physical problems, and understanding biochemical processes. By using quantum walks to simulate the behaviour of these systems, one can gain in-depth knowledge about their properties and behaviours.


\begin{figure}[htbp]
     \centering
     \includegraphics[width=1.0\textwidth]{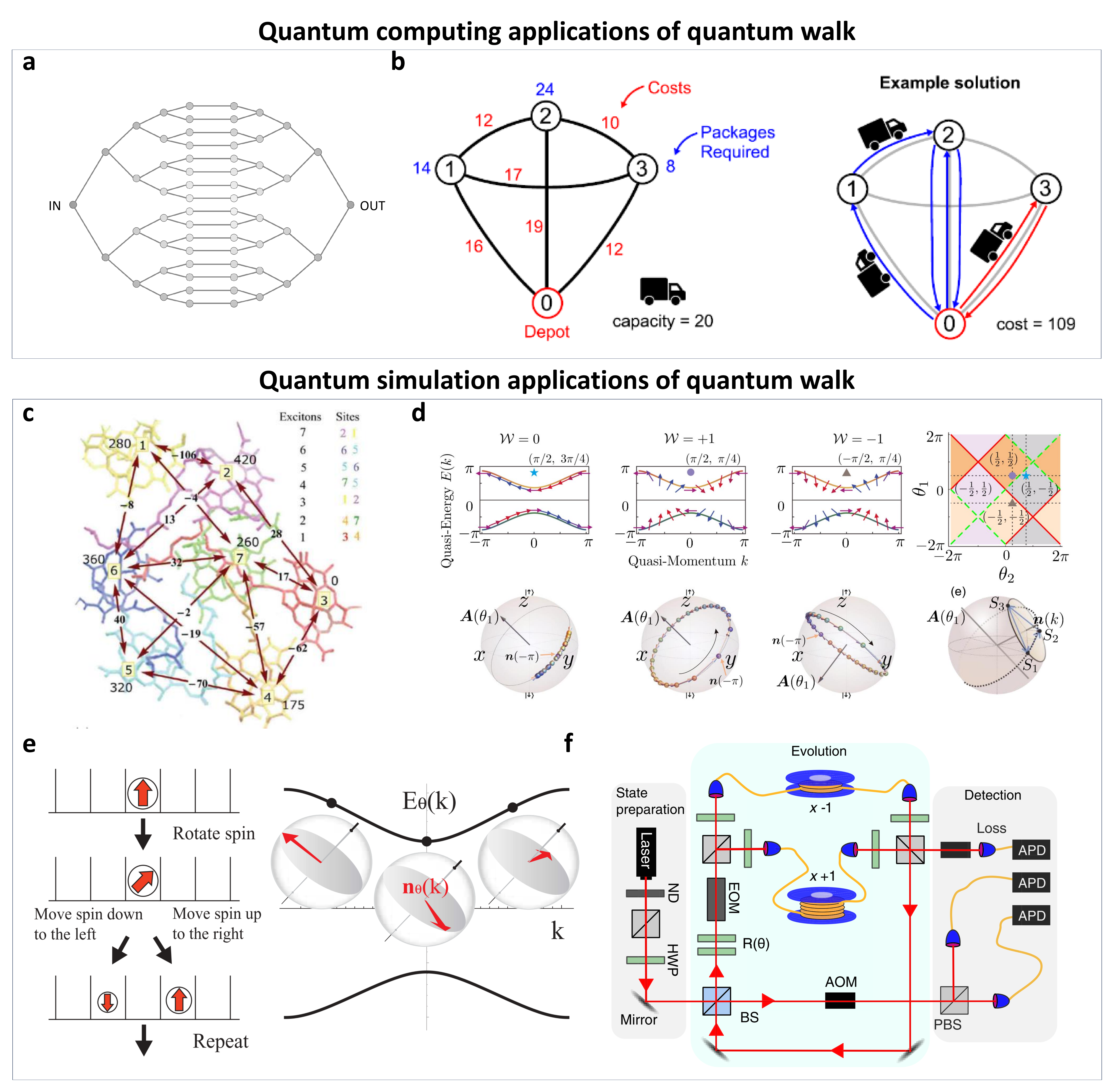}
     \caption{\small{\textbf{Quantum computing and quantum simulation applications of quantum walk.} \textbf{a}, Schematic diagram of the glued tree problem for which CTQW achieves exponential speedup over classical walk~\cite{childs2003exponential}. \textbf{b}, An example of a Capacitated Vehicle Routing Problem of size $n=3$ and with vehicle capacity $V=20$. Reprinted from Ref~\cite{bennett_quantum_2021}Copyright\copyright~2021 Bennett, Matwiejew, Marsh and Wang. \textbf{c}, The structure of the FMO protein. Quantum walk can enhance the excitation energy transfer for the FMO complex. Reprinted figure with permission from Ref~\cite{mohseni2008environment} Copyright \copyright2008 by the American Physical Society. \textbf{d}, The topological properties of a split-step discrete-time quantum walk in the standard time frame. Reprinted figure with permission from Ref~\cite{xu_measuring_2018} Copyright \copyright2018 by the American Physical Society. \textbf{e}, The realization of one-dimensional topological phases in discrete-time quantum walks. Reprinted figure with permission from Ref~\cite{Kitagawa_exploring_2010} Copyright \copyright2010 by the American Physical Society. \textbf{f}, A time-multiplexed implementation of the photonic quantum walk to simulate non-Hermitian quasicrystals. Reprinted figure with permission from Ref~\cite{lin_topological_2022} Copyright \copyright2022 by the American Physical Society.}}
     \label{fig:7}
 \end{figure}

Quantum walk evolution on its own provides a path for exploring multi-particle quantum dynamics and various characteristics. The applications of multi-particle quantum walks can depend on various factors, including the number, exchange symmetry, and indistinguishability of the particles involved, and the underlying graph structures where they move. Furthermore, entangled photons can be used to simulate multi-particle quantum walks with bosonic, fermionic or anyonic statistics by way of an entanglement-based scheme. 
This method was first used in Refs~\cite{sansoni2012two, matthews2013observing}, where both CTQW and DTQW of two particles have been experimentally demonstrated. Specifically, Sansoni et al. demonstrated the implementation of a two-particle bosonic-fermionic quantum walk using integrated photonics~\cite{sansoni2012two}. Matthews et al. demonstrated the simulation of fermionic statistics with photons in one-dimensional CTQW evolution~\cite{matthews2013observing}. These experiments show that differences in particle exchange symmetry can affect the quantum walk dynamics, due to the Pauli exclusion principle (Fermi-Dirac statistics) or Bose-Einstein statistics. With an extended entanglement-based protocol, Qiang et al. showed that the CTQW evolutions of multiple particles with both tunable particle exchange symmetry and indistinguishability by adding control over the entangled photons~\cite{qiang_implementing_2021}. In 2022, Wang et al. realized a large-scale full-programmable quantum walk~\cite{wang_large-scale_2022} where they demonstrated the use of this platform to simulate the behaviours of multi-particle systems and explore topological phases in quantum systems using quantum walks, including higher-order topological insulators. On the other hand, a larger quantum walk can be used to simulate the behaviours of multi-particle quantum walks on a smaller graph to investigate multiple particle dynamics. In 2012, Schreiber et al. demonstrated the use of a two-dimensional quantum walk on a larger graph to simulate the dynamics of two interacting particles on a one-dimensional graph~\cite{schreiber20122d}. 

Quantum walk can act as a general approach for simulating quantum Hamiltonian dynamics and further investigating their various properties. Berry et al. proposed algorithms for simulating non-sparse Hamiltonians with nearly optimal dependence on all parameters by implementing a combination of quantum walk steps with fractional-query simulation~\cite{berry2015hamiltonian}. They also proposed an algorithm for simulating Hamiltonians using a black-box approach~\cite{berry2012black}, where the Hamiltonian is treated as an oracle. These algorithms aim to improve the efficiency of Hamiltonian simulation, especially for non-sparse Hamiltonians. Furthermore, Berry and Novo proposed a method to simulate Hamiltonian evolution by repeatedly using a superposition of steps of a quantum walk, and then applying a correction to the weightings for the numbers of steps of the quantum walk~\cite{Berry_corrected_2016}. In conventional quantum mechanics, the Hamiltonian describing a quantum system must be Hermitian, and the corresponding evolution is a unitary process. However, in practice the physical quantum systems are typically open, meaning that they interact with the environment and exchange energy and information with them, resulting in non-Hermitian systems. One approach to simulating non-Hermitian systems using quantum walks is to introduce a complex potential into the walk~\cite{Rudner_topological_2009, Weidemann_topological_2022, lin_topological_2022}, which breaks the symmetry of the system and allows for the simulation of non-Hermitian behaviour. Another approach is to implement a non-unitary quantum walk~\cite{Mittal_persistence_2021, lin_observation_2022}, which allows for the simulation of non-Hermitian systems without requiring the introduction of a complex potential. Parity-time (PT) symmetric systems are a special type of non-Hermitian systems, where eigenenergies of the PT-symmetric non-Hermitian Hamiltonian are entirely real. In 2017, Xiao et. al proposed a quantum walk model that is suitable for PT-symmetric systems, and they first experimentally realized PT-symmetric discrete-time quantum walks with single photons~\cite{xiao_observation_2017}.

Topological phases have unique properties that challenge our understanding of phases and phase transitions. Instead of being characterized by local order parameters, these phases are defined by non-local topological invariants that determine the existence and number of topological edge states at the interface between the bulk and boundary regions, through what is known as the bulk-boundary correspondence. Quantum walks provide new insights into how topological phenomena can be simulated and studied in quantum dynamics, including exploring topological phase transitions~\cite{wang_simulating_2019}, measuring topological invariant~\cite{zhan_detecting_2017}, and simulating topological phases and edge states~\cite{xiao_observation_2017}. For example, Kitagawa et al. observed topologically protected bound states in photonic quantum walks~\cite{Kitagawa_observation_2012}, Asbóth explored the role of symmetries in one-dimensional quantum walks~\cite{Asboth_symmetries_2012}, while Panahiyan and Fritzsche simulated multiphase configurations, phase transitions, and edge states with quantum walks by utilizing a step-dependent coin~\cite{Panahiyan_simulation_2019, Panahiyan_controllable_2020}. Quantum walks can also be used to simulate topological phases in non-unitary systems. Mittal et al. investigated the persistence of topological phases in non-Hermitian quantum walks~\cite{Mittal_persistence_2021}. They extended the one-dimensional split-step quantum walk model to a non-unitary quantum walk and showed the persistence of topological phases. In addition, non-unitary Floquet operators have been used to explore topological properties of non-Hermitian systems. By imposing PT symmetry, Xiao et al. realized and investigated Floquet topological phases driven by PT-symmetric quantum walks~\cite{xiao_observation_2017}.

Quantum walks are also used for simulating and understanding complex biochemical processes, such as photosynthetic energy transfer, the behaviour of biological molecules like proteins and DNA, and chemical reactions. Mohseni et al. used the quantum walk to simulate the energy transfer process in a photosynthetic system~\cite{mohseni2008environment}. Quantum walks can enhance the efficiency of energy transfer and conversion ~\cite{Karafyllidis_quantum_2017}, which is crucial for energy harvesting and utilization in biological systems, while Hoyer et al. showed that there is a limit to the quantum speedup in these processes~\cite{hoyer2010limits}. Quantum walks can be also used to simulate the protein-DNA target search and DNA sequence assembly process~\cite{ D'Acunto_protein_2021, Varsamis_quantum_2023}, and it was shown that they can improve the efficiency and accuracy of the search and assembly. Varsamis and Karafyllidis proposed a quantum walks-assisted algorithm for peptide and protein folding prediction~\cite{Varsamis_quantum_2023}, which is important for understanding protein function and further drug design. The essence of chemical reactions is the process of interactions between molecules, and quantum walks have been used for simulating the interactions between molecules, in order to obtain a better understanding of the mechanism and properties of chemical reactions. Chia et al. explored the use of quantum walks in coherent chemical kinetics~\cite{Chia_Coherent1_2016, chia2016coherent}, and developed reaction operators for radical pairs and found that quantum walks can be used to model and predict the behaviour of radical pairs in this system. Overall, the use of quantum walks for biochemical simulations can lead to new insights and possible breakthroughs in fields such as drug discovery and bioengineering.

\subsection{Quantum information processing}

Quantum walks have been extensively studied in the field of quantum information processing and have been used as a toolbox for the preparation, manipulation, characterization, and transmission of quantum states (information). Furthermore, various protocols for quantum communication and routing based on quantum walks are developed. One of the key advantages of using quantum walk in quantum information processing is the ability to encode and process quantum information in a highly controlled and tunable manner. Quantum walks have thus been explored in various tasks of quantum information processing, including entanglement generation~\cite{Innocenti_quantum_2017,Vieira_dynamically_2013,Moulieras_entanglement_2013}, perfect state transfer, quantum state measurement~\cite{bian_experimental_2014,zhao_experimental_2015}, quantum teleportation~\cite{li_new_2019}, quantum communication~\cite{yang_quantum_2018,chen_quantum_2019}, and quantum cryptography~\cite{abd_encryption_2019}. 

\begin{figure}[htbp]
    \centering
    \includegraphics[width=1.0\textwidth]{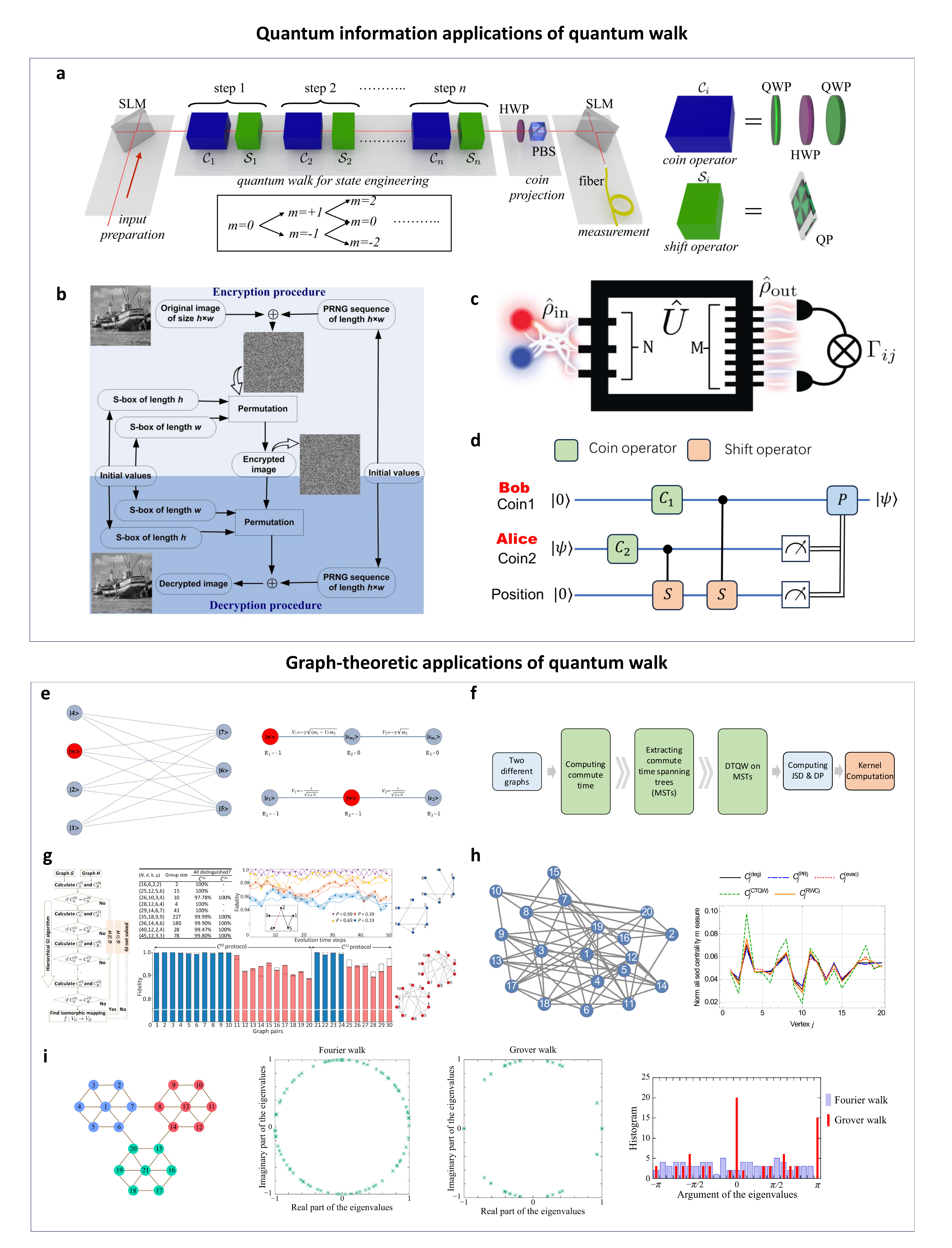}
    \caption{\small{\textbf{Examples of quantum information processing and graph-theoretic applications of quantum walk.} \textbf{a}, Scheme for the proposed implementation of the quantum walk state engineering protocol}}
\label{fig:4}
\end{figure}
\begin{figure}
\caption*{using orbital angular momentum (OAM) and spin angular momentum of a photon. Reprinted figure with permission from Ref~\cite{Innocenti_quantum_2017} Copyright \copyright2017 by the American Physical Society. \textbf{b}, General framework for quantum-inspired cascaded DTQW image encryption technique, reproduced from Ref~\cite{Abd_quantum_2020}. \textbf{c}, Diagram of a hybrid quantum walk for two-photon tomography, reproduced from Ref~\cite{Titchener_two_2016}. \textbf{d}, Quantum teleportation with two-coin quantum walks on a line~\cite{wang_generalized_2017}. \textbf{e}, Quantum walk based spatial search on complete bipartite graph $K_{4,3}$ with solution node $ |w\rangle $, reproduced from Ref~\cite{novo_systematic_2015}. \textbf{f}, The framework to compute the similarity between two complete weighted graphs using quantum walk based algorithm~\cite{lu_quantum_2020}. \textbf{g}, Experimental demonstration of quantum walk based GI algorithm. From~\cite{qiang_implementing_2021}. \copyright The Authors, some rights reserved; exclusive licensee AAAS. Distributed under a CC BY-NC 4.0 license http://creativecommons.org/licenses/by-nc/4.0/”.  Reprinted with permission from AAAS. \textbf{h}, Vertex centrality measure of randomly generated Erdős-Rényi graph $G(20,0.3)$. Reprinted figure with permission from Ref~\cite{izaac2017centrality} Copyright \copyright2017 by the American Physical Society. \textbf{i}, DTQW on complex networks for community detection on a prototypical three-community network, reproduced from Ref~\cite{Mukai_discrete_2020}. }
\end{figure}

Quantum entanglement is a key resource for quantum information processing. Quantum walks have been shown to be a powerful tool for generating and manipulating quantum entanglement. By controlling the evolution of quantum walks, necessary entanglement can be generated via quantum walks. In 2013, Vieira et al. proposed a dynamically disordered quantum walk that is designed to maximize the entanglement between two particles by introducing randomness into the evolution of the quantum walk~\cite{Vieira_dynamically_2013}, which can generate more entanglement than other types of quantum walks and meanwhile the entanglement can be controlled by adjusting the parameters of the walk. Gratsea et al. proposed a scheme for preparing hybrid maximally entangled states of two qubits by controlling the evolution of a quantum walk on a one-dimensional lattice~\cite{Gratsea_generation_2020}, which involves encoding the target state into the initial state of the walk and controlling the walk evolution using a sequence of unitary operations. Using a discrete-time quantum walk, one can also prepare arbitrary superposition states of two qubits~\cite{Innocenti_quantum_2017}. Additionally, quantum walks were used to control the degree of entanglement between particles and to generate entangled states with specific properties, such as maximal entanglement~\cite{Moulieras_entanglement_2013, Vieira_dynamically_2013}. Furthermore, quantum walks can be used to study the properties of entangled states and to test the principles of quantum mechanics. For example, Rohde et al. studied the entanglement dynamics of discrete-time quantum walks and showed that the entanglement between the walker and the environment can exhibit quasi-periodic behaviour~\cite{Rohde_entanglement_2012}. Melnikov and Fedichkin have also studied the entanglement properties of quantum walks of interacting fermions on a cycle graph~\cite{Rohde_entanglement_2012}.

The ability to accurately determine the quantum state of a system is crucial for many applications in quantum information processing. Quantum walks possess significant potential and advantages in implementing the characterization and measurement of quantum states. Firstly, quantum walks can realize generalized measurements and non-local quantum state measurements. In 2013, Kurzynski and Wojcik proposed that quantum walks can be viewed as a generalized measurement device~\cite{Kurzynski_quantum_2013} and they showed that a one-dimensional discrete-time quantum walk can be used to implement a generalized measurement in terms of a positive operator value measure (POVM) on a single qubit. Bian et al.~\cite{bian_experimental_2014} and Zhao et al.~\cite{zhao_experimental_2015} respectively used one-dimensional photonic quantum walks to realize POVM measurements of a single qubit. Additionally, quantum walks can be used to achieve efficient quantum state reconstruction and tomography. Titchener et al. used two-photon Thomson scattering and quantum walks to implement tomography of two-photon states~\cite{Titchener_two_2016}. They employed a hybrid quantum walk within on-chip coupled waveguide arrays as the measurement process and successfully reconstructed the input two-photon density matrix. Hou et al. achieved deterministic collective measurements using photonic quantum walks~\cite{hou_deterministic_2018}, where they used a collective symmetric informationally POVM that is realized through a photonic quantum walk, to perform qubit state tomography with unprecedented accuracy.

Quantum walks have also been explored as a tool for quantum state transmission, and further for applications of quantum communication, including quantum networks, quantum teleportation, and quantum key distribution. By encoding information in the state of a quantum walker, it is possible to transmit quantum states over long distances. Targeting quantum communication applications, Yang et al.~\cite{yang_quantum_2018} and Chen et al.~\cite{chen_quantum_2019} proposed a novel discrete-time quantum walk approach for quantum network communication, and presented a quantum multi-unicast communication scheme and perfect state transfer scheme over the butterfly network and the inverted crown network, respectively. Shang et al. proposed quantum communication protocols based on quantum walks with two coins~\cite{shang_quantum_2018}, where the information is encoded in the coin state and transferred to any target position using two coin operators. Srikara and Chandrashekar developed two quantum direct communication protocols, namely quantum secure direct communication protocol and controlled quantum dialogue protocol, using DTQWs on a cycle graph~\cite{Srikara_quantum_2020}. Panda et al. proposed a quantum secure direct communication protocol based on the recurrence phenomenon of k-cycle DTQW and the entanglement generated during the evolution in an optical setup~\cite{panda_quantum_2023}. In quantum teleportation applications, Wang et al. proposed a generalized quantum teleportation protocol based on quantum walks with two coins~\cite{wang_generalized_2017}, where an unknown qubit state can be teleported using quantum walks on a line and quantum walks on a cycle with four vertices. This protocol can transmit any unknown $d$-dimensional states without the prior entangled state, making it more flexible than existing $d$-dimensional state teleportation protocols. In addition to single-qubit teleportation, quantum walks can also realize multi-qubit teleportation. Li et al. proposed a multi-particle quantum walk based teleportation scheme~\cite{li_new_2019}, where the entanglement between walkers and their respective coins is generated by the step operators in multi-particle QWs. This protocol allows for flexibly choosing the transmission path and enables more efficient transmission of quantum states. Moreover, quantum walks can also be used to implement a quantum key distribution (QKD) protocol. Vlachou et al. proposed a QKD scheme based on quantum walks~\cite{Vlachou_quantum_2018}, which includes verification procedures against full man-in-the-middle attacks and has provable unconditional security.

In applications of quantum cryptography and security, quantum walks have been extensively explored for designing novel secure schemes and also attacking various cryptographic schemes. For example, quantum walks have been used in quantum image encryption~\cite{abd_encryption_2019, abd_optical_2021, su_robust_2022, su_quantum_2023}, steganography~\cite{abd_robust_2020, abd_secret_2020, abd_novel_2019}, generating high-quality random numbers~\cite{Abd_quantum_2020,abd_controlled_2020}, quantum cryptanalysis~\cite{Kaplan_quantum_2016} and attacking lattice-based cryptography~\cite{Chailloux_lattice_2021}. Quantum walk can be used for encrypting colour images using controlled one-particle quantum walks~\cite{abd_encryption_2019}, two-particle quantum walks~\cite{su_quantum_2023}, two-dimensional quantum walks and quantum coding. It can also be used in combination with Gray code and Henon map for cloud-based image encryption~\cite{abd_novel_2023}. Quantum walk has also been applied in steganography~\cite{abd_robust_2020, abd_secret_2020, abd_novel_2019}, which can achieve high security and can be used for cloud-based image steganography and e-healthcare platforms. Several Refs~\cite{abd_quantum-inspired_2021, abd_secure_2020, Abd_efficient_2019} propose security protocols based on quantum walks, which can be used to protect data and information in various scenarios, such as 5G networks and IoT-based smart cities. On the other hand, quantum walks can be used for attacking cryptographic schemes. Kaplan et al. proposed a quantum-based differential and linear cryptanalysis method that can be used to break some traditional symmetric cryptography algorithms~\cite{Kaplan_quantum_2016}. Lattice-based cryptography is one of the leading candidates for post-quantum cryptography. The Shortest Vector Problem (SVP) is a crucial problem for the cryptanalysis of lattice-based cryptography. Chailloux and Loyer have proposed a heuristic algorithm based on quantum walks for solving the SVP, which significantly improves the running time of the algorithm[216]. Another common problem in cryptanalysis is collision finding, which has been extensively studied using both classical and quantum algorithms. To improve the efficiency of the algorithm for finding multiple collision pairs, Bonnetain et al. have introduced a chained quantum walk algorithm that significantly reduces the algorithm's complexity~\cite{bonnetain2023finding}.

 \subsection{Graph-theoretic applications}

 The underlying Hamiltonian that determines the evolution of quantum walks is directly associated with the structure of a graph, where, for example, in CTQW each vertex of the graph encodes a quantum basis state and each edge represents a possible transition between two quantum basis states. Quantum walks thus become a promising tool for solving graph-theoretic problems and further various network applications via mapping them into related graph-theoretic problems. Quantum walks possess distinct characteristics on a graph compared to classical walks and enable the exploration of graph structure more efficiently, such as searching marked vertices and ranking vertex importance. Quantum walks show particular effectiveness in identifying structural differences between graphs and have been intensively studied for graph comparison applications. Furthermore, a variety of network applications have been studied by mapping onto graph-theoretic problems and further addressed by using quantum walks. Here we summarize the recent advancement of quantum walks used in graph-theoretic and network applications.  

Compared to classical random walks, quantum walks are more efficient in exploring graph characteristics by performing quantum walk evolutions on the graph. It has been shown that it is possible to achieve faster hitting time and mixing time compared to classical random walks~\cite{aharonov2001quantum, Godsil_discrete_2019}. This further enables a more accurate estimation of the spectrum and characterization of related properties such as connectivity and expansion property. The graph's spectrum generally indicates essential features and properties of the graph, such as its diameter, degree distribution, and clustering coefficient~\cite{zhan_perfect_2014}. Based on this, quantum walk based algorithms are proposed for exploring graph structural characteristics, such as searching marked vertices and ranking vertex centrality, which can outperform their classical counterparts in different aspects. In 2003, Childs and Goldstone first proposed a CTQW-based algorithm for spatial search problems, i.e., searching marked vertices in graphs, and they showed that quantum walks can achieve a quadratic speedup over classical algorithms for searching a marked vertex on an unweighted graph of $N$ vertices~\cite{childs2004spatial}, where the complexity of quantum walk algorithm is $O(\sqrt{N})$ but the classical algorithms are $O(N)$. A list of works has been shown afterward for further investigation of CTQWs used in the spatial search problems~\cite{Janmark_global_2014, Wong2015Grover, Chakraborty2020Finding, Apers2022Quadratic, novo_systematic_2015}. For example, Janmark et al. demonstrated that global symmetry is not necessary for fast quantum search using CTQWs~\cite{Janmark_global_2014}. They proposed a new algorithm that achieves the same speedup as the standard Grover search algorithm using a non-symmetric Hamiltonian. Wong and Thomas proposed a variant of Grover search using lackadaisical CTQWs that allows the quantum walker to rest occasionally during the search, reducing resource requirements~\cite{Wong2015Grover}. Furthermore, quantum walk based algorithms were used for searching in marked vertices in weighted graphs~\cite{wong_faster_2015, wong_coined_2017, wang_optimal_2020}, dynamic graphs~\cite{chakraborty2017optimal, Herrman_continuous_2019, cattaneo2018quantum}, and also searching marked edges in graphs~\cite{Teixeira_walking_2023}. Apart from CTQW, the discrete quantum walk models, including Szegedy's walk, staggered quantum walk, and coin-based quantum walk, have also been shown to solve the spatial searching problem~\cite{santos2016szegedy, portugal2016staggered-0, portugal2016staggered-1, santha2008quantum}. Quantum walk based search is attractive because it is expected to have useful properties with regard to their robustness to noise and ease of physical implementation~\cite{cattaneo2018quantum}, and various experimental demonstrations have been presented. For example, by using a bulk optics quantum system~\cite{qu2022deterministic}, Qu et al. experimentally demonstrated deterministic search on star graphs via quantum walks. With a programmable silicon photonic chip, Qiang et al. implemented quantum walk based search on various graphs for both single marked vertices and multiple marked vertices~\cite{qiang_implementing_2021}, and they further experimentally demonstrated, on a large-scale full-programmable quantum walk system, quantum walk based search on graphs with 10 to 210 vertices~\cite{wang_large-scale_2022}.

In addition to searching marked vertices, quantum walks show significant potential in the problem of ranking vertex centrality. Vertex centrality is an integral tool in graph theory and network analysis, and it can be used for quantifying the importance of each vertex. A higher vertex centrality measure means more importance of the corresponding vertex in the graph. Vertex centrality measure has been widely used in ranking the web pages on the Internet which is well-known as PageRank algorithms, and other applications like identifying critical infrastructure nodes in social and urban networks. The PageRank algorithm was first proposed by Paparo et al. based on Szegedy quantum walk models~\cite{paparo2012google,paparo2013quantum}, and it shows more sensitivity compared to the classical algorithm, which is capable of highlighting the secondary hubs and resolving the degeneracy of low-lying nodes in the network~\cite{paparo2013quantum}. By using CTQW models, an alternative quantum centrality measure was also proposed~\cite{izaac2017centrality}, and it requires a smaller Hilbert space compared to the SQW-based algorithm. The CTQW-based vertex centrality ranking algorithm is furthermore extended to the situations of centrality measure on directed networks~\cite{izaac2017quantum} and weighted graphs~\cite{Wang2022ContinuoustimeQW}. These algorithms have also been experimentally demonstrated, including ranking vertices on directed graphs using bulk optics quantum systems~\cite{wang2020experimental,wu2020experimental}, and testing quantum centrality ranking for scale-free networks of 55 vertices using a fully programmable silicon quantum photonic processor~\cite{wang_large-scale_2022} where the experimental results achieved high fidelities with the theoretical predictions.

Quantum walks have also demonstrated their effectiveness in identifying structural differences between graphs, and are further used for several kinds of graph comparison problems, such as distinguishing non-isomorphic graph pairs, measuring similarity between graphs, and solving graph classification. 
Graph isomorphism (GI) problem is to determine whether two graphs have the same structure. It is thought to be a neither NP-complete nor polynomial time problem, and the best current proven general algorithm is in quasi-polynomial time~\cite{babai2016graph}. A list of quantum walk based GI algorithms has been proposed, showing the potential of quantum walk in solving GI problems~\cite{shiau2005physically,emms2006matrix,douglas2008classical,emms2009graph-2,emms2009graph,gamble2010two,berry2011two,rudinger2012noninteracting,qiang2012enhanced,smith2012algebraic,wang2015graph,qiang_implementing_2021}. These algorithms are mainly different in the aspects of the used models of quantum walks (i.e., discrete-time or continuous-time), the number of the particles involved, the presence of particle interactions, localized inhomogeneities, and the ways of constructing graph certificates. 
For CTQW, the single-particle model has achieved an improved distinguishing ability for SRGs with 64 vertices by adding phases to edges~\cite{qiang_implementing_2021}, while both interacting and non-interacting multi-particle CTQW, especially like two bosons with interaction~\cite{gamble2010two} and four fermions with non-interaction~\cite{rudinger2012noninteracting}, can also distinguish non-isomorphic SRGs with same scale.
Strong distinguishability also appears in the single-particle discrete model, for example, as the level of search-based signature increases, the isomorphism algorithm can obtain a stronger distinguishing ability even for SRGs up to 64 vertices~\cite{wang2015graph}. It is noteworthy that although none of the proposed quantum walk based GI algorithms have been proven analytically to be able to distinguish all graphs in polynomial time, their capability of distinguishing nonisomorphic graphs has been tested by abundant numerical simulations on wide classes of graphs. Moreover, some of the algorithms, e.g., CTQW-based GI algorithm have been extensively experimentally demonstrated on a fully programmable silicon photonic quantum walk processors~\cite{qiang_implementing_2021,wang_large-scale_2022}, where the sizes of experimentally tested graphs are up to 210 vertices~\cite{wang_large-scale_2022}.
Based on the distinguishability of nonisomorphic graphs, quantum walks can further provide an effective measure of the similarity between graphs. There are two ways to use quantum walks to measure graph similarity: one is to use a similar way of calculating the fidelity between the graph certificates that are directly constructed through the output probability distribution of the evolved states, and the other is to use the complete quantum state to compute graph kernels~\cite{bai2015quantum}. Quantum Jensen-Shannon divergence and related methods~\cite{lamberti2008metric,zhang2020r} show strong separating ability and have become a popular method of computing graph kernel, which has applied to both DTQW~\cite{bai2015quantum} and CTQW~\cite{bai2013quantum} models. Furthermore, computing the similarity matrix of the graph set as a key procedure can be directly used in classical classification models, leading to a better performance in graph classification problems~\cite{bai2015quantum,zhang2020r}.

Based on the studies of graph-theoretic quantum walk algorithms, more applications of practical interests in various fields can be further solved using quantum walk algorithms, such as network routing~\cite{zhan_perfect_2014}, social network analysis~\cite{zhang_quantum_2021}, community detection~\cite{Mukai_discrete_2020}, and information propagation modeling~\cite{yan_information_2022}. These applications in general can be modelled into graph-theoretic problems, and quantum walk algorithms are used to find optimal solutions with speedup over classical algorithms or explore complex dynamics and in-depth structural analysis via quantum walk dynamics. 
In the application of network routing, the network can be modelled into a graph where vertices in the graph represent network routers or switches and edges represent the physical links connecting these devices, and then the scheme for realizing quantum routing was based on controllable perfect state transfer via discrete-time quantum walks~\cite{zhan_perfect_2014}. 
In the field of social networks, by modelling vertices as users and edges as social relationships between users, complicated social networks can be mapped into large-scale graphs and quantum walks can be applied to explore and analyze users' behaviours and social trends~\cite{zhang_quantum_2021}, which further helps to conduct in-depth analysis of the structure and dynamics of social networks, for example, quantum walks have been shown to define specific centrality metrics in terms of node occupation on single-layer and multi-layer network, and further to identify leaders in criminal networks~\cite{ficara_classical_nodate}. 
Apart from information networks and social networks, quantum walk based algorithms are also proposed for solving problems in other networks including traffic networks~\cite{hu_analyzing_2022}, unmanned aerial vehicle (UAV) communications~\cite{liang2022hadamard}, and sensor networks~\cite{singh_localization_2022}. Specifically, a method named multi-scale characteristic analysis for online car-hailing traffic volume with quantum walks has been proposed, demonstrating the application of quantum walk algorithms in traffic optimizations~\cite{hu_analyzing_2022}. A Hadamard coin-driven quantum walk model~\cite{liang2022hadamard} is proposed to identify the important edges of undirected complex networks and is further deployed in a dynamic complex network involving a typical communication scenario found in UAV swarms to select significant UAV nodes in a dynamic network.

\section{Conclusion and outlook}
In this review, we have summarized the theories and characteristics of quantum walks, introduced the progress of physical implementations of quantum walks, and described the quantum walk based algorithms and applications. Quantum walks present a promising theoretical framework for application-specific quantum computing while holding the potential for universal quantum computing. The power of quantum walk computing systems over classical computers has been well demonstrated via the boson sampling task --- as a particular multi-particle quantum walk sampling task. Moreover, with the recent developments of the physical implementations of quantum walks, particularly in the integrated quantum photonics systems, significant advances in both scale and programmability are demonstrated for quantum walks simulations and applications---most of which are targeting the problems of practical interests. This suggests that quantum walk models show a feasible path for building a practical and useful NISQ computer in the near future. 

Despite this rapid progress, there are some challenges for the realization of practical quantum walk computing systems. One major obstacle is to devise quantum walk algorithms. This requires further investigation of various quantum walk models and in-depth research of their characteristics, e.g., to understand the role they play in achieving quantum computational advantages. Another challenge is continuously scaling up the physical implementations of quantum walks, while meanwhile improving the programmability. For example, in integrated quantum photonics systems, the scale-up of quantum walks depends on both the number of evolving photons and the size of Hamiltonian, and thus increasing the photon numbers becomes critical, and large-scale but low-loss reconfigurable optical networks would also be core requirements, together with high-efficiency photon detection. Furthermore, although quantum walks may not be used for targeting fault-tolerant quantum computing at the current stage, methods for implementing quantum walks with error correction or fault tolerance are necessary, considering that the potential large-scale quantum walks for practical interests would require simulations with high precision. Towards this, it may on the one hand investigate error mitigation or error correction techniques in implementing quantum walks, and on the other hand, explore quantum walk algorithms in noisy conditions. Although these challenges require continuous efforts to solve, the field of quantum walk computing is vigorously developing, and it walks the way to a foreseen future for quantum computing.

\section*{Acknowledgments}
This work is supported by the National Natural Science Foundation of China (Grant No.62075243) and the National Natural Science Foundation of China ``Mathematical Basic Theory of Quantum Computing'' special project (Grant No.12341103). X.Q. acknowledges support from the National Young Talents Program.

\bibliography{main}

\begin{thebibliography}{100}

\bibitem{shor1997polynomial}
Peter~W Shor.
\newblock Polynomial-time algorithms for prime factorization and discrete logarithms on a quantum computer.
\newblock {\em SIAM journal on computing}, 26(5):1484--1509, 1997.

\bibitem{grover1996fast}
Lov~K Grover.
\newblock A fast quantum mechanical algorithm for database search.
\newblock In {\em Proceedings of the Twenty-Eighth Annual ACM Symposium on Theory of Computing}, STOC '96, pages 212--219. Association for Computing Machinery, 1996.

\bibitem{aaronson2011computational}
Scott Aaronson and Alex Arkhipov.
\newblock The computational complexity of linear optics.
\newblock In {\em Proceedings of the Forty-Third Annual ACM Symposium on Theory of Computing}, STOC '11, pages 333--342. Association for Computing Machinery, 2011.

\bibitem{farhi1998quantum}
Edward Farhi and Sam Gutmann.
\newblock Quantum computation and decision trees.
\newblock {\em Physical Review A}, 58(2):915--928, 1998.

\bibitem{kempe2003quantum}
Julia Kempe.
\newblock Quantum random walks: an introductory overview.
\newblock {\em Contemporary Physics}, 44(4):\ 307--327, 2003.

\bibitem{childs2013universal}
Andrew~M. Childs, David Gosset, and Zak Webb.
\newblock Universal computation by multiparticle quantum walk.
\newblock {\em Science}, 339(6121):791--794, 2013.

\bibitem{childs2004spatial}
Andrew~M. Childs and Jeffrey Goldstone.
\newblock Spatial search by quantum walk.
\newblock {\em Physical Review A}, 70(2):\ 022314, 2004.

\bibitem{douglas2008classical}
Brendan~L. Douglas and Jingbo~B Wang.
\newblock A classical approach to the graph isomorphism problem using quantum walks.
\newblock {\em Journal of Physics A: Mathematical and Theoretical}, 41(7):075303, 2008.

\bibitem{gamble2010two}
John~King Gamble, Mark Friesen, Dong Zhou, Robert Joynt, and S.~N. Coppersmith.
\newblock Two-particle quantum walks applied to the graph isomorphism problem.
\newblock {\em Physical Review A}, 81(5):052313, 2010.

\bibitem{berry2011two}
Scott~D Berry and Jingbo~B. Wang.
\newblock Two-particle quantum walks: Entanglement and graph isomorphism testing.
\newblock {\em Physical Review A}, 83(4):042317, 2011.

\bibitem{berry2010quantum}
Scott~D. Berry and Jingbo~B. Wang.
\newblock Quantum-walk-based search and centrality.
\newblock {\em Physical Review A}, 82(4):042333, 2010.

\bibitem{sanchez2012quantum}
Eduardo S{\'a}nchez-Burillo, Jordi Duch, Jes{\'u}s G{\'o}mez-Garde{\~n}es, and David Zueco.
\newblock Quantum navigation and ranking in complex networks.
\newblock {\em Scientific Reports}, 2:605, 2012.

\bibitem{lloyd1996universal}
Seth Lloyd.
\newblock Universal quantum simulators.
\newblock {\em Science}, 273(5278):1073, 1996.

\bibitem{mohseni2008environment}
Masoud Mohseni, Patrick Rebentrost, Seth Lloyd, and Alan Aspuru-Guzik.
\newblock Environment-assisted quantum walks in photosynthetic energy transfer.
\newblock {\em The Journal of chemical physics}, 129(17):174106, 2008.

\bibitem{berry2012black}
Dominic~W. Berry and Andrew~M. Childs.
\newblock Black-box hamiltonian simulation and unitary implementation.
\newblock {\em Quantum Information \& Computation}, 12(1-2):29--62, 2012.

\bibitem{qiang2016efficient}
Xiaogang Qiang, Thomas Loke, Ashley Montanaro, Kanin Aungskunsiri, Xiaoqi Zhou, Jeremy~L. O'Brien, Jingbo~B. Wang, and Jonathan C.~F. Matthews.
\newblock Efficient quantum walk on a quantum processor.
\newblock {\em Nature Communications}, 7:11511, 2016.

\bibitem{menssen2017distinguishability}
Adrian~J. Menssen, Alex~E. Jones, Benjamin~J. Metcalf, Malte~C. Tichy, Stefanie Barz, W.~Steven Kolthammer, and Ian~A. Walmsley.
\newblock Distinguishability and many-particle interference.
\newblock {\em Physical Review Letters}, 118(15):153603, 2017.

\bibitem{melnikov2016quantum}
Alexey~A. Melnikov and Leonid~E. Fedichkin.
\newblock Quantum walks of interacting fermions on a cycle graph.
\newblock {\em Scientific Reports}, 6:34226, 2016.

\bibitem{brennen_anyonic_2010}
Gavin~K. Brennen, Demosthenes Ellinas, Viv Kendon, Jiannis~K. Pachos, Ioannis Tsohantjis, and Zhenghan Wang.
\newblock Anyonic quantum walks.
\newblock {\em Annals of Physics}, 325(3):664 -- 681, 2010.

\bibitem{lehman2011quantum}
Lauri Lehman, Vaclav Zatloukal, Gavin~K Brennen, Jiannis~K Pachos, and Zhenghan Wang.
\newblock Quantum walks with non-abelian anyons.
\newblock {\em Physical Review Letters}, 106(23):230404, 2011.

\bibitem{liang2022hadamard}
Wen Liang, Fei Yan, Abdullah~M Iliyasu, Ahmed~S Salama, and Kaoru Hirota.
\newblock A hadamard walk model and its application in identification of important edges in complex networks.
\newblock {\em Computer Communications}, 193:378--387, 2022.

\bibitem{di2011mimicking}
Carlo Di~Franco, M~Mc~Gettrick, and Th~Busch.
\newblock Mimicking the probability distribution of a two-dimensional grover walk with a single-qubit coin.
\newblock {\em Physical review letters}, 106(8):080502, 2011.

\bibitem{ambainis2004coins}
Andris Ambainis, Julia Kempe, and Alexander Rivosh.
\newblock Coins make quantum walks faster.
\newblock In {\em Proceedings of the Sixteenth Annual ACM-SIAM Symposium on Discrete Algorithms}, SODA '05, pages 1099--1108. Society for Industrial and Applied Mathematics, 2005.

\bibitem{wong2018faster}
Thomas~G Wong.
\newblock Faster search by lackadaisical quantum walk.
\newblock {\em Quantum Information Processing}, 17:1--9, 2018.

\bibitem{giri2020lackadaisical}
Pulak~Ranjan Giri and Vladimir Korepin.
\newblock Lackadaisical quantum walk for spatial search.
\newblock {\em Modern Physics Letters A}, 35(08):2050043, 2020.

\bibitem{szegedy2004quantum}
Mario Szegedy.
\newblock Quantum speed-up of markov chain based algorithms.
\newblock In {\em 45th Annual IEEE symposium on foundations of computer science}, pages 32--41. IEEE, 2004.

\bibitem{portugal2016staggered-0}
Renato Portugal, Raqueline~AM Santos, Tharso~D Fernandes, and Demerson~N Gon{\c{c}}alves.
\newblock The staggered quantum walk model.
\newblock {\em Quantum Information Processing}, 15:85--101, 2016.

\bibitem{portugal2016staggered-1}
Renato Portugal.
\newblock Staggered quantum walks on graphs.
\newblock {\em Physical Review A}, 93(6):062335, 2016.

\bibitem{childs2003exponential}
Andrew~M Childs, Richard Cleve, Enrico Deotto, Edward Farhi, Sam Gutmann, and Daniel~A Spielman.
\newblock Exponential algorithmic speedup by a quantum walk.
\newblock In {\em Proceedings of the 35th Annual ACM Symposium on Theory of Computing}, pages 59--68, 2003.

\bibitem{christandl2004perfect}
Matthias Christandl, Nilanjana Datta, Artur Ekert, and Andrew~J Landahl.
\newblock Perfect state transfer in quantum spin networks.
\newblock {\em Physical review letters}, 92(18):187902, 2004.

\bibitem{whitfield2010quantum}
James~D Whitfield, C{\'e}sar~A Rodr{\'\i}guez-Rosario, and Al{\'a}n Aspuru-Guzik.
\newblock Quantum stochastic walks: A generalization of classical random walks and quantum walks.
\newblock {\em Physical Review A}, 81(2):022323, 2010.

\bibitem{attal2012open}
Stephane Attal, Francesco Petruccione, Christophe Sabot, and Ilya Sinayskiy.
\newblock Open quantum random walks.
\newblock {\em Journal of Statistical Physics}, 147(4):832--852, 2012.

\bibitem{aharonov1993quantum}
Yakir Aharonov, Luiz Davidovich, and Nicim Zagury.
\newblock Quantum random walks.
\newblock {\em Physical Review A}, 48(2):1687, 1993.

\bibitem{omar2006quantum}
Y~Omar, N~Paunkovi{\'c}, L~Sheridan, and S~Bose.
\newblock Quantum walk on a line with two entangled particles.
\newblock {\em Physical Review A}, 74(4):042304, 2006.

\bibitem{tregenna2003controlling}
Ben Tregenna, Will Flanagan, Rik Maile, and Viv Kendon.
\newblock Controlling discrete quantum walks: coins and initial states.
\newblock {\em New Journal of Physics}, 5(1):83, 2003.

\bibitem{ambainis2001one}
Andris Ambainis, Eric Bach, Ashwin Nayak, Ashvin Vishwanath, and John Watrous.
\newblock One-dimensional quantum walks.
\newblock In {\em Proceedings of the thirty-third annual ACM symposium on Theory of computing}, STOC '01, pages 37--49, New York, NY, USA, 2001. Association for Computing Machinery.

\bibitem{nayak2000quantum}
Ashwin Nayak and Ashvin Vishwanath.
\newblock Quantum walk on the line.
\newblock Technical report, 2000.

\bibitem{konno2008quantum}
Norie Konno.
\newblock {\em Quantum walks}, pages 309--452.
\newblock Springer, Berlin, 2008.

\bibitem{chandrashekar2008optimizing}
C~Madaiah Chandrashekar, Radhakrishna Srikanth, and Raymond Laflamme.
\newblock Optimizing the discrete time quantum walk using a su (2) coin.
\newblock {\em Physical Review A}, 77(3):032326, 2008.

\bibitem{loke2017efficient}
Thomas Loke and J.~B. Wang.
\newblock Efficient quantum circuits for szegedy quantum walks.
\newblock {\em Annals of Physics}, 382:64--84, 2017.

\bibitem{santha2008quantum}
Miklos Santha.
\newblock Quantum walk based search algorithms.
\newblock In {\em International Conference on Theory and Applications of Models of Computation}, TAMC 2008, pages 31--46. Springer, 2008.

\bibitem{magniez2007search}
Fr\'{e}d\'{e}ric Magniez, Ashwin Nayak, J\'{e}r\'{e}mie Roland, and Miklos Santha.
\newblock Search via quantum walk.
\newblock {\em SIAM Journal on Computing}, 40(1):142--164, 2011.

\bibitem{portugal2013quantum}
Renato Portugal.
\newblock {\em Staggered model}.
\newblock Springer International Publishing, Cham, 2013.

\bibitem{konno2018spectral}
Norio Konno, Yusuke Ide, and Iwao Sato.
\newblock The spectral analysis of the unitary matrix of a 2-tessellable staggered quantum walk on a graph.
\newblock {\em Linear Algebra and its Applications}, 545:207--225, 2018.

\bibitem{konno2018partition}
Norio Konno, Renato Portugal, Iwao Sato, and Etsuo Segawa.
\newblock Partition-based discrete-time quantum walks.
\newblock {\em Quantum Information Processing}, 17:1--35, 2018.

\bibitem{portugal2017staggered}
Renato Portugal, Marcos~Cesar de~Oliveira, and Jalil~Khatibi Moqadam.
\newblock Staggered quantum walks with hamiltonians.
\newblock {\em Physical Review A}, 95(1):012328, 2017.

\bibitem{coutinho2019discretization}
Gabriel Coutinho and Renato Portugal.
\newblock Discretization of continuous-time quantum walks via the staggered model with hamiltonians.
\newblock {\em Natural Computing}, 18(2):403--409, 2019.

\bibitem{santos2022decoherence}
Raqueline~AM Santos and Franklin de~L Marquezino.
\newblock Decoherence on staggered quantum walks.
\newblock {\em Physical Review A}, 105(3):032452, 2022.

\bibitem{chandrashekar2010relationship}
CM~Chandrashekar, Subhashish Banerjee, and R~Srikanth.
\newblock Relationship between quantum walks and relativistic quantum mechanics.
\newblock {\em Physical Review A}, 81(6):062340, 2010.

\bibitem{qiang_implementing_2021}
Xiaogang Qiang, Yizhi Wang, Shichuan Xue, Renyou Ge, Lifeng Chen, Yingwen Liu, Anqi Huang, Xiang Fu, Ping Xu, Teng Yi, Fufang Xu, Mingtang Deng, Jingbo~B. Wang, Jasmin D.~A. Meinecke, Jonathan C.~F. Matthews, Xinlun Cai, Xuejun Yang, and Junjie Wu.
\newblock Implementing graph-theoretic quantum algorithms on a silicon photonic quantum walk processor.
\newblock {\em Science Advances}, 7(9):eabb8375, 2021.

\bibitem{Chakraborty2020Finding}
Shantanav Chakraborty, Leonardo Novo, and J\'er\'emie Roland.
\newblock Finding a marked node on any graph via continuous-time quantum walks.
\newblock {\em Physical Review A}, 102:022227, 2020.

\bibitem{Apers2022Quadratic}
Simon Apers, Shantanav Chakraborty, Leonardo Novo, and J\'er\'emie Roland.
\newblock Quadratic speedup for spatial search by continuous-time quantum walk.
\newblock {\em Physical Review Letter}, 129:160502, 2022.

\bibitem{Underwood_universal_2010}
Michael~S. Underwood and David~L. Feder.
\newblock Universal quantum computation by discontinuous quantum walk.
\newblock {\em Physical Review A}, 82(4):042304, 2010.

\bibitem{govia2017quantum}
Luke~CG Govia, Bruno~G Taketani, Peter~K Schuhmacher, and Frank~K Wilhelm.
\newblock Quantum simulation of a quantum stochastic walk.
\newblock {\em Quantum Science and Technology}, 2(1):015002, 2017.

\bibitem{dudhe2022testing}
Naini Dudhe, Pratyush~Kumar Sahoo, and Colin Benjamin.
\newblock Testing quantum speedups in exciton transport through a photosynthetic complex using quantum stochastic walks.
\newblock {\em Physical Chemistry Chemical Physics}, 24(4):2601--2613, 2022.

\bibitem{dudhe2022resolving}
Colin Benjamin and Naini Dudhe.
\newblock Resolving degeneracies in google search via quantum stochastic walks.
\newblock {\em Journal of Statistical Mechanics: Theory and Experiment}, 2024(1):013402, 2024.

\bibitem{wang2022implementation}
Luji Wang, Jiayi Lin, and Shengjun Wu.
\newblock Implementation of quantum stochastic walks for function approximation, two-dimensional data classification, and sequence classification.
\newblock {\em Physical Review Research}, 4(2):023058, 2022.

\bibitem{dhahri2019open}
Ameur Dhahri and Farrukh Mukhamedov.
\newblock Open quantum random walks, quantum markov chains and recurrence.
\newblock {\em Reviews in Mathematical Physics}, 31(07):1950020, 2019.

\bibitem{dhahri2019quantum}
Ameur Dhahri, Chul~Ki Ko, and Hyun~Jae Yoo.
\newblock Quantum markov chains associated with open quantum random walks.
\newblock {\em Journal of Statistical Physics}, 176:1272--1295, 2019.

\bibitem{souissi2023structure}
Abdessatar Souissi, Tarek Hamdi, Farrukh Mukhamedov, and Amenallah Andolsi.
\newblock On the structure of quantum markov chains on cayley trees associated with open quantum random walks.
\newblock {\em Axioms}, 12(9):864, 2023.

\bibitem{kang2023markov}
Yuanbao Kang.
\newblock Markov properties of partially open quantum random walks.
\newblock {\em Journal of Mathematical Physics}, 64(3), 2023.

\bibitem{konno2013limit}
Norio Konno and Hyun~Jae Yoo.
\newblock Limit theorems for open quantum random walks.
\newblock {\em Journal of Statistical Physics}, 150:299--319, 2013.

\bibitem{attal2015central}
St{\'e}phane Attal, Nadine Guillotin-Plantard, and Christophe Sabot.
\newblock Central limit theorems for open quantum random walks and quantum measurement records.
\newblock {\em Annales Henri Poincar{\'e}}, 16(1):15--43, 2015.

\bibitem{lardizabal2016open}
Carlos~F. Lardizabal.
\newblock Open quantum random walks and the mean hitting time formula.
\newblock {\em Quantum Infomation \& Computation}, 17(1-2):79--105, 2017.

\bibitem{konno2002quantum}
Norio Konno.
\newblock Quantum random walks in one dimension.
\newblock {\em Quantum Information Processing}, 1:345--354, 2002.

\bibitem{grimmett2004weak}
Geoffrey Grimmett, Svante Janson, and Petra~F Scudo.
\newblock Weak limits for quantum random walks.
\newblock {\em Physical Review E}, 69(2):026119, 2004.

\bibitem{konno2005limit}
Norio Konno.
\newblock Limit theorem for continuous-time quantum walk on the line.
\newblock {\em Physical Review E}, 72(2):026113, 2005.

\bibitem{konno2005new}
Norio Konno.
\newblock A new type of limit theorems for the one-dimensional quantum random walk.
\newblock {\em Journal of the Mathematical Society of Japan}, 57(4):1179--1195, 2005.

\bibitem{lovett2012spatial}
Neil~B Lovett, Matthew Everitt, Matthew Trevers, Daniel Mosby, Dan Stockton, and Viv Kendon.
\newblock Spatial search using the discrete time quantum walk.
\newblock {\em Natural Computing}, 11(1):23--35, 2012.

\bibitem{rudinger2013comparing}
Kenneth Rudinger, John~King Gamble, Eric Bach, Mark Friesen, Robert Joynt, and SN~Coppersmith.
\newblock Comparing algorithms for graph isomorphism using discrete-and continuous-time quantum random walks.
\newblock {\em Journal of Computational and Theoretical Nanoscience}, 10(7):1653--1661, 2013.

\bibitem{childs2009universal}
Andrew~M Childs.
\newblock Universal computation by quantum walk.
\newblock {\em Physical Review Letters}, 102(18):180501, 2009.

\bibitem{lovett2010universal}
Neil~B Lovett, Sally Cooper, Matthew Everitt, Matthew Trevers, and Viv Kendon.
\newblock Universal quantum computation using the discrete-time quantum walk.
\newblock {\em Physical Review A}, 81(4):042330, 2010.

\bibitem{strauch2006connecting}
Frederick~W Strauch.
\newblock Connecting the discrete-and continuous-time quantum walks.
\newblock {\em Physical Review A}, 74(3):030301, 2006.

\bibitem{childs2010relationship}
Andrew~M. Childs.
\newblock On the relationship between continuous-and discrete-time quantum walk.
\newblock {\em Communications in Mathematical Physics}, 294(2):581--603, 2010.

\bibitem{portugal2016establishing}
Renato Portugal.
\newblock Establishing the equivalence between szegedy's and coined quantum walks using the staggered model.
\newblock {\em Quantum Information Processing}, 15:1387--1409, 2016.

\bibitem{portugal2017connecting}
Renato Portugal and Etsuo Segawa.
\newblock Connecting coined quantum walks with szegedy's model.
\newblock {\em Interdisciplinary Information Sciences}, 23(1):119--125, 2017.

\bibitem{marquezino2008mixing}
Franklin~L Marquezino, Renato Portugal, Gonzalo Abal, and Raul Donangelo.
\newblock Mixing times in quantum walks on the hypercube.
\newblock {\em Physical Review A}, 77(4):042312, 2008.

\bibitem{boito2023quantum}
Paola Boito and GM~Del~Corso.
\newblock Quantum hitting time according to a given distribution, 2023.

\bibitem{childs2002example}
Andrew~M Childs, Edward Farhi, and Sam Gutmann.
\newblock An example of the difference between quantum and classical random walks.
\newblock {\em Quantum Information Processing}, 1:35--43, 2002.

\bibitem{tang_experimental_2018}
Hao Tang, Carlo Di~Franco, Ziyu Shi, Tianshen He, Zhen Feng, Jun Gao, Ke~Sun, Zhanming Li, Zhiqiang Jiao, Tianyu Wang, M.~S. Kim, and Xianmin Jin.
\newblock Experimental quantum fast hitting on hexagonal graphs.
\newblock {\em Nature Photonics}, 12(12):754--758, 2018.

\bibitem{wang_large-scale_2022}
Yizhi Wang, Yingwen Liu, Junwei Zhan, Shichuan Xue, Yuzhen Zheng, Ru~Zeng, Zhihao Wu, Zihao Wang, Qilin Zheng, Dongyang Wang, Weixu Shi, Xiang Fu, Ping Xu, Yang Wang, Yong Liu, Jiangfang Ding, Guangyao Huang, Chunlin Yu, Anqi Huang, Xiaogang Qiang, Mingtang Deng, Weixia Xu, Kai Lu, Xuejun Yang, and Junjie Wu.
\newblock Large-scale full-programmable quantum walk and its applications, 2022.

\bibitem{santos2010quantum}
Raqueline Azevedo~Medeiros Santos and Renato Portugal.
\newblock Quantum hitting time on the complete graph.
\newblock {\em International Journal of Quantum Information}, 8(05):881--894, 2010.

\bibitem{emms2007graph}
David Emms, Richard~C Wilson, and Edwin Hancock.
\newblock Graph embedding using quantum commute times.
\newblock In {\em Graph-Based Representations in Pattern Recognition}, pages 371--382, Berlin, Heidelberg, 2007. Springer.

\bibitem{jonasson1998cover}
Johan Jonasson.
\newblock On the cover time for random walks on random graphs.
\newblock {\em Combinatorics, Probability and Computing}, 7(3):265--279, 1998.

\bibitem{chupeau2015cover}
Marie Chupeau, Olivier B{\'e}nichou, and Rapha{\"e}l Voituriez.
\newblock Cover times of random searches.
\newblock {\em Nature Physics}, 11(10):844--847, 2015.

\bibitem{anderson1958absence}
Philip~W Anderson.
\newblock Absence of diffusion in certain random lattices.
\newblock {\em Physical Review}, 109(5):1492, 1958.

\bibitem{ying2016anderson}
Tianping Ying, Yueqiang Gu, Xiao Chen, Xinbo Wang, Shifeng Jin, Linlin Zhao, Wei Zhang, and Xiaolong Chen.
\newblock Anderson localization of electrons in single crystals: Li x fe7se8.
\newblock {\em Science advances}, 2(2):e1501283, 2016.

\bibitem{dominguez2004simple}
F~Dom{\i}nguez-Adame and VA~Malyshev.
\newblock A simple approach to anderson localization in one-dimensional disordered lattices.
\newblock {\em American Journal of Physics}, 72(2):226--230, 2004.

\bibitem{hundertmark2008short}
Dirk Hundertmark.
\newblock A short introduction to anderson localization.
\newblock {\em Analysis and stochastics of growth processes and interface models}, 1:194--219, 2008.

\bibitem{ortuno2009random}
M~Ortu{\~n}o, AM~Somoza, and JT~Chalker.
\newblock Random walks and anderson localization in a three-dimensional class c network model.
\newblock {\em Physical review letters}, 102(7):070603, 2009.

\bibitem{crespi_anderson_2013}
Andrea Crespi, Roberto Osellame, Roberta Ramponi, Vittorio Giovannetti, Rosario Fazio, Linda Sansoni, Francesco De~Nicola, Fabio Sciarrino, and Paolo Mataloni.
\newblock Anderson localization of entangled photons in an integrated quantum walk.
\newblock {\em Nature Photonics}, 7(4):322--328, 2013.

\bibitem{duda2023quantum}
Rostislav Duda, Moein~N Ivaki, Isac Sahlberg, Kim P{\"o}yh{\"o}nen, and Teemu Ojanen.
\newblock Quantum walks on random lattices: Diffusion, localization, and the absence of parametric quantum speedup.
\newblock {\em Physical Review Research}, 5(2):023150, 2023.

\bibitem{karamlou2022quantum}
Amir~H Karamlou, Jochen Braum{\"u}ller, Yariv Yanay, Agustin Di~Paolo, Patrick~M Harrington, Bharath Kannan, David Kim, Morten Kjaergaard, Alexander Melville, Sarah Muschinske, et~al.
\newblock Quantum transport and localization in 1d and 2d tight-binding lattices.
\newblock {\em npj Quantum Information}, 8(1):35, 2022.

\bibitem{ghosh2014simulating}
Joydip Ghosh.
\newblock Simulating anderson localization via a quantum walk on a one-dimensional lattice of superconducting qubits.
\newblock {\em Physical Review A}, 89(2):022309, 2014.

\bibitem{muraleedharan2019quantum}
Gopikrishnan Muraleedharan, Akimasa Miyake, and Ivan~H Deutsch.
\newblock Quantum computational supremacy in the sampling of bosonic random walkers on a one-dimensional lattice.
\newblock {\em New Journal of Physics}, 21(5):055003, 2019.

\bibitem{chandrashekar2012quantum}
CM~Chandrashekar and Th~Busch.
\newblock Quantum walk on distinguishable non-interacting many-particles and indistinguishable two-particle.
\newblock {\em Quantum Information Processing}, 11:1287--1299, 2012.

\bibitem{rudinger2012noninteracting}
Kenneth Rudinger, John~King Gamble, Mark Wellons, Eric Bach, Mark Friesen, Robert Joynt, and SN~Coppersmith.
\newblock Noninteracting multiparticle quantum random walks applied to the graph isomorphism problem for strongly regular graphs.
\newblock {\em Physical Review A}, 86(2):022334, 2012.

\bibitem{sansoni2012two}
Linda Sansoni, Fabio Sciarrino, Giuseppe Vallone, Paolo Mataloni, Andrea Crespi, Roberta Ramponi, and Roberto Osellame.
\newblock Two-particle bosonic-fermionic quantum walk via integrated photonics.
\newblock {\em Physical Review Letters}, 108(1):010502, 2012.

\bibitem{bremner2011classical}
Michael~J Bremner, Richard Jozsa, and Dan~J Shepherd.
\newblock Classical simulation of commuting quantum computations implies collapse of the polynomial hierarchy.
\newblock {\em Proceedings of the Royal Society A: Mathematical, Physical and Engineering Sciences}, 467(2126):459--472, 2011.

\bibitem{lund2017quantum}
Austin~P Lund, Michael~J Bremner, and Timothy~C Ralph.
\newblock Quantum sampling problems, bosonsampling and quantum supremacy.
\newblock {\em npj Quantum Information}, 3(1):15, 2017.

\bibitem{brod2021bosons}
Daniel~Jost Brod.
\newblock Bosons vs. fermions--a computational complexity perspective.
\newblock {\em Revista Brasileira de Ensino de F{\'\i}sica}, 43, 2021.

\bibitem{takeuchi2016ancilla}
Yuki Takeuchi and Yasuhiro Takahashi.
\newblock Ancilla-driven instantaneous quantum polynomial time circuit for quantum supremacy.
\newblock {\em Physical Review A}, 94(6):062336, 2016.

\bibitem{bremner2016average}
Michael~J Bremner, Ashley Montanaro, and Dan~J Shepherd.
\newblock Average-case complexity versus approximate simulation of commuting quantum computations.
\newblock {\em Physical Review Letters}, 117(8):080501, 2016.

\bibitem{douglas2009efficient}
BL~Douglas and JB~Wang.
\newblock Efficient quantum circuit implementation of quantum walks.
\newblock {\em Physical Review A}, 79(5):052335, 2009.

\bibitem{loke2012efficient}
T~Loke and J.~B. Wang.
\newblock Efficient circuit implementation of quantum walks on non-degree-regular graphs.
\newblock {\em Physical Review A}, 86(4):042338, 2012.

\bibitem{loke_efficient_2017}
T~Loke and J.~B. Wang.
\newblock Efficient quantum circuits for continuous-time quantum walks on composite graphs.
\newblock {\em Journal of Physics A: Mathematical and Theoretical}, 50(5):055303, 2017.

\bibitem{qiang2018large}
Xiaogang Qiang, Xiaoqi Zhou, Jianwei Wang, Callum~M. Wilkes, Thomas Loke, Sean O'Gara, Laurent Kling, Graham~D. Marshall, Raffaele Santagati, Timothy~C. Ralph, et~al.
\newblock Large-scale silicon quantum photonics implementing arbitrary two-qubit processing.
\newblock {\em Nature Photonics}, 12(9):534, 2018.

\bibitem{du2003experimental}
Jiangfeng Du, Hui Li, Xiaodong Xu, Mingjun Shi, Jihui Wu, Xianyi Zhou, and Rongdian Han.
\newblock Experimental implementation of the quantum random-walk algorithm.
\newblock {\em Physical Review A}, 67(4):042316, 2003.

\bibitem{ryan2005experimental}
Colm~A Ryan, Martin Laforest, Jean-Christian Boileau, and Raymond Laflamme.
\newblock Experimental implementation of a discrete-time quantum random walk on an nmr quantum-information processor.
\newblock {\em Physical Review A}, 72(6):062317, 2005.

\bibitem{ramasesh_direct_2017}
V.~V. Ramasesh, E.~Flurin, M.~Rudner, I.~Siddiqi, and N.~Y. Yao.
\newblock Direct probe of topological invariants using bloch oscillating quantum walks.
\newblock {\em Physical Review Letters}, 118(13):130501, 2017.

\bibitem{yan_strongly_2019}
Zhiguang Yan, Yu-Ran Zhang, Ming Gong, Yulin Wu, Yarui Zheng, Shaowei Li, Can Wang, Futian Liang, Jin Lin, Yu~Xu, Cheng Guo, Lihua Sun, Cheng-Zhi Peng, Keyu Xia, Hui Deng, Hao Rong, J.~Q. You, Franco Nori, Heng Fan, Xiaobo Zhu, and Jian-Wei Pan.
\newblock Strongly correlated quantum walks with a 12-qubit superconducting processor.
\newblock {\em Science}, page eaaw1611, 2019.

\bibitem{gong_quantum_2021}
Ming Gong, Shiyu Wang, Chen Zha, Ming-Cheng Chen, He-Liang Huang, Yulin Wu, Qingling Zhu, Youwei Zhao, Shaowei Li, Shaojun Guo, Haoran Qian, Yangsen Ye, Fusheng Chen, Chong Ying, Jiale Yu, Daojin Fan, Dachao Wu, Hong Su, Hui Deng, Hao Rong, Kaili Zhang, Sirui Cao, Jin Lin, Yu~Xu, Lihua Sun, Cheng Guo, Na~Li, Futian Liang, V.~M. Bastidas, Kae Nemoto, W.~J. Munro, Yong-Heng Huo, Chao-Yang Lu, Cheng-Zhi Peng, Xiaobo Zhu, and Jian-Wei Pan.
\newblock Quantum walks on a programmable two-dimensional 62-qubit superconducting processor.
\newblock {\em Science}, 372(6545):948--952, 2021.

\bibitem{schmitz2009quantum}
Hector Schmitz, Robert Matjeschk, Ch. Schneider, Jan Glueckert, Martin Enderlein, Thomas Huber, and Tobias Schaetz.
\newblock Quantum walk of a trapped ion in phase space.
\newblock {\em Physical Review Letters}, 103(9):090504, 2009.

\bibitem{zahringer2010realization}
F~Z{\"a}hringer, G~Kirchmair, R~Gerritsma, E~Solano, R~Blatt, and CF~Roos.
\newblock Realization of a quantum walk with one and two trapped ions.
\newblock {\em Physical Review Letters}, 104(10):100503, 2010.

\bibitem{matjeschk_experimental_2012}
Robert Matjeschk, Ch~Schneider, Martin Enderlein, Thomas Huber, Hector Schmitz, Jan Glueckert, and Tobias Schaetz.
\newblock Experimental simulation and limitations of quantum walks with trapped ions.
\newblock {\em New Journal of Physics}, 14(3):035012, 2012.

\bibitem{karski2009quantum}
Michal Karski, Leonid F{\"o}rster, Jai-Min Choi, Andreas Steffen, Wolfgang Alt, Dieter Meschede, and Artur Widera.
\newblock Quantum walk in position space with single optically trapped atoms.
\newblock {\em Science}, 325(5937):174--177, 2009.

\bibitem{barkhofen_supersymmetric_2018}
Sonja Barkhofen, Lennart Lorz, Thomas Nitsche, Christine Silberhorn, and Henning Schomerus.
\newblock Supersymmetric polarization anomaly in photonic discrete-time quantum walks.
\newblock {\em Physical Review Letters}, 121(26):260501, 2018.

\bibitem{lorz_photonic_2019}
Lennart Lorz, Evan Meyer-Scott, Thomas Nitsche, V{\'a}clav Poto{\v{c}}ek, Aur{\'e}l G{\'a}bris, Sonja Barkhofen, Igor Jex, and Christine Silberhorn.
\newblock Photonic quantum walks with four-dimensional coins.
\newblock {\em Physical Review Research}, 1(3):033036, 2019.

\bibitem{xu_measuring_2018}
Xiaoye Xu, QinQin Wang, WeiWei Pan, Kai Sun, JinShi Xu, Geng Chen, JianShun Tang, Ming Gong, YongJian Han, ChuanFeng Li, et~al.
\newblock Measuring the winding number in a large-scale chiral quantum walk.
\newblock {\em Physical Review Letters}, 120(26):260501, 2018.

\bibitem{wang_simulating_2019}
Kunkun Wang, Xingze Qiu, Lei Xiao, Xiang Zhan, Zhihao Bian, Wei Yi, and Peng Xue.
\newblock Simulating dynamic quantum phase transitions in photonic quantum walks.
\newblock {\em Physical Review Letters}, 122(2):020501, 2019.

\bibitem{zhan_detecting_2017}
Xiang Zhan, Lei Xiao, Zhihao Bian, Kunkun Wang, Xingze Qiu, Barry~C Sanders, Wei Yi, and Peng Xue.
\newblock Detecting topological invariants in nonunitary discrete-time quantum walks.
\newblock {\em Physical Review Letters}, 119(13):130501, 2017.

\bibitem{xue_experimental_2015}
Peng Xue, Rong Zhang, Hao Qin, Xiang Zhan, ZH~Bian, Jian Li, and Barry~C Sanders.
\newblock Experimental quantum-walk revival with a time-dependent coin.
\newblock {\em Physical Review Letters}, 114(14):140502, 2015.

\bibitem{xiao_observation_2017}
L~Xiao, X~Zhan, ZH~Bian, KK~Wang, X~Zhang, XP~Wang, J~Li, K~Mochizuki, D~Kim, N~Kawakami, et~al.
\newblock Observation of topological edge states in parity--time-symmetric quantum walks.
\newblock {\em Nature Physics}, 13(11):1117--1123, 2017.

\bibitem{wang_generalized_2023}
Xiaowei Wang, Xiang Zhan, Yulin Li, Lei Xiao, Gaoyan Zhu, Dengke Qu, Quan Lin, Yue Yu, and Peng Xue.
\newblock Generalized quantum measurements on a higher-dimensional system via quantum walks.
\newblock {\em Physical Review Letters}, 131(15):150803, 2023.

\bibitem{zhang_implementation_2010}
Pei Zhang, BiHeng Liu, RuiFeng Liu, HongRong Li, FuLi Li, and GuangCan Guo.
\newblock Implementation of one-dimensional quantum walks on spin-orbital angular momentum space of photons.
\newblock {\em Physical Review A}, 81(5):052322, 2010.

\bibitem{goyal_implementing_2013}
Sandeep~K Goyal, Filippus~S Roux, Andrew Forbes, and Thomas Konrad.
\newblock Implementing quantum walks using orbital angular momentum of classical light.
\newblock {\em Physical Review Letters}, 110(26):263602, 2013.

\bibitem{giordani_experimental_2019}
Taira Giordani, Emanuele Polino, Sabrina Emiliani, Alessia Suprano, Luca Innocenti, Helena Majury, Lorenzo Marrucci, Mauro Paternostro, Alessandro Ferraro, Nicol{\`o} Spagnolo, et~al.
\newblock Experimental engineering of arbitrary qudit states with discrete-time quantum walks.
\newblock {\em Physical Review Letters}, 122(2):020503, 2019.

\bibitem{qu2022deterministic}
Dengke Qu, Samuel Marsh, Kunkun Wang, Lei Xiao, Jingbo Wang, and Peng Xue.
\newblock Deterministic search on star graphs via quantum walks.
\newblock {\em Physical Review Letters}, 128(5):050501, 2022.

\bibitem{schreiber2010photons}
Andreas Schreiber, Katiuscia~N. Cassemiro, V.~Poto{\v{c}}ek, Aur{\'e}l G{\'a}bris, Peter~J. Mosley, Erika Andersson, Igor Jex, and Ch. Silberhorn.
\newblock Photons walking the line: a quantum walk with adjustable coin operations.
\newblock {\em Physical Review Letters}, 104(5):050502, 2010.

\bibitem{nguyen_quantum_2020}
Dan~T Nguyen, Thien~An Nguyen, Rostislav Khrapko, Daniel~A Nolan, and Nicholas~F Borrelli.
\newblock Quantum walks in periodic and quasiperiodic fibonacci fibers.
\newblock {\em Scientific Reports}, 10(1):7156, 2020.

\bibitem{defienne_two-photon_2016}
Hugo Defienne, Marco Barbieri, Ian~A Walmsley, Brian~J Smith, and Sylvain Gigan.
\newblock Two-photon quantum walk in a multimode fiber.
\newblock {\em Science Advances}, 2(1):e1501054, 2016.

\bibitem{peruzzo2010quantum}
Alberto Peruzzo, Mirko Lobino, Jonathan C.~F. Matthews, Nobuyuki Matsuda, Alberto Politi, Konstantinos Poulios, Xiao-Qi Zhou, Yoav Lahini, Nur Ismail, Kerstin W{\"o}rhoff, et~al.
\newblock Quantum walks of correlated photons.
\newblock {\em Science}, 329(5998):1500--1503, 2010.

\bibitem{poulios2014quantum}
Konstantinos Poulios, Robert Keil, Daniel Fry, Jasmin D.~A. Meinecke, Jonathan C.~F. Matthews, Alberto Politi, Mirko Lobino, Markus Gr{\"a}fe, Matthias Heinrich, Stefan Nolte, et~al.
\newblock Quantum walks of correlated photon pairs in two-dimensional waveguide arrays.
\newblock {\em Physical Review Letters}, 112(14):143604, 2014.

\bibitem{benedetti_quantum_2021}
Claudia Benedetti, Dario Tamascelli, Matteo~GA Paris, and Andrea Crespi.
\newblock Quantum spatial search in two-dimensional waveguide arrays.
\newblock {\em Physical Review Applied}, 16(5):054036, 2021.

\bibitem{crespi_suppression_2016}
Andrea Crespi, Roberto Osellame, Roberta Ramponi, Marco Bentivegna, Fulvio Flamini, Nicol{\`o} Spagnolo, Niko Viggianiello, Luca Innocenti, Paolo Mataloni, and Fabio Sciarrino.
\newblock Suppression law of quantum states in a 3d photonic fast fourier transform chip.
\newblock {\em Nature Communications}, 7(1):10469, 2016.

\bibitem{xu_quantum_2021}
XiaoYun Xu, XiaoWei Wang, DanYang Chen, C.~Morais Smith, and XianMin Jin.
\newblock Quantum transport in fractal networks.
\newblock {\em Nature Photonics}, 15(9):703--710, 2021.

\bibitem{jiao_two-dimensional_2020}
ZhiQiang Jiao, Jun Gao, WenHao Zhou, XiaoWei Wang, RuoJing Ren, XiaoYun Xu, LuFeng Qiao, Yao Wang, and XianMin Jin.
\newblock Two-dimensional quantum walks of correlated photons.
\newblock {\em Optica}, 8(9):1129--1135, 2021.

\bibitem{tang_experimental_2018-1}
Hao Tang, XiaoFeng Lin, Zhen Feng, JingYuan Chen, Jun Gao, Ke~Sun, ChaoYue Wang, PengCheng Lai, XiaoYun Xu, Yao Wang, LuFeng Qiao, AiLin Yang, and XianMin Jin.
\newblock Experimental two-dimensional quantum walk on a photonic chip.
\newblock {\em Science Advances}, 4(5):eaat3174, 2018.

\bibitem{wang_experimental_2022}
Yang Wang, Xinyao Yu, Shichuan Xue, Yizhi Wang, Junwei Zhan, Chao Wu, Pingyu Zhu, Qilin Zheng, Miaomiao Yu, Yingwen Liu, Xiaogang Qiang, Junjie Wu, Xuejun Yang, and Ping Xu.
\newblock Experimental demonstration of quantum transport enhancement using time-reversal symmetry breaking on a silicon photonic chip.
\newblock {\em Science China Physics, Mechanics \& Astronomy}, 65(10):100362, 2022.

\bibitem{garcia2003speed}
JJ~Garcia-Ripoll, P~Zoller, and JI~Cirac.
\newblock Speed optimized two-qubit gates with laser coherent control techniques for ion trap quantum computing.
\newblock {\em Physical Review Letters}, 91(15):157901, 2003.

\bibitem{garcia-ripoll_coherent_2005}
J.~J. Garc\'{i}a-Ripoll, P.~Zoller, and J.~I. Cirac.
\newblock Coherent control of trapped ions using off-resonant lasers.
\newblock {\em Physical Review A}, 71(6):062309, 2005.

\bibitem{carolan2015universal}
Jacques Carolan, Christopher Harrold, Chris Sparrow, Enrique Mart{\'\i}n-L{\'o}pez, Nicholas~J. Russell, Joshua~W. Silverstone, Peter~J. Shadbolt, Nobuyuki Matsuda, Manabu Oguma, Mikitaka Itoh, et~al.
\newblock Universal linear optics.
\newblock {\em Science}, 349(6249):711--716, 2015.

\bibitem{caruso2016fast}
Filippo Caruso, Andrea Crespi, Anna~Gabriella Ciriolo, Fabio Sciarrino, and Roberto Osellame.
\newblock Fast escape of a quantum walker from an integrated photonic maze.
\newblock {\em Nature communications}, 7(1):11682, 2016.

\bibitem{tang2022generating}
Hao Tang, Leonardo Banchi, Tian-Yu Wang, Xiao-Wen Shang, Xi~Tan, Wen-Hao Zhou, Zhen Feng, Anurag Pal, Hang Li, Cheng-Qiu Hu, et~al.
\newblock Generating haar-uniform randomness using stochastic quantum walks on a photonic chip.
\newblock {\em Physical Review Letters}, 128(5):050503, 2022.

\bibitem{bao_very-large-scale_2023}
Jueming Bao, Zhaorong Fu, Tanumoy Pramanik, Jun Mao, Yulin Chi, Yingkang Cao, Chonghao Zhai, Yifei Mao, Tianxiang Dai, Xiaojiong Chen, Xinyu Jia, Leshi Zhao, Yun Zheng, Bo~Tang, Zhihua Li, Jun Luo, Wenwu Wang, Yan Yang, Yingying Peng, Dajian Liu, Daoxin Dai, Qiongyi He, Alif~Laila Muthali, Leif~K. Oxenløwe, Caterina Vigliar, Stefano Paesani, Huili Hou, Raffaele Santagati, Joshua~W. Silverstone, Anthony Laing, Mark~G. Thompson, Jeremy~L. O'Brien, Yunhong Ding, Qihuang Gong, and Jianwei Wang.
\newblock Very-large-scale integrated quantum graph photonics.
\newblock {\em Nature Photonics}, 17:573--581, 2023.

\bibitem{wang2018multidimensional}
Jianwei Wang, Stefano Paesani, Yunhong Ding, Raffaele Santagati, Paul Skrzypczyk, Alexia Salavrakos, Jordi Tura, Remigiusz Augusiak, Laura Man{\v{c}}inska, Davide Bacco, et~al.
\newblock Multidimensional quantum entanglement with large-scale integrated optics.
\newblock {\em Science}, 360(6386):285--291, 2018.

\bibitem{chi_programmable_2022}
Yulin Chi, Jieshan Huang, Zhanchuan Zhang, Jun Mao, Zinan Zhou, Xiaojiong Chen, Chonghao Zhai, Jueming Bao, Tianxiang Dai, Huihong Yuan, Ming Zhang, Daoxin Dai, Bo~Tang, Yan Yang, Zhihua Li, Yunhong Ding, Leif~K. Oxenløwe, Mark~G. Thompson, Jeremy~L. O’Brien, Yan Li, Qihuang Gong, and Jianwei Wang.
\newblock A programmable qudit-based quantum processor.
\newblock {\em Nature Communications}, 13(1):1166, 2022.

\bibitem{bartolucci_fusion-based_2023}
Sara Bartolucci, Patrick Birchall, Hector Bombín, Hugo Cable, Chris Dawson, Mercedes Gimeno-Segovia, Eric Johnston, Konrad Kieling, Naomi Nickerson, Mihir Pant, Fernando Pastawski, Terry Rudolph, and Chris Sparrow.
\newblock Fusion-based quantum computation.
\newblock {\em Nature Communications}, 14(1):912, 2023.

\bibitem{matthews2013observing}
Jonathan C.~F. Matthews, Konstantinos Poulios, Jasmin D.~A. Meinecke, Alberto Politi, Alberto Peruzzo, Nur Ismail, Kerstin W{\"o}rhoff, Mark~G. Thompson, and Jeremy~L. O'Brien.
\newblock Observing fermionic statistics with photons in arbitrary processes.
\newblock {\em Scientific Reports}, 3:1539, 2013.

\bibitem{shi2020quantum}
Zi-Yu Shi, Hao Tang, Zhen Feng, Yao Wang, Zhan-Ming Li, Jun Gao, Yi-Jun Chang, Tian-Yu Wang, Jian-Peng Dou, Zhe-Yong Zhang, et~al.
\newblock Quantum fast hitting on glued trees mapped on a photonic chip.
\newblock {\em Optica}, 7(6):613--618, 2020.

\bibitem{childs2007quantum}
Andrew~M Childs, Leonard~J Schulman, and Umesh~V Vazirani.
\newblock Quantum algorithms for hidden nonlinear structures.
\newblock In {\em 48th Annual IEEE Symposium on Foundations of Computer Science}, FOCS'07, pages 395--404. IEEE, 2007.

\bibitem{ambainis2007quantum}
Andris Ambainis.
\newblock Quantum walk algorithm for element distinctness.
\newblock {\em SIAM Journal on Computing}, 37(1):210--239, 2007.

\bibitem{Reichardt_span_2008}
Ben~W. Reichardt and Robert Spalek.
\newblock Span-program-based quantum algorithm for evaluating formulas.
\newblock In {\em Proceedings of the Fortieth Annual ACM Symposium on Theory of Computing}, STOC '08, pages 103--112. Association for Computing Machinery, 2008.

\bibitem{Buhrman_quantum_2006}
Harry Buhrman and Robert Spalek.
\newblock Quantum verification of matrix products.
\newblock In {\em Proceedings of the Seventeenth Annual ACM-SIAM Symposium on Discrete Algorithm}, SODA '06, pages 880--889. Society for Industrial and Applied Mathematics, 2006.

\bibitem{Magniez_quantum_2007}
Frederic Magniez and Ashwin Nayak.
\newblock Quantum complexity of testing group commutativity.
\newblock {\em Algorithmica}, 48(3):221--232, 2007.

\bibitem{Becker_improved_2011}
Anja Becker, Jean-Sebastien Coron, and Antoine Joux.
\newblock Improved generic algorithms for hard knapsacks.
\newblock In {\em Advances in Cryptology -- EUROCRYPT 2011}, Lecture Notes in Computer Science, pages 364--385. Springer, 2011.

\bibitem{magniez2007quantum}
Fr{\'e}d{\'e}ric Magniez, Miklos Santha, and Mario Szegedy.
\newblock Quantum algorithms for the triangle problem.
\newblock {\em SIAM Journal on Computing}, 37(2):413--424, 2007.

\bibitem{dernbach_quantum_2019}
Stefan Dernbach, Arman Mohseni-Kabir, Siddharth Pal, Miles Gepner, and Don Towsley.
\newblock Quantum walk neural networks with feature dependent coins.
\newblock {\em Applied Network Science}, 4(1):76, 2019.

\bibitem{schuld_quantum_2014}
Maria Schuld, Ilya Sinayskiy, and Francesco Petruccione.
\newblock Quantum walks on graphs representing the firing patterns of a quantum neural network.
\newblock {\em Physical Review A}, 89(3):032333, 2014.

\bibitem{de_souza_classical_2022}
Luciano~S. de~Souza, Jonathan H.~A. de~Carvalho, and Tiago A.~E. Ferreira.
\newblock Classical artificial neural network training using quantum walks as a search procedure.
\newblock {\em IEEE Transactions on Computers}, 71(2):378--389, 2022.

\bibitem{marsh_quantum_2019}
S.~Marsh and J.~B. Wang.
\newblock A quantum walk-assisted approximate algorithm for bounded {NP} optimisation problems.
\newblock {\em Quantum Information Processing}, 18(3):61, 2019.

\bibitem{marsh_combinatorial_2020}
S.~Marsh and J.~B. Wang.
\newblock Combinatorial optimization via highly efficient quantum walks.
\newblock {\em Physical Review Research}, 2(2):023302, 2020.

\bibitem{slate_quantum_2021}
Nicholas Slate, Edric Matwiejew, Samuel Marsh, and Jingbo Wang.
\newblock Quantum walk-based portfolio optimisation.
\newblock {\em Quantum}, 5:513, 2021.

\bibitem{bennett_quantum_2021}
T.~Bennett, E.~Matwiejew, S.~Marsh, and J.~B. Wang.
\newblock Quantum walk-based vehicle routing optimisation.
\newblock {\em Frontiers in Physics}, 9:730856, 2021.

\bibitem{yumin_novel_2014}
Yumin Dong and Shufen Xiao.
\newblock A novel algorithm of quantum random walk in server traffic control and task scheduling.
\newblock {\em Journal of Applied Mathematics}, 2014:818479, 2014.

\bibitem{montanaro_quantum-walk_2018}
Ashley Montanaro.
\newblock Quantum-walk speedup of backtracking algorithms.
\newblock {\em Theory of Computing}, 14(15):1--24, 2018.

\bibitem{Kitagawa_exploring_2010}
Takuya Kitagawa, Mark~S. Rudner, Erez Berg, and Eugene Demler.
\newblock Exploring topological phases with quantum walks.
\newblock {\em Physical Review A}, 82(3), 2010.

\bibitem{lin_topological_2022}
Quan Lin, Tianyu Li, Lei Xiao, Kunkun Wang, Wei Yi, and Peng Xue.
\newblock Topological phase transitions and mobility edges in non-hermitian quasicrystals.
\newblock {\em Physical Review Letters}, 129(11):113601, 2022.

\bibitem{schreiber20122d}
Andreas Schreiber, Aur{\'e}l G{\'a}bris, Peter~P Rohde, Kaisa Laiho, Martin {\v{S}}tefa{\v{n}}{\'a}k, V{\'a}clav Poto{\v{c}}ek, Craig Hamilton, Igor Jex, and Christine Silberhorn.
\newblock A 2d quantum walk simulation of two-particle dynamics.
\newblock {\em Science}, 336(6077):55--58, 2012.

\bibitem{berry2015hamiltonian}
Dominic~W Berry, Andrew~M Childs, and Robin Kothari.
\newblock Hamiltonian simulation with nearly optimal dependence on all parameters.
\newblock In {\em 2015 IEEE 56th Annual Symposium on Foundations of Computer Science}, pages 792--809, 2015.

\bibitem{Berry_corrected_2016}
Dominic~W. Berry and Leonardo Novo.
\newblock Corrected quantum walk for optimal hamiltonian simulation.
\newblock {\em Quantum Information \& Computation}, 16(15-16):1295--1317, 2016.

\bibitem{Rudner_topological_2009}
M.~S. Rudner and L.~S. Levitov.
\newblock Topological transition in a non-hermitian quantum walk.
\newblock {\em Physical Review Letters}, 102(6):065703, 2009.

\bibitem{Weidemann_topological_2022}
Sebastian Weidemann, Mark Kremer, Stefano Longhi, and Alexander Szameit.
\newblock Topological triple phase transition in non-hermitian floquet quasicrystals.
\newblock {\em Nature}, 601(7893):354--359, 2022.

\bibitem{Mittal_persistence_2021}
Vikash Mittal, Aswathy Raj, Sanjib Dey, and Sandeep~K. Goyal.
\newblock Persistence of topological phases in non-hermitian quantum walks.
\newblock {\em Scientific Reports}, 11(1):10262, 2021.

\bibitem{lin_observation_2022}
Quan Lin, Tianyu Li, Lei Xiao, Kunkun Wang, Wei Yi, and Peng Xue.
\newblock Observation of non-hermitian topological anderson insulator in quantum dynamics.
\newblock {\em Nature Communications}, 13(1):3229, 2022.

\bibitem{Kitagawa_observation_2012}
Takuya Kitagawa, Matthew~A. Broome, Alessandro Fedrizzi, Mark~S. Rudner, Erez Berg, Ivan Kassal, Alan Aspuru-Guzik, Eugene Demler, and Andrew~G. White.
\newblock Observation of topologically protected bound states in photonic quantum walks.
\newblock {\em Nature Communications}, 3:882, 2012.

\bibitem{Asboth_symmetries_2012}
J.~K. Asboth.
\newblock Symmetries, topological phases, and bound states in the one-dimensional quantum walk.
\newblock {\em Physical Review B}, 86(19):195414, 2012.

\bibitem{Panahiyan_simulation_2019}
S.~Panahiyan and S.~Fritzsche.
\newblock Simulation of the multiphase configuration and phase transitions with quantum walks utilizing a step-dependent coin.
\newblock {\em Physical Review A}, 100(6):062115, 2019.

\bibitem{Panahiyan_controllable_2020}
S.~Panahiyan and S.~Fritzsche.
\newblock Controllable simulation of topological phases and edge states with quantum walk.
\newblock {\em Physics Letters A}, 384(32):126828, 2020.

\bibitem{Karafyllidis_quantum_2017}
Ioannis~G. Karafyllidis.
\newblock Quantum transport in the fmo photosynthetic light-harvesting complex.
\newblock {\em Journal of Biological Physics}, 43(2):239--245, 2017.

\bibitem{hoyer2010limits}
Stephan Hoyer, Mohan Sarovar, and K~Birgitta Whaley.
\newblock Limits of quantum speedup in photosynthetic light harvesting.
\newblock {\em New Journal of Physics}, 12(6):065041, 2010.

\bibitem{D'Acunto_protein_2021}
Mario D'Acunto.
\newblock Protein-dna target search relies on quantum walk.
\newblock {\em Biosystems}, 201:104340, 2021.

\bibitem{Varsamis_quantum_2023}
G.~D. Varsamis, I.~G. Karafyllidis, K.~M. Gilkes, U.~Arranz, R.~Martin-Cuevas, G.~Calleja, J.~Wong, H.~C. Jessen, P.~Dimitrakis, P.~Kolovos, and R.~Sandaltzopoulos.
\newblock Quantum algorithm for de novo dna sequence assembly based on quantum walks on graphs.
\newblock {\em Biosystems}, 233:105037, 2023.

\bibitem{Chia_Coherent1_2016}
A.~Chia, K.~C. Tan, L.~Pawela, P.~Kurzynski, T.~Paterek, and D.~Kaszlikowski.
\newblock Coherent chemical kinetics as quantum walks. i. reaction operators for radical pairs.
\newblock {\em Physical Review E}, 93(3):032407, 2016.

\bibitem{chia2016coherent}
A~Chia, Agnieszka G{\'o}recka, P~Kurzy{\'n}ski, T~Paterek, and D~Kaszlikowski.
\newblock Coherent chemical kinetics as quantum walks. ii. radical-pair reactions in arabidopsis thaliana.
\newblock {\em Physical Review E}, 93(3):032408, 2016.

\bibitem{Innocenti_quantum_2017}
Luca Innocenti, Helena Majury, Taira Giordani, Nicolo Spagnolo, Fabio Sciarrino, Mauro Paternostro, and Alessandro Ferraro.
\newblock Quantum state engineering using one-dimensional discrete-time quantum walks.
\newblock {\em Physical Review A}, 96(6):062326, 2017.

\bibitem{Vieira_dynamically_2013}
Rafael Vieira, Edgard P.~M. Amorim, and Gustavo Rigolin.
\newblock Dynamically disordered quantum walk as a maximal entanglement generator.
\newblock {\em Physical Review Letters}, 111(18):180503, 2013.

\bibitem{Moulieras_entanglement_2013}
Simon Moulieras, Maciej Lewenstein, and Graciana Puentes.
\newblock Entanglement engineering and topological protection by discrete-time quantum walks.
\newblock {\em Journal of Physics B: Atomic, Molecular and Optical Physics}, 46(10, SI):104005, 2013.

\bibitem{bian_experimental_2014}
Zhihao Bian, Jian Li, Hao Qin, Xiang Zhan, and Peng Xue.
\newblock Experimental realization of a single qubit sic povm on via a one-dimensional photonic quantum walk, 2014.

\bibitem{zhao_experimental_2015}
Yuanyuan Zhao, Nengkun Yu, Pawel Kurzynski, Guoyong Xiang, ChuanFeng Li, and GuangCan Guo.
\newblock Experimental realization of generalized qubit measurements based on quantum walks.
\newblock {\em Physical Review A}, 91(4):042101, 2015.

\bibitem{li_new_2019}
HengJi Li, XiuBo Chen, YaLan Wang, YanYan Hou, and Jian Li.
\newblock A new kind of flexible quantum teleportation of an arbitrary multi-qubit state by multi-walker quantum walks.
\newblock {\em Quantum Information Processing}, 18(9):266, 2019.

\bibitem{yang_quantum_2018}
Yuguang Yang, Jiajie Yang, Yihua Zhou, Weimin Shi, Xiubo Chen, Jian Li, and Huijuan Zuo.
\newblock Quantum network communication: a discrete-time quantum-walk approach.
\newblock {\em Science China Information Sciences}, 61(4):042501, 2018.

\bibitem{chen_quantum_2019}
XiuBo Chen, YaLan Wang, Gang Xu, and YiXian Yang.
\newblock Quantum network communication with a novel discrete-time quantum walk.
\newblock {\em IEEE Access}, 7:13634--13642, 2019.

\bibitem{abd_encryption_2019}
Bassem Abd-El-Atty, Ahmed~A. Abd El-Latif, and Salvador~E. Venegas-Andraca.
\newblock An encryption protocol for neqr images based on one-particle quantum walks on a circle.
\newblock {\em Quantum Information Processing}, 18(9):272, 2019.

\bibitem{Abd_quantum_2020}
Ahmed~A. Abd El-Latif, Bassem Abd-El-Atty, Mohamed Amin, and Abdullah~M. Iliyasu.
\newblock Quantum-inspired cascaded discrete-time quantum walks with induced chaotic dynamics and cryptographic applications.
\newblock {\em Scientific Reports}, 10(1):1930, 2020.

\bibitem{Titchener_two_2016}
James~G. Titchener, Alexander~S. Solntsev, and Andrey~A. Sukhorukov.
\newblock Two-photon tomography using on-chip quantum walks.
\newblock {\em Optics Letters}, 41(17):4079--4082, 2016.

\bibitem{wang_generalized_2017}
Yu~Wang, Yun Shang, and Peng Xue.
\newblock Generalized teleportation by quantum walks.
\newblock {\em Quantum Information Processing}, 16(9):221, 2017.

\bibitem{novo_systematic_2015}
Leonardo Novo, Shantanav Chakraborty, Masoud Mohseni, Hartmut Neven, and Yasser Omar.
\newblock Systematic dimensionality reduction for quantum walks: optimal spatial search and transport on non-regular graphs.
\newblock {\em Scientific Reports}, 5:13304, 2015.

\bibitem{lu_quantum_2020}
Lu~Bai, Luca Rossi, Lixin Cui, Jian Cheng, and Edwin~R. Hancock.
\newblock A quantum-inspired similarity measure for the analysis of complete weighted graphs.
\newblock {\em IEEE Transactions on Cybernetics}, 50(3):1264--1277, 2020.

\bibitem{izaac2017centrality}
Josh~A Izaac, Xiang Zhan, Zhihao Bian, Kunkun Wang, Jian Li, Jingbo~B Wang, and Peng Xue.
\newblock Centrality measure based on continuous-time quantum walks and experimental realization.
\newblock {\em Physical Review A}, 95(3):032318, 2017.

\bibitem{Mukai_discrete_2020}
Kanae Mukai and Naomichi Hatano.
\newblock Discrete-time quantum walk on complex networks for community detection.
\newblock {\em Physical Review Research}, 2:023378, 2020.

\bibitem{Gratsea_generation_2020}
Aikaterini Gratsea, Maciej Lewenstein, and Alexandre Dauphin.
\newblock Generation of hybrid maximally entangled states in a one-dimensional quantum walk.
\newblock {\em Quantum Science and Technology}, 5(2):025002, 2020.

\bibitem{Rohde_entanglement_2012}
Peter~P. Rohde, Alessandro Fedrizzi, and Timothy~C. Ralph.
\newblock Entanglement dynamics and quasi-periodicity in discrete quantum walks.
\newblock {\em Journal of Modern Optics}, 59(8):710--720, 2012.

\bibitem{Kurzynski_quantum_2013}
Pawel Kurzynski and Antoni Wojcik.
\newblock Quantum walk as a generalized measuring device.
\newblock {\em Physical Review Letters}, 110(20):200404, 2013.

\bibitem{hou_deterministic_2018}
Zhibo Hou, JunFeng Tang, Jiangwei Shang, Huangjun Zhu, Jian Li, Yuan Yuan, KangDa Wu, GuoYong Xiang, Chuan-Feng Li, and Guang-Can Guo.
\newblock Deterministic realization of collective measurements via photonic quantum walks.
\newblock {\em Nature Communications}, 9:1414, 2018.

\bibitem{shang_quantum_2018}
Yun Shang, Yu~Wang, Meng Li, and Ruqian Lu.
\newblock Quantum communication protocols by quantum walks with two coins.
\newblock {\em Europhysics Letters}, 124(6):60009, 2019.

\bibitem{Srikara_quantum_2020}
S.~Srikara and C.~M. Chandrashekar.
\newblock Quantum direct communication protocols using discrete-time quantum walk.
\newblock {\em Quantum Information Processing}, 19(9):295, 2020.

\bibitem{panda_quantum_2023}
Sanjeet~Swaroop Panda, P.~A.~Ameen Yasir, and C.~M. Chandrashekar.
\newblock Quantum direct communication protocol using recurrence in k-cycle quantum walks.
\newblock {\em Physical Review A}, 107(2):022611, 2023.

\bibitem{Vlachou_quantum_2018}
Chrysoula Vlachou, Walter Krawec, Paulo Mateus, Nikola Paunkovic, and Andre Souto.
\newblock Quantum key distribution with quantum walks.
\newblock {\em Quantum Information Processing}, 17(11):288, 2018.

\bibitem{abd_optical_2021}
Bassem Abd-El-Atty, Abdullah~M. Iliyasu, Ahmad Alanezi, and Ahmed~A. Abd El-latif.
\newblock Optical image encryption based on quantum walks.
\newblock {\em Optics and Lasers in Engineering}, 138:106403, 2021.

\bibitem{su_robust_2022}
Yining Su and Xingyuan Wang.
\newblock A robust visual image encryption scheme based on controlled quantum walks.
\newblock {\em Physica A: Statistical Mechanics and its Applications}, 587:126529, 2022.

\bibitem{su_quantum_2023}
Yining Su and Xingyuan Wang.
\newblock Quantum color image encryption based on controlled two-particle quantum walks.
\newblock {\em Multimedia Tools and Applications}, 82:42679--42697, 2023.

\bibitem{abd_robust_2020}
Bassem Abd-El-Atty, Abdullah~M. Iliyasu, Haya Alaskar, and Ahmed~A. Abd El-Latif.
\newblock A robust quasi-quantum walks-based steganography protocol for secure transmission of images on cloud-based e-healthcare platforms.
\newblock {\em Sensors}, 20(11):3108, 2020.

\bibitem{abd_secret_2020}
Ahmed~A. Abd El-Latif, Bassem Abd-El-Atty, Sherif Elseuofi, Hany~S. Khalifa, Ahmed~S. Alghamdi, Kemal Polat, and Mohamed Amin.
\newblock Secret images transfer in cloud system based on investigating quantum walks in steganography approaches.
\newblock {\em Physica A: Statistical Mechanics and its Applications}, 541:123687, 2020.

\bibitem{abd_novel_2019}
Ahmed~A. Abd EL-Latif, Bassem Abd-El-Atty, and Salvador~E. Venegas-Andraca.
\newblock A novel image steganography technique based on quantum substitution boxes.
\newblock {\em Optics and Laser Technology}, 116:92--102, 2019.

\bibitem{abd_controlled_2020}
Ahmed~A. Abd EL-Latif, Bassem Abd-El-Atty, and Salvador~E. Venegas-Andraca.
\newblock Controlled alternate quantum walk-based pseudo-random number generator and its application to quantum color image encryption.
\newblock {\em Physica A: Statistical Mechanics and its Applications}, 547:123869, 2020.

\bibitem{Kaplan_quantum_2016}
Marc Kaplan, Ga\"{e}tan Leurent, Anthony Leverrier, and Mar{\'{\i} }a Naya-Plasencia.
\newblock Quantum differential and linear cryptanalysis.
\newblock {\em {IACR} Transactions on Symmetric Cryptology}, 1:71--94, 2016.

\bibitem{Chailloux_lattice_2021}
Andre Chailloux and Johanna Loyer.
\newblock Lattice sieving via quantum random walks.
\newblock In M~Tibouchi and H~Wang, editors, {\em Advances in Cryptology -- ASIACRYPT 2021}, volume 13093, pages 63--91. Springer International Publishing, 2021.

\bibitem{abd_novel_2023}
Bassem Abd-El-Atty, Mohammed ElAffendi, and Ahmed~A. Abd El-Latif.
\newblock A novel image cryptosystem using gray code, quantum walks, and henon map for cloud applications.
\newblock {\em Complex \& Intelligent Systems}, 9(1):609--624, 2023.

\bibitem{abd_quantum-inspired_2021}
Ahmed~A. Abd El-Latif, Bassem Abd-El-Atty, Irfan Mehmood, Khan Muhammad, Salvador~E. Venegas-Andraca, and Jialiang Peng.
\newblock Quantum-inspired blockchain-based cybersecurity: securing smart edge utilities in iot-based smart cities.
\newblock {\em Information Processing \& Management}, 58(4):102549, 2021.

\bibitem{abd_secure_2020}
Ahmed~A. Abd El-Latif, Bassem Abd-El-Atty, Wojciech Mazurczyk, Carol Fung, and Salvador~E. Venegas-Andraca.
\newblock Secure data encryption based on quantum walks for 5g internet of things scenario.
\newblock {\em IEEE Transactions on Network and Service Management}, 17(1):118--131, 2020.

\bibitem{Abd_efficient_2019}
Ahmed~A. Abd EL-Latif, Bassem Abd-El-Atty, Salvador~E. Venegas-Andraca, and Wojciech Mazurczyk.
\newblock Efficient quantum-based security protocols for information sharing and data protection in 5g networks.
\newblock {\em Future Generation Computer Systems-The International Journal of eScience}, 100:893--906, 2019.

\bibitem{bonnetain2023finding}
Xavier Bonnetain, Andr{\'e} Chailloux, Andr{\'e} Schrottenloher, and Yixin Shen.
\newblock Finding many collisions via reusable quantum walks.
\newblock In {\em Advances in Cryptology -- EUROCRYPT 2023}, pages 221--251, Cham, 2023. Springer Nature Switzerland.

\bibitem{aharonov2001quantum}
Dorit Aharonov, Andris Ambainis, Julia Kempe, and Umesh Vazirani.
\newblock Quantum walks on graphs.
\newblock In {\em Proceedings of the thirty-third annual ACM symposium on Theory of computing}, pages 50--59. Association for Computing Machinery, 2001.

\bibitem{Godsil_discrete_2019}
Chris Godsil and Hanmeng Zhan.
\newblock Discrete-time quantum walks and graph structures.
\newblock {\em Journal of Combinatorial Theory, Series A}, 167:181--212, 2019.

\bibitem{zhan_perfect_2014}
Xiang Zhan, Hao Qin, Zhi-hao Bian, Jian Li, and Peng Xue.
\newblock Perfect state transfer and efficient quantum routing: A discrete-time quantum-walk approach.
\newblock {\em Physical Review A}, 90(1):012331, 2014.

\bibitem{Janmark_global_2014}
Jonatan Janmark, David~A. Meyer, and Thomas~G. Wong.
\newblock Global symmetry is unnecessary for fast quantum search.
\newblock {\em Physical Review Letters}, 112(21):210502, 2014.

\bibitem{Wong2015Grover}
Thomas~G Wong.
\newblock Grover search with lackadaisical quantum walks.
\newblock {\em Journal of Physics A: Mathematical and Theoretical}, 48(43):435304, 2015.

\bibitem{wong_faster_2015}
Thomas~G. Wong.
\newblock Faster quantum walk search on a weighted graph.
\newblock {\em Physical Review A}, 92(3):032320, 2015.

\bibitem{wong_coined_2017}
Thomas~G. Wong.
\newblock Coined quantum walks on weighted graphs.
\newblock {\em Journal of Physics A: Mathematical and Theoretical}, 50(47):475301, 2017.

\bibitem{wang_optimal_2020}
Yunkai Wang, Shengjun Wu, and Wei Wang.
\newblock Optimal quantum search on truncated simplex lattices.
\newblock {\em Physical Review A}, 101(6):062333, 2020.

\bibitem{chakraborty2017optimal}
Shantanav Chakraborty, Leonardo Novo, Serena Di~Giorgio, and Yasser Omar.
\newblock Optimal quantum spatial search on random temporal networks.
\newblock {\em Physical Review Letters}, 119(22):220503, 2017.

\bibitem{Herrman_continuous_2019}
Rebekah Herrman and Travis~S. Humble.
\newblock Continuous-time quantum walks on dynamic graphs.
\newblock {\em Physical Review A}, 100(1):012306, 2019.

\bibitem{cattaneo2018quantum}
Marco Cattaneo, Matteo A.~C. Rossi, Matteo G.~A. Paris, and Sabrina Maniscalco.
\newblock Quantum spatial search on graphs subject to dynamical noise.
\newblock {\em Physical Review A}, 98(5):052347, 2018.

\bibitem{Teixeira_walking_2023}
Caue~F. Teixeira~da Silva, Daniel Posner, and Renato Portugal.
\newblock Walking on vertices and edges by continuous-time quantum walk.
\newblock {\em Quantum Information Processing}, 22(2):93, 2023.

\bibitem{santos2016szegedy}
Raqueline~AM Santos.
\newblock Szegedy's quantum walk with queries.
\newblock {\em Quantum Information Processing}, 15(11):4461--4475, 2016.

\bibitem{paparo2012google}
Giuseppe~Davide Paparo and MA~Martin-Delgado.
\newblock Google in a quantum network.
\newblock {\em Scientific Reports}, 2(1):1--12, 2012.

\bibitem{paparo2013quantum}
Giuseppe~Davide Paparo, Markus M{\"u}ller, Francesc Comellas, and Miguel~Angel Martin-Delgado.
\newblock Quantum google in a complex network.
\newblock {\em Scientific Reports}, 3(1):1--16, 2013.

\bibitem{izaac2017quantum}
JA~Izaac, JB~Wang, PC~Abbott, and XS~Ma.
\newblock Quantum centrality testing on directed graphs via p t-symmetric quantum walks.
\newblock {\em Physical Review A}, 96(3):032305, 2017.

\bibitem{Wang2022ContinuoustimeQW}
Yang Wang, Shichuan Xue, Junjie Wu, and Ping Xu.
\newblock Continuous-time quantum walk based centrality testing on weighted graphs.
\newblock {\em Scientific Reports}, 12, 2022.

\bibitem{wang2020experimental}
Kunkun Wang, Yuhao Shi, Lei Xiao, Jingbo Wang, Yogesh~N Joglekar, and Peng Xue.
\newblock Experimental realization of continuous-time quantum walks on directed graphs and their application in pagerank.
\newblock {\em Optica}, 7(11):1524--1530, 2020.

\bibitem{wu2020experimental}
Tong Wu, JA~Izaac, Zi-Xi Li, Kai Wang, ZhaoZhong Chen, Shining Zhu, Jingbo Wang, Xiao-Song Ma, et~al.
\newblock Experimental parity-time symmetric quantum walks for centrality ranking on directed graphs.
\newblock {\em Physical Review Letters}, 125(24):240501, 2020.

\bibitem{babai2016graph}
L{\'a}szl{\'o} Babai.
\newblock Graph isomorphism in quasipolynomial time.
\newblock In {\em Proceedings of the forty-eighth annual ACM symposium on Theory of Computing}, STOC '16, pages 684--697. Association for Computing Machinery, 2016.

\bibitem{shiau2005physically}
S-Y Shiau, R~Joynt, and SN~Coppersmith.
\newblock Physically-motivated dynamical algorithms for the graph isomorphism problem.
\newblock {\em Quantum Information and Computation}, 5(6):492--506, 2005.

\bibitem{emms2006matrix}
David Emms, Edwin~R Hancock, Simone Severini, and Richard~C Wilson.
\newblock A matrix representation of graphs and its spectrum as a graph invariant.
\newblock {\em The Electronic Journal of Combinatorics}, 13(1):R34, 2006.

\bibitem{emms2009graph-2}
David Emms, Richard~C Wilson, and Edwin~R Hancock.
\newblock Graph matching using the interference of discrete-time quantum walks.
\newblock {\em Image and Vision Computing}, 27(7):934--949, 2009.

\bibitem{emms2009graph}
David Emms, Richard~C Wilson, and Edwin~R Hancock.
\newblock Graph matching using the interference of continuous-time quantum walks.
\newblock {\em Pattern Recognition}, 42(5):985--1002, 2009.

\bibitem{qiang2012enhanced}
Xiaogang Qiang, Xuejun Yang, Junjie Wu, and Xuan Zhu.
\newblock An enhanced classical approach to graph isomorphism using continuous-time quantum walk.
\newblock {\em Journal of Physics A: Mathematical and Theoretical}, 45(4):045305, 2012.

\bibitem{smith2012algebraic}
Jamie Smith.
\newblock {\em Algebraic aspects of multi-particle quantum walks}.
\newblock PhD thesis, University of Waterloo, 2012.

\bibitem{wang2015graph}
Huiquan Wang, Junjie Wu, Xuejun Yang, and Xun Yi.
\newblock A graph isomorphism algorithm using signatures computed via quantum walk search model.
\newblock {\em Journal of Physics A: Mathematical and Theoretical}, 48(11):115302, 2015.

\bibitem{bai2015quantum}
Lu~Bai, Luca Rossi, Peng Ren, Zhihong Zhang, and Edwin~R Hancock.
\newblock A quantum jensen-shannon graph kernel using discrete-time quantum walks.
\newblock In {\em Graph-Based Representations in Pattern Recognition}, GbRPR 2015, pages 252--261. Springer International Publishing, 2015.

\bibitem{lamberti2008metric}
Pedro~W Lamberti, Ana~P Majtey, Antoni Borras, Montserrat Casas, and Angel Plastino.
\newblock Metric character of the quantum jensen-shannon divergence.
\newblock {\em Physical Review A}, 77(5):052311, 2008.

\bibitem{zhang2020r}
Yi~Zhang, Lulu Wang, Richard~C Wilson, and Kai Liu.
\newblock An r-convolution graph kernel based on fast discrete-time quantum walk.
\newblock {\em IEEE transactions on neural networks and learning systems}, 33(1):292--303, 2020.

\bibitem{bai2013quantum}
Lu~Bai, Edwin~R Hancock, Andrea Torsello, and Luca Rossi.
\newblock A quantum jensen-shannon graph kernel using the continuous-time quantum walk.
\newblock In {\em Graph-Based Representations in Pattern Recognition}, GbRPR 2013, pages 121--131. Springer International Publishing, 2013.

\bibitem{zhang_quantum_2021}
Qizi Zhang and Jerome Busemeyer.
\newblock A quantum walk model for idea propagation in social network and group decision making.
\newblock {\em Entropy}, 23(5):622, 2021.

\bibitem{yan_information_2022}
Fei Yan, Wen Liang, and Kaoru Hirota.
\newblock An information propagation model for social networks based on continuous-time quantum walk.
\newblock {\em Neural Computing \& Applications}, 34(16, SI):13455--13468, 2022.

\bibitem{ficara_classical_nodate}
Annamaria Ficara, Giacomo Fiumara, Pasquale De~Meo, and Salvatore Catanese.
\newblock Classical and quantum random walks to identify leaders in criminal networks.
\newblock In {\em Complex Networks and Their Applications XI}, pages 190--201, Cham, 2023. Springer International Publishing.

\bibitem{hu_analyzing_2022}
Xu~Hu, Xiaoyu Niu, Lingxin Qian, Binghuang Pan, and Zhaoyuan Yu.
\newblock Analyzing the multi-scale characteristic for online car-hailing traffic volume with quantum walk.
\newblock {\em IET Intelligent Transport Systems}, 16(10):1328--1341, 2022.

\bibitem{singh_localization_2022}
Om~Mehta and Seema Mahajan.
\newblock Localization of sensor node by novel quantum walk-pathfinding rider optimization (qwpro) by mobile anchor node.
\newblock In {\em Futuristic Trends in Networks and Computing Technologies}, pages 141--164, Singapore, 2022. Springer Nature Singapore.

\end{thebibliography}

\end{document}